\documentstyle[12pt]{article}

\def\bs{\begin{subequations}}
\def\es{\end{subequations}} 

\def\dd{_{\textbf{\raisebox{.3ex}{.}}}} 
\def\dj{_{\textbf{\raisebox{.5ex}{.}}}} 
\def\dqa{_{\textbf{\raisebox{.3ex}{.}} \alpha}} 
\def\dqb{_{\textbf{\raisebox{.3ex}{.}} \beta}}
\def\dqi{_{\textbf{\raisebox{.3ex}{.}} i}}

\catcode`\@=11

\newtoks\@stequation
\def\subequations{\refstepcounter{equation}
  \edef\@savedequation{\the\c@equation}%
  \@stequation=\expandafter{\theequation}
  \edef\@savedtheequation{\the\@stequation}
  \edef\oldtheequation{\theequation}%
  \setcounter{equation}{0}%
  \def\theequation{\oldtheequation\alph{equation}}}

\def\endsubequations{\setcounter{equation}{\@savedequation}%
  \@stequation=\expandafter{\@savedtheequation}%
  \edef\theequation{\the\@stequation}\global\@ignoretrue}

\catcode`\@=12

\catcode`\@=11
\def\vereq#1#2{\lower3pt\vbox{\baselineskip1.5pt \lineskip1.5pt
\ialign{$\m@th#1\hfill##\hfil$\crcr#2\crcr\sim\crcr}}}
\catcode`\@=12

\makeatletter
        \renewcommand{\theequation}{\thesection.\arabic{equation}}%
        \@addtoreset{equation}{section}%
\makeatother

\renewcommand{\thefootnote}{\fnsymbol{footnote}}

\begin{document}
\begin{titlepage}
\begin{center}
September 28, 1998              \hfill    UCB-PTH-98/42 \\
                                   \hfill    LBNL-42212 \\
                                \hfill    hep-th/9809197 \\

\vskip .25in

{\large \bf The Algebras of Large N Matrix Mechanics \\}

\vskip 0.3in

M.B. Halpern$^{a,b}$\footnote{E-mail: halpern@physics.berkeley.edu}
 and C. Schwartz$^a$

\vskip 0.15in

$^a${\em Department of Physics,
     University of California\\
     Berkeley, California 94720}\\
and\\
$^b${\em Theoretical Physics Group\\
     Ernest Orlando Lawrence Berkeley National Laboratory\\
     University of California,
     Berkeley, California 94720}
        
\end{center}

\vskip .3in

\vfill

\begin{abstract}
Extending early work, we formulate the large N matrix mechanics of 
general bosonic, fermionic and supersymmetric matrix models, including
Matrix theory: The Hamiltonian framework of large N matrix mechanics
provides a natural setting in which to study the algebras of the
large N limit, including (reduced)
Lie algebras, (reduced) supersymmetry algebras and free algebras. We find
in particular a broad array of new free algebras which we call
 symmetric Cuntz algebras, interacting symmetric Cuntz algebras,
symmetric Bose/Fermi/Cuntz algebras and symmetric Cuntz superalgebras, and
we discuss the role of these algebras in solving the large N theory. 
Most important, the interacting Cuntz algebras are associated to a set 
of new (hidden!) local quantities which are generically conserved only 
at large N. A number of other new large N phenomena are also observed, 
including the intrinsic nonlocality of the (reduced) trace class  
operators of the theory and a closely related large N field identification 
phenomenon which is  associated to another set (this time nonlocal) of new
 conserved quantities at large N.
\end{abstract}

\vfill

\end{titlepage}

\renewcommand{\thefootnote}{\arabic{footnote}}
\setcounter{footnote}{0}
\renewcommand{\thepage}{\arabic{page}}
\setcounter{page}{1}

\section{Introduction}

Studies of the large N limit of matrix models have included many
intertwined directions, among which we mention:
\vspace{12pt}
\\$\bullet$   Planar diagram summation \cite{tHooft}
\\$\bullet$   Integration~[2-5] 
\\$\bullet$   Schwinger-Dyson methods~[6-11] 
\\$\bullet$   Euclidean master fields \cite{Haan, Gopa}
\\$\bullet$   Solution of the Schrodinger equation \cite{Brezin, Marc}
\\$\bullet$   Phase space master fields~[14-20] 
\\$\bullet$   Large N matrix mechanics~[17-19] 
\\$\bullet$   Reduced integral and stochastic formulations~[21-25]
\\$\bullet$   Stochastic master fields \cite{Green, Doug2}
\\$\bullet$   Microcanonical master fields \cite{Halp1}
\\$\bullet$   Free algebras [7, 16-19, 8, 28, 12, 11, 26, 20, 29, 30].
         
\vspace{12pt}         
\noindent Other references and approaches can be found in the partial reviews of
Refs.~[31-34]. The independent observations in physics
~[7, 16-19, 8] of free or 
Cuntz algebras in the large N limit are also intertwined chronologically
with the  development of these algebras in mathematics~[35-39].

In this paper, we focus on large N matrix mechanics~[17-19], which 
was originally introduced to systematize closely related ongoing work on
phase space master fields~[14-16]. The approach 
through matrix mechanics 
was interrupted,  however, in the early 1980's after the solution of the
one hermitian matrix model \cite{me81,us} and the (one polygon) unitary matrix 
model \cite{me82}. 

Extending this early work,  we study here the large N matrix mechanics
of general bosonic, fermionic and supersymmetric matrix models, including
gauged matrix models such as the n=16 supersymmetric gauge quantum 
mechanics \cite{Claud}, now called Matrix theory \cite{Bank}. Because 
it is a Hamiltonian approach, large N matrix mechanics is an ideal laboratory 
for studying the algebras of the large N limit, including (reduced) Lie algebras,
(reduced) supersymmetry 
algebras and free algebras. In particular, we will find a broad array
of new  free algebras in the large N limit, and we will discuss
the role of these algebras in solving the various theories. To aid the reader 
in making the transition from the early work,
we give here a brief review of the approach.

Large N matrix mechanics, 
which follows Heisenberg's original development \cite{Heis}, is based on 
the large N completeness relation \cite{me81}
\begin{equation}
\textbf{1}\dd\; \stackrel{_{\textstyle =}}{_{_{N}}}  \;\; \mid 0\dd \rangle
 \langle  \dd 0 \mid
 + \sum_{rs,A} \mid 
rs,A \rangle \langle  rs,A \mid  , \;\;\;\;\;\; r,s = 1 . . .  N
\end{equation}
for the states which saturate the traced  Wightman functions of the theory.
Here $\mid 0\dd \rangle$ is the
ground state of the theory (which dominates the invariant channels by
large N factorization) and $\{ \mid rs,A \rangle \}$ is a set of 
dominant adjoint
eigenstates of the Hamiltonian. The large N dynamics is then formulated
in terms of reduced states and operators, using Bardakci's reduced matrix
elements \cite{Bard}. For example, the reduced completeness relation reads  
\begin{equation}
\textbf{1} =\; \mid 0 \rangle  \langle  0 \mid + \sum_{A} \mid A 
\rangle \langle  A \mid  \label{ia}
\end{equation}
where  $\mid 0 \rangle$ is the reduced ground state and the states 
 $\{ \mid A \rangle \}$ are the corresponding reduced adjoint eigenstates.

The master fields \cite{Witt} of the theory are the set of reduced
matrix elements of the reduced fields, and have the translation-covariant
form \cite{me81} 
\bs 
\begin{equation}
M_{\mu \nu} (x) = exp(i  p_{\mu \nu} \cdot x)  M_{\mu \nu} (0)
\end{equation}
\begin{equation} 
 p_{\mu \nu} =  p_{\mu} -  p_{\nu}, \;\;\;\;\;  
\mu = (0,A) , \; \; \; \; \;  \nu = (0,B)
\end{equation}
\es 
where $\{ p_{\mu \nu} \}$ is the set of energy-momentum differences of the 
various
reduced states in (\ref{ia}). The master fields also satisfy the 
large N classical
equations of motion and a set of equal-time ``constraints''[15-19], which
follow directly by taking matrix elements of the reduced equations of
motion and the reduced equal-time algebra of the theory [16-19]. In the 1-matrix
model, for example, the reduced equal-time algebra takes the
``semiclassical'' form [7, 16-18] 
\begin{equation}
[ \phi  ,  \pi  ] = i  \mid 0 \rangle  \langle  0 \mid \label{ib}
\end{equation}
for any potential. More generally, this gives the large N correspondences:
\vspace{12pt}
\\$\bullet$  master fields $\leftrightarrow$ reduced fields
\\$\bullet$  large N classical equations of motion $\leftrightarrow$ reduced 
equations of motion
\\$\bullet$  equal-time constraints $\leftrightarrow$ reduced equal-time algebra. 

\vspace{12pt}
\noindent In this paper, we shall prefer the equivalent terminology on the right
side of this list.

\vspace{12pt}
Here is an overview of our main conclusions:

\vspace{12pt}
\noindent 1) Unified formulation. \\
Large N matrix mechanics provides a unified large N Hamiltonian 
formulation of bosonic (see Secs.~2, 3 and 4), fermionic (see Secs.~2, 5 
 and 6) and supersymmetric matrix models (see Secs.~5 and 6), as well as 
gauged matrix models (see Secs.~2 and 6): The special case of Matrix 
theory \cite{Claud, Bank} is discussed explicitly in Sec.~6.  The 
algebraic structures discussed below can be straightforwardly 
generalized to higher dimensional large N Hamiltonian quantum field 
theory~\cite{Bard, me81}, and we expect that the same algebraic structures can 
also be found in large N Euclidean quantum field theory~\cite{Haan, Gopa}.

\vspace{12pt}
\noindent 2) Generalized free algebras. \\
The reduced large N theories come equipped with
their own reduced equal-time algebras (see Subsec.~2.5), which generalize
Eq.~(\ref{ib}). These equal-time algebras are new free algebras in their own
 right, and, with the help
of the reduced equations of motion, one sees that the equal-time algebras 
are closely related to the Cuntz algebra\footnote{More precisely, the 
algebra (\ref{km}), which appeared independently in Refs.~\cite{ExC, Haan}, 
 is called the extended Cuntz algebra in mathematics. A Kronecker-delta  
 realization of the one-dimensional algebra was also seen 
 independently in large N matrix mechanics~\cite{me81}.} 
\begin{equation}
a_{m} a^{\dagger}_{n} = \delta_{mn}, \;\;\;\;\;\; a_{m} \mid 0 \rangle = 0,
 \label{km} \label{kr}
 \;\;\;\;\;\; 
a^{\dagger}_{m} a_{m} = 1 - \mid 0 \rangle \langle 0 \mid 
\end{equation}
and generalizations thereof. The Cuntz algebra (\ref{km}) arises in the special
case of large N bosonic oscillators (see Subsec.~3.1), while other cases
show a broad array of generalizations of the Cuntz algebra, which we call:
\vspace{12pt}
\\$\bullet$   symmetric Cuntz algebras (Subsec.~3.1; see Eq.~(\ref{sg}))
\\$\bullet$   interacting symmetric Cuntz algebras (Sec.~4; see Eq.~(\ref{sl}))
\\$\bullet$   symmetric Bose/Fermi/Cuntz algebras (Subsec.~5.1; see 
Eq.~(\ref{sj}))
\\$\bullet$   symmetric Cuntz superalgebras (Subsec.~5.2; see 
Eqs.~(\ref{cz}) and (\ref{sk})). 

\vspace{12pt}
\noindent Symmetric Cuntz algebras contain two Cuntz subalgebras 
((\ref{km}) and a tilde version of (\ref{km})) which 
act respectively at the beginning or the end of large N words (see, for 
example, Eq.~(\ref{kq})). We  remark in particular  on the interacting 
Cuntz algebras,
\begin{equation}
A_{m} A^{\dagger}_{n} = C_{mn}(\phi), \;\;\;\;\;\; A_{m} \mid 0 \rangle = 0,
 \;\;\;\;\;\;
A^{\dagger}_{m}(C^{-1}(\phi))_{mn} A_{n} = 1 - \mid 0 \rangle \langle 0 \mid 
\label{ky}
\end{equation}
where the reduced operator $C_{mn}(\phi)$, which is a function only 
of the reduced coordinates $\phi$,  is determined by the potential. 

The interacting Cuntz algebras are a central result of this paper, 
in part because they imply  a number of new local 
 conserved quantities at large N (see Subsec.~4.5), including
\begin{equation}
{\cal J} = A^{\dagger}_{m} (C^{-1}(\phi))_{mn} A_{n}, \;\;\;\;\;\;
 \frac{d}{dt} {\cal J} = {\cal J} \mid 0 \rangle = 0 
\end{equation}
which follows directly from (\ref{ky}).  In the original unreduced theory, 
these  quantities correspond to new (hidden) local  but 
nonpolynomial operators which are generically conserved only at large N.

\vspace{12pt}
\noindent 3) Conserved nonlocal reduced operators. \\
The local conserved  trace class
operators of the theory, such as the Hamiltonian, the angular momenta  
and the supercharges, are represented at large N by reduced
conserved operators called the reduced Hamiltonian, the reduced 
angular momenta and the reduced supercharges. 
These reduced operators satisfy reduced algebras (see Subsecs.~2.3, 
2.6, 3.3, 5.3, 5.4 and Sec.~6)
which are closely related to the unreduced algebras of the theory. As an
example, the reduced Hamiltonian still controls, in the
normal fashion, the time evolution of all reduced operators (see
Subsec.~2.6), although the form of the 
reduced supersymmetry algebras can be surprisingly different from 
their unreduced form in the case
of a gauge theory such as Matrix theory (see Eq.~(\ref{kp})). 

The explicit composite forms of the reduced trace class operators can in
 principle be 
determined by solving their reduced algebraic relations, and this is one of 
the central problems of large N matrix mechanics.
The generalized free algebras above are seen to play an
important role in the construction of these reduced operators.

What is  most interesting here is that
 the reduced trace class operators are intrinsically nonlocal 
(see Subsecs 3.2, 3.3, 4.6, 5.3 and 5.5). Early examples of this 
general phenomenon were seen in Refs.~\cite{us, me82}.
  
\vspace{12pt}
\noindent 4) Large N field identification. \\
Because each  reduced trace class operator $T$ (corresponding to a 
conserved local trace
class operator $T\dd$) is nonlocal, we find that there exists,   
universally for each $T\dd$,  another nonlocal operator $D_{rs}$ in the 
unreduced theory which also corresponds
at large N to the same nonlocal reduced operator
\bs 
\begin{equation}
\left. \begin{array}{c} T\dd\;(local) \\ D_{rs}\;(nonlocal) \end{array}
\right\rangle \begin{array}{c} \; \\ {\longrightarrow} \\ ^{N} \end{array} 
T \; ( nonlocal )
\end{equation}
\begin{equation}
\frac{d}{dt} T\dd = \frac{d}{dt} T = 0, \;\;\;\;\;\; \frac{d}{dt} 
D_{rs} \;\stackrel{_{\textstyle =}}{_{_{N}}} \; 0
\end{equation}
\es
(see Subsecs.~3.5, 4.6, 5.3 and 5.5).  The new large N-conserved 
nonlocal operators $D_{rs}$ are closely related to the densities of the original
local trace class operators $T\dd$, and provide us in principle with 
another class of unreduced operators which are generically conserved 
only at large N. 

\vspace{12pt}
\noindent 5) Large N fermions and bosons. \\
Large N fermions and bosons are surprisingly similar, exhibiting some
aspects of a Bose-Fermi equivalence (see Subsecs.~5.2, 5.3 and 5.5). This
equivalence, which is explicit in the Cuntz superalgebras above, is another
example of the classical nature of the large N limit. In particular,
large N fermions and bosons both satisfy the same classical or Boltzmann
statistics, and the Pauli principle is lost for large N fermions. The
equivalence also makes possible certain large N bosonic constructions
of supersymmetry (see Subsec.~5.4).

The interacting Cuntz algebras have not yet been extended to matrix 
models with fermions, although we believe that they can be. Further 
study in this direction is particularly important for Matrix theory, 
where the associated new large N-conserved quantities (local and 
nonlocal) may be related 
to the question of hidden 11-dimensional symmetry~\cite{Bank}.

\section{The Setup}

\subsection{SU(N)-Invariant Hamiltonian Systems}

In this section, we  establish our notation for a large class of 
SU(N)-invariant  matrix Hamiltonian systems, where the symmetry can be 
global or local.  In the course of this discussion, we will 
often loosely refer to the group SU(N) as the gauge group of the theory,  
whether the symmetry is gauged or not. 

We begin  with a canonical set of  B hermitian  adjoint bosons and f hermitian 
adjoint fermions
\bs
\begin{equation}
[  \phi^{m}_{a}  ,  \pi^{n}_{b}  ] = i  \delta_{ab} 
\delta^{mn} ,\; \; \; \; \; \;     [  \Lambda_{\alpha a}  ,  
\Lambda_{\beta b}  ]_{+} = \delta_{ab}  \delta_{\alpha \beta}
\end{equation}
\begin{equation}
\phi ^{m \dagger}_{a}
  = \phi ^{m}_{a} , \; \; \; \; \; \;   \pi ^{m \dagger}_{a}
  = \pi ^{m}_{a} , \; \; \; \; \; \;   \Lambda _{\alpha a} 
  ^{\dagger}  = \Lambda _{\alpha a}  
\end{equation}
\begin{equation}
a = 1. . .  N^{2} , \; \; \; \; \; \;     m=1. . .  B ,  
\; \; \; \; \; \;  \alpha = 1. . .  f.
\end{equation}
\es 
 The generators of SU(N), 
sometimes called the gauge generators,  are
\begin{equation}
G_{a} = G^{\dagger}_{a} = f_{abc}  (\phi^{m}_{b}  \pi^{m}_{c}   -\frac{i}{2}  
\Lambda_{\alpha b}  \Lambda_{\alpha c}  ) .
\end{equation}
The dynamics of the system is described by an invariant
 Hamiltonian  $H\dd$,\footnote{This $H\dd$ is the ordinary Hamiltonian of the 
 system in the Heisenberg picture; the purpose of the dot subscript is to 
 distinguish certain objects in the original, unreduced theory from their reduced 
 counterparts.}
\bs
\begin{equation}
\dot{\rho} = i [H\dd, \rho ] , \;\;\;\;\;\; \rho = \phi,\; \pi\; or\; 
\Lambda
\end{equation}
\begin{equation}
[ G_{a}, H\dd ] = 0 
\end{equation}
\es 
which  is constructed from the canonical operators.

To go over to a matrix notation, we also introduce a set of  NxN matrices 
in the fundamental representation of the gauge group\footnote{The 
normalization in (\ref{te}) corresponds to $\alpha^{2} = 2$ for any root 
$\alpha$ of SU(N).}
\bs \label{te}
\begin{equation}
[T_{a},T_{b}] = i f_{abc} T_{c}, \; \; \; \; \; \; T^{\dagger}_{a} = T_{a},  
  \; \; \; \; \; \;   Tr \, T_{a}  T_{b} = \delta_{ab} \label{z3}
\end{equation}
\begin{equation}
Tr \, T_{a} = \sqrt N  \delta_{a, N^{2}}, \;\;\;\;\;\;
(T_{a})_{rs}  (T_{a})_{uv} = \delta_{su}  \delta_{rv}, \;\;\;\;\;\;
 r,s = 1 \ldots N
\end{equation}
\es 
and define the adjoint  matrix fields as
\bs 
\begin{equation}
\phi^{m} = \phi^{m}_{a}  T_{a} , \; \; \; \; \; \;    \pi^{m} = 
\pi^{m}_{a}  T_{a} , \; \; \; \; \; \;    \Lambda_{\alpha} = 
\Lambda_{\alpha a}  T_{a} 
\end{equation}
\begin{equation}
(\phi ^{m}_{rs})^{\dagger}
  = \phi ^{m}_{sr} , \; \; \; \; \; \;   (\pi ^{m}_{rs})^{\dagger}
  = \pi ^{m}_{sr} , \; \; \; \; \; \;   ((\Lambda _{\alpha}  )_{rs}
  )^{\dagger}  = (\Lambda _{\alpha}  )_{sr}
\end{equation}
\begin{equation}
[ \phi ^{m}_{rs}  ,  \pi ^{n}_{uv}  ] = i  \delta^{mn} 
\delta_{su}  \delta_{rv}, \;\;\;\;\;\; 
[  (\Lambda_{\alpha})_{rs}  ,  (\Lambda_{\beta})_{uv}  ]_{+}
 =  \delta_{\alpha \beta}  \delta_{su}  \delta_{rv} .
\end{equation}
\es
The corresponding form of the  gauge generators is 
\bs \label{st}
\begin{equation}
G_{rs} = G_{a} (T_{a})_{rs} = (-i  [\phi^{m}  ,  \pi^{m}  ] -  
\Lambda_{\alpha}  \Lambda_{\alpha}  +  (F-B)N )_{rs} \label{o}
\end{equation}
\begin{equation}
G^{\dagger}_{rs} = G_{sr}, \;\;\;\;\;\;Tr \, G = 0 , \; \; \; \; \; \;   
 F = \frac{f}{2}
\end{equation}
\begin{equation}
[G_{rs}, H\dd] = 0 . \label{lj}
\end{equation}
\es 
See Sec.~6 for the corresponding forms when the matrix fields are 
traceless.

We consider next the traced Wightman functions
\begin{equation}
\langle \dd 0 \mid Tr(\rho_{1}(t_{1}) \ldots \rho_{n}(t_{n}))\mid 0\dd 
\rangle, \;\;\;\;\;\; \rho = \phi, \; \pi \; or \; \Lambda
\end{equation}
where $\mid 0\dd \rangle$ is the vacuum or ground state of the theory 
(see below for the case of degenerate ground states).  For gauged 
matrix models, these are the invariant  Wightman functions in the 
temporal gauge, where the missing factors $T \textup{exp}(i \int dt A_{0}(t))$ 
are unity. 

The channels of the traced Wightman functions are defined as 
\begin{equation}
\langle \dd 0 \mid (\rho_{1}(t_{1}) \ldots \rho_{i}(t_{i}))_{rs} \;\; 
(\rho_{i+1}(t_{i+1}) \ldots \rho_{n}(t_{n}))_{sr}\mid 0\dd \rangle
\end{equation}
and the subset of Hamiltonian eigenstates which saturate the channels 
  span a Hilbert space. At finite N,  these 
states are the set of all invariant  and adjoint eigenstates.  The Hilbert 
space simplifies, however,  at large N because large N factorization 
tells us that  the  ground state $\mid 0\dd \rangle$ dominates among 
the invariant states. Moreover, a certain dynamically-determined  subset 
of the adjoint states \mbox{$\mid rs,A \rangle$} may dominate at large N.  This 
situation is summarized by the completeness relation~\cite{me81}
\begin{equation}
\textbf{1}\dd \;\stackrel{_{\textstyle =}}{_{_{N}}} \;\; \mid 0\dd \rangle \langle  
\dd 0 \mid + \sum_{rs,A} \mid rs,A \rangle \langle  rs,A \mid  \label{b}
\end{equation}
for the large N traced Wightman functions. In further detail, we may 
specify the properties of these time-independent states as
\bs \label{td}
\begin{equation}
H\dd  \mid  0\dd  \rangle = E_{0}  \mid  0\dd  \rangle , \; \; \; \; \; \;     
H\dd   \mid  rs,A  \rangle = E_{A}  \mid  rs,A  \rangle 
\end{equation}
\begin{equation}
E_{0} = O(N^{2}), \;\;\;\;\;\; E_{\mu} - E_{\nu} = O(N^{0}), \;\;\;\;\;\; 
\mu = (0,A), \;\;\; \nu = (0,B)
\end{equation}
\begin{equation}
G_{rs}  \mid  0\dd  \rangle = 0     \label{p}, \;\;\;\;\;\;    
G_{rs}  \mid  pq,A  \rangle =  \delta_{rq}  \mid  ps,A  \rangle
 - \delta_{sp}  \mid  rq,A  \rangle \label{q}
\end{equation}
\begin{equation}
\mid  rr,A  \rangle = 0 , \; \; \; \; \; \;  \langle  rs,A  
\mid  0\dd  \rangle = 0 , \;\;\;\;\;\; 
\langle  rs,A  \mid pq,B \rangle = P_{sr,pq}  \delta_{AB}  
\end{equation}
\begin{equation}
P_{sr,pq} = \delta_{rp}  \delta_{sq}  -  \frac{1}{N}  \delta_{sr} 
 \delta_{pq}, \;\;\;\;\;\; 
P_{rs,pq}  P_{qp,tu} = P_{rs,tu}  \label{a}
\end{equation}
\es 
where $P$ in  (\ref{a}) is a projector.  In the large N matrix 
mechanics of gauge theories, the ground state $\mid 0\dd \rangle$ is 
the only state which satisfies the Gauss law in (\ref{p}), although the 
adjoint eigenstates $\mid rs, A \rangle$ are needed \cite{Bard, me82} 
as well to saturate the channels of the invariant Wightman functions.

For bosonic systems, one expects that the ground state $\mid 0\dd 
\rangle$ is unique, but for  systems with fermions the 
completeness relation (\ref{b}) should generally include the sum over  a set 
of possibly degenerate ground states $\{\mid 0\dd \rangle _{i}\}$,
\bs \label{ss}
\begin{equation}
\textbf{1}\dd\; \stackrel{_{\textstyle =}}{_{_{N}}}  \;\; \mid 0\dd \rangle_{i} 
\; _{i}\langle  \dd 0 \mid
 + \sum_{rs,A} \mid 
rs,A \rangle \langle  rs,A \mid  \label{c}
\end{equation}
\begin{equation}
G_{rs} \mid 0\dd \rangle_{i} = 0, \;\;\;\;\;\; 
H\dd \mid 0\dd \rangle _{i} = E_{0} \mid 0\dd \rangle _{i}, \;\;\;\;\;\;
 _{i}\langle \dd 0 \mid 0\dd \rangle _{j} = \delta_{ij}
 \end{equation}
\es 
to be dynamically determined as well. For simplicity  we will continue
to treat the ground state as unique, and the explicit examples 
of this paper are limited to cases where the ground state is unique or 
is believed to be unique, as in  Matrix theory. Our discussion below 
goes through as well, however, for degenerate vacua, and  the reader 
can obtain the corresponding results by appending a subscript $i$ to each vacuum 
state with the summation convention of (\ref{c}).

\subsection{Reduced Formulation}

The large N theory can be reformulated in terms of  reduced operators 
which act in a reduced Hilbert space~[16-19].  The reduced ground
 state $\mid 0 
\rangle$ and the dominant reduced adjoint eigenstates $\mid A \rangle$ are in 
 correspondence with the true ground state and dominant adjoint 
eigenstates $\mid 0\dd \rangle$ and $\mid rs,A \rangle$. These  
 time-independent states satisfy the reduced completeness relation 
\begin{equation} \label{kn}
{\textbf 1} = \sum_{\mu = (0,A)} \; \mid \mu \rangle \langle \mu \mid \; =
 \; \mid 0 \rangle \langle 0 \mid + \sum_{A} \;  \mid A \rangle \langle A \mid
\end{equation}
in the reduced Hilbert space. The reduced quantities are related to 
the original, unreduced quantities of the theory via reduced matrix elements 
which we illustrate first for the bosonic fields $\phi$:
\bs \label{sa}
\begin{equation} 
\langle \dd 0  \mid \frac{\phi^{m}_{rs}(t)}{\sqrt N}  \mid 0\dd 
\rangle = \delta_{rs}  \langle 0  \mid \phi_{m}(t)  \mid 0 
\rangle  \label{d}
\end{equation}
\begin{equation}
\langle \dd 0  \mid \frac{\phi^{m}_{rs}}{\sqrt N}  \mid pq,A 
\rangle =  f(N) P_{rs,pq}  \langle 0  \mid \phi_{m}  
\mid A \rangle
\end{equation}
\begin{equation}
\langle pq,A  \mid \frac{\phi^{m}_{rs}}{\sqrt N}  \mid 0\dd
\rangle =  f(N) P_{qp,rs}  \langle A  \mid \phi_{m}  
\mid 0 \rangle
\end{equation}
\begin{equation}
\langle pq,A  \mid \frac{\phi^{m}_{rs}}{\sqrt N}  \mid 
st,B \rangle = \langle sq,A  \mid \frac{\phi^{m}_{sp}}
{\sqrt N}  \mid rt,B \rangle =
P_{qp,rt}   \langle A  \mid \phi_{m}  \mid B \rangle  \label{e}
\end{equation}
\begin{equation}
\langle pq,A  \mid \frac{\phi^{m}_{rs}}{\sqrt N}  \mid 
tr,B \rangle = \langle pr,A  \mid \frac{\phi^{m}_{qr}}
{\sqrt N}  \mid ts,B \rangle =
P_{qp,ts}   \langle A  \mid \tilde{\phi}_{m}  \mid B \rangle  \label{f}
\end{equation}
\begin{equation}
\tilde{\phi}_{m} \mid 0 \rangle \equiv \phi_{m} \mid 0 \rangle
   , \; \; \; \; \; \;     \langle 0  \mid \tilde{\phi}_{m}
  \equiv  \langle 0  \mid \phi_{m}  \label{g}
\end{equation}
\begin{equation}
f(N) = (N-\frac{1}{N})^{-\frac{1}{2}} 
\end{equation}
\es
where $P$ is the projector defined in (\ref{a}).
All the operators above are evaluated at time t, 
although we have written this explicitly only in (\ref{d}).
The same definitions apply for $\pi$ (take time derivatives of all 
definitions in (\ref{sa})) and also for $\Lambda$,  which defines a map from 
the original operators to the reduced operators
\begin{equation}
\frac{\phi^{m}_{rs}}{\sqrt N}, \;\; \frac{\pi^{m}_{rs}}{\sqrt N} , \;\; 
\frac{(\Lambda_{\alpha})_{rs}}{\sqrt N} \;\;
\longrightarrow \;\; \left\{
\begin{array}{l}
\phi_{m}, \;\; \pi_{m}, \;\; \Lambda_{\alpha} \\
\tilde{\phi}_{m}, \;\; \tilde{\pi}_{m}, \;\;\tilde{\Lambda}_{\alpha} .
\end{array} \right.
\end{equation}
It follows for example that
\bs 
\begin{equation}
\rho^{\dagger} = \rho, \;\;\;\;\;\; \tilde{\rho}^{\dagger} = \tilde{\rho}
, \;\;\;\;\;\; \rho = \phi, \; \pi, \; or \; \Lambda 
\end{equation}
\begin{equation}
\tilde{\rho}\mid 0 \rangle = \rho \mid 0 \rangle, \;\;\;\;\;\;  
\langle 0 \mid \tilde{\rho} = \langle 0 \mid \rho, \;\;\;\;\;\;  
\dot{\tilde{\rho}} \mid 0 \rangle = \dot{\rho} \mid 0 \rangle. \label{x}
\end{equation}
\es
The reduced matrix elements 
(2.13a-d) were introduced by Bardakci in the first paper of Ref.~[16], 
and further studied in  
 Refs.~[16-18],  but  the 
reduced matrix elements which define the reduced  tilde 
operators $\tilde{\rho}$ in 
(\ref{f}), (\ref{g}) are new and the new tilde operators 
will play a central role in this paper.

According to (\ref{e}), (\ref{f}),  the existence of two distinct 
reduced operators $\rho$ and $\tilde{\rho}$ for each unreduced $\rho_{rs}$ 
corresponds to the presence of a symmetric and an antisymmetric adjoint 
representation 
\begin{equation}
(\textup{adjoint}) \otimes (\textup{adjoint}) = (\textup{singlet}) \oplus 
(\textup{adjoint}) \oplus (\textup{adjoint})' \oplus \ldots \; \; 
\end{equation}
in the product of two adjoint representations of SU(N). 
A related interpretation of the tilde operators is noted in App.~A.

We consider next the evaluation of  matrix elements of matrix 
products at equal time. Using the matrix elements of the canonical operators 
and the 
 completeness relations (\ref{b}), (\ref{kn}) one finds  that 
\bs \label{sb1}
\begin{eqnarray}
\langle \dd 0  \mid  ( \frac{\phi^{m_{1}}(t)}{\sqrt N} \ldots 
\frac{\phi^{m_{n}}(t)}{\sqrt N})_{rs} \mid 0\dd \rangle =
\delta_{rs}  \langle 0  \mid \phi_{m_{1}}(t) \ldots \phi_{m_{n}}(t) \mid 0 
\rangle  \label{i} \\
= \delta_{rs}  \langle 0  \mid \tilde{\phi}_{m_{n}}(t)
 \ldots \tilde{\phi}_{m_{1}}(t) \mid 0 \rangle 
\end{eqnarray}
\es
\bs
\begin{eqnarray}
\langle \dd 0  \mid  ( \frac{\phi^{m_{1}}}{\sqrt N} \ldots 
\frac{\phi^{m_{n}}}{\sqrt N} )_{rs} \mid pq,A \rangle = f(N) 
P_{rs,pq}  \langle 0  \mid \phi_{m_{1}} \ldots \phi_{m_{n}} 
\mid A \rangle \label{fy} \\
= f(N) P_{rs,pq}  \langle 0  \mid \tilde
{\phi}_{m_{n}} \ldots \tilde{\phi}_{m_{1}} \mid A \rangle 
\end{eqnarray} 
\es
\bs
\begin{eqnarray}
\langle pq,A  \mid  ( \frac{\phi^{m_{1}}}{\sqrt N} \ldots 
\frac{\phi^{m_{n}}}{\sqrt N} )_{rs}\mid 0\dd \rangle = f(N) 
P_{qp,rs}  \langle A  \mid \phi_{m_{1}} \ldots \phi_{m_{n}} \mid 0 
\rangle \label{fx} \\
= f(N) P_{qp,rs}  \langle A  \mid \tilde{\phi}_{m_{n}}
 \ldots \tilde{\phi}_{m_{1}} \mid 0 \rangle
\end{eqnarray} 
\es
\bs \label{sb4}
\begin{eqnarray}
\langle pq,A \mid (\frac{\rho_{1}}{\sqrt N} \frac{\rho_{2}}{\sqrt N} 
\frac{\rho_{3}}{\sqrt 
N})_{rs} \mid st,B \rangle  & = & \langle sq,A \mid (\frac
{\rho_{1}}{\sqrt N} \frac{\rho_{2}}
{\sqrt N} \frac{\rho_{3}}{\sqrt N})_{sp} \mid rt,B\rangle \nonumber \\
& = & P_{qp,rt} \langle A \mid \rho_{1} \rho_{2} \rho_{3} \mid B \rangle 
\label{fz}
\end{eqnarray} 
\begin{eqnarray}
\langle pq,A \mid (\frac{\rho_{1}}{\sqrt N} \frac{\rho_{2}}{\sqrt N} \frac
{\rho_{3}}{\sqrt 
N})_{rs} \mid tr,B \rangle & = & \langle pr,A \mid (\frac
{\rho_{1}}{\sqrt N} \frac{\rho_{2}}
{\sqrt N} \frac{\rho_{3}}{\sqrt N})_{qr} \mid ts,B \rangle  \nonumber \\
& = & P_{qp,ts} \langle A \mid \widetilde{\rho_{1}\rho_{2}\rho_{3}} \mid B 
\rangle  \label{j}
\end{eqnarray} 
\es
\begin{equation}
\widetilde{\rho_{1}\rho_{2}\rho_{3}} \mid 0 \rangle = 
\rho_{1}\rho_{2}\rho_{3} \mid 0 \rangle  \label{h}
\end{equation}
where $\rho_{1}, \; \rho_{2}, \; \rho_{3}$ can be any of the canonical operators.
All the operators above are evaluated at time $t$, although this is
 explicit only in (\ref{sb1}).

The  (a) parts of each of these results can be extended  
to the  product  $\rho_1(t_{1}) \ldots \rho_n(t_{n})$  of any number of 
operators at arbitrary times, for example,
\begin{equation}
\langle \dd 0 \mid Tr( \frac{\rho_{1}(t_{1})}{\sqrt N} \ldots  
\frac{\rho_{n}(t_{n})}{\sqrt N} 
) \mid 0\dd \rangle = N \langle 0 \mid \rho_{1}(t_{1}) \ldots 
\rho_{n}(t_{n}) \mid 0 \rangle
 \label{m}
\end{equation}
so that all the traces  of the theory are 
computable in terms of the reduced quantities. Recall
that, for locally invariant 
theories, these are the invariant Wightman functions in the temporal 
gauge.  For globally invariant theories, a broader class of invariant 
Wightman functions is discussed in App.~B.

The (b) parts of these results and (\ref{h}) can also be extended to 
define the tilde of the composite operator $\rho_{1} \ldots \rho_{n}$:
 When all the operators in the original, unreduced 
matrix product commute, one finds that  the tilde of the reduced 
product is just the product of the tilde operators in the opposite order. 
This applies for example to the case of a general equal-time function $R$  of the 
$\phi$ fields,
\begin{equation}
R(\phi) = \sum_{n=0}^{\infty} r^{(n)}_{m_{1} \ldots m_{n}} 
\phi_{m_{1}} \ldots \phi_{m_{n}}, \;\;\;\;\;\; \tilde{R}(\tilde{\phi})
 = \sum_{n=0}^{\infty} r^{(n)}_{m_{1} \ldots m_{n}} 
\tilde{\phi}_{m_{n}} \ldots \tilde{\phi}_{m_{1}}  \label{k}
\end{equation}
where $\{ r \}$ are arbitrary coefficients and $\widetilde{R(\phi)} = 
\tilde{R}(\tilde{\phi})$. Note that this relation is consistent 
with the (b) parts of (\ref{sb1}) through (\ref{sb4}).  The same relations (\ref{k}) 
hold when all the $\phi$'s are replaced by $\pi$'s at equal time.

The tilde forms do not, however, extend in such a simple manner when 
the unreduced operators fail to commute, including, for example, a 
mixed product of $\phi$'s and $\pi$'s at equal time or a product of 
$\phi$'s at different times.  The tilde of a general equal-time  product 
 is determined in Subsec.~2.5 and App.~D, but, owing to 
the complexity of many-time commutators, we will not discuss the 
tilde of many-time composite operators. Unless specified otherwise, 
all the operator products below are taken at equal time. 

This completes the definition of the general equal-time reduced operators $R$, 
$\tilde{R}$ which correspond to the general equal-time matrix products 
(and sums of products) $R_{rs}$ 
of the original canonical operators:
\bs \label{ln2}
\begin{equation}
\langle \dd 0 \mid R_{rs}(\frac{\rho}{\sqrt N}) \mid 0\dd \rangle = 
\delta_{rs}  \langle 0 \mid R(\rho) \mid 0 \rangle \label{ln1} 
\end{equation}
\begin{equation}
\langle\dd 0 \mid R_{rs}(\frac{\rho}{\sqrt N}) \mid pq,A \rangle = 
 f(N) P_{rs,pq} \langle 0 \mid R(\rho) \mid A \rangle 
\end{equation}
\begin{equation}
\langle pq,A \mid R_{rs}(\frac{\rho}{\sqrt N}) \mid 0\dd \rangle = 
 f(N) P_{qp,rs} \langle A \mid R(\rho) \mid 0 \rangle 
\end{equation}
\begin{equation}
\langle pq,A \mid R_{rs}(\frac{\rho}{\sqrt N}) \mid st,B \rangle = 
P_{qp,rt}  \langle A \mid R(\rho) \mid B \rangle 
\end{equation} 
\begin{equation}
 \langle pq,A \mid R_{rs}(\frac{\rho}{\sqrt N}) \mid tr,B \rangle = 
P_{qp,ts}  \langle A \mid \tilde{R}(\tilde{\rho}) \mid B \rangle 
\end{equation}
\begin{equation} 
 \widetilde{R(\rho)} = \tilde{R} (\tilde{\rho}), \;\;\;\;\;\; 
 \tilde{R} (\tilde{\rho}) \mid 0 \rangle = R(\rho) \mid 0 \rangle 
 , \;\;\;\;\;\; 
\langle 0 \mid \tilde{R} (\tilde{\rho})  = \langle 0 \mid R(\rho) 
 \label{y}
\end{equation}
\es
where $\rho$ is the set of canonical variables.
 In what follows, we shall refer to any operator of this 
type as a \textit{density class operator} or simply a \textit{density}. 

\subsection{Trace Class Operators}

A \textit{trace class operator} $T\dd$ is an invariant operator which is the 
trace of a density, such as the Hamiltonian, the angular momentum 
generators or the supercharges of the theory.  The reduced matrix 
elements and reduced operators $T$ which correspond to any trace class 
operator $T\dd$ are defined as
\bs 
\begin{equation}
T\dd = C(N)\; Tr (t( \frac{\phi}{\sqrt N}  ,  \frac{\pi}{\sqrt N}  , 
\frac{\Lambda}{\sqrt N}  ))  \label{ab}
\end{equation}
\begin{eqnarray}
\langle \dd 0  \mid  T\dd  \mid 0\dd \rangle = \langle 0  \mid  T
  \mid 0 \rangle & = & N\; C(N)\; \langle 0  \mid   t(\phi  ,   \pi  , 
\Lambda )  \mid 0 \rangle  \nonumber \\ 
& = & N\; C(N)\; \langle 0  \mid   \tilde{t}
(\tilde{\phi}  ,   \tilde{\pi}  , \tilde{\Lambda} )  \mid 0 \rangle 
\label{l}
\end{eqnarray}
\begin{equation}
\langle A  \mid  T  \mid 0 \rangle  \equiv  \langle 0  \mid  T
  \mid A \rangle  \equiv  0 \label{n1} 
\end{equation}
\begin{equation}
\langle pq,A  \mid T\dd  \mid rs,B \rangle  \equiv  P_{qp,rs}
\langle A  \mid  T  \mid B \rangle  \label{n}
\end{equation}
\es 
where (\ref{l}) is nothing but the trace of (\ref{ln1}) above. 
The definitions in (\ref{n1}),(\ref{n}) are required for the 
consistency of multiplication of reduced trace class operators, and 
are also consistent with (\ref{ln2}). We leave the 
constant $C(N)$ undetermined here, but we will see below that 
$C(N) = N$ is selected for the usual Hamiltonian, supercharges and 
angular momenta. 

The unreduced operators $T\dd$  are  important  quantities 
  which generate the dynamics and  internal 
 symmetries of the theory, and we will see below that their reduced 
counterparts $T$ still generate the same important transformations 
in the reduced theory at large N. The reduced Hamiltonian $H$, which 
generates the time translations of the reduced theory, was constructed 
 in Refs.~\cite{us, me82} for the 
one hermitian and the (one polygon) unitary matrix models.  

For  all such  quantities, we encounter here an ÒopacityÓ 
phenomenon which was seen but not emphasized 
  in the examples of Refs.~\cite{us, me82}: The 
composite structure of  the reduced operator $T$  is apparently  
computable at this level only for the vacuum expectation value, as 
given in (\ref{l}), but not for the adjoint matrix elements in 
(\ref{n}).   
Technically,  the  reason 
 is that  the trace class operators are formed from adjoint operators 
but an adjoint operator
acting on an adjoint state generally contains higher representations 
than singlet and adjoint,
so that we cannot straightforwardly saturate the adjoint matrix 
elements of a trace class operator with
the states of our reduced space.

So, the opacity phenomenon means that 
we do not yet know the composite structure of  reduced operators $T$ 
 (only that the vacuum expectation value of $T$ must equal the forms  
shown in (\ref{l})).  Nevertheless, it is known from Refs.~\cite{us, me82}  that
the composite structure of these reduced operators can be found by 
solving their reduced algebraic relations, and we will place special emphasis on
the construction of these operators below (see Subsecs.~3.2, 3.3, 4.6 and Sec.~5).
 In particular, we will see that the reduced 
trace class operators are intrinsically nonlocal, in accord with the 
early examples~\cite{us, me82}.

\subsection{Derived Maps}

In this section, we use the formalism above to  
infer a number of derived maps into  the reduced  space. 
Applications of these maps are given later.

\vspace{12pt}
\noindent A.  Gauge generators.  Considering the matrix elements of the gauge 
generators $G_{rs}$ in (\ref{o}) we find from  (\ref{p}) that 
\begin{equation}
 [\phi_{m} , \pi_{m} ] - i 
\Lambda_{\alpha} \Lambda_{\alpha}  =  [\tilde{\phi}_{m} , \tilde{\pi}_{m} ] - i 
\tilde{\Lambda}_{\alpha} \tilde{\Lambda}_{\alpha} 
 =  i [ (B-F-1) + \mid 0 \rangle \langle 0 \mid ]  \label{t} .
\end{equation}
These relations are written as part of the reduced equal-time algebra of the 
theory, according to
the original interpretation in Ref.~\cite{us}. Because $G_{rs}$ is a density, 
however, these relations 
can be equivalently  understood as the action of the reduced symmetry  
generators $G,\; \tilde{G}$ on the states,
\bs \label{sy}
\begin{equation}
G \equiv -i[\phi_{m},\pi_{m}] - \Lambda_{\alpha} \Lambda_{\alpha} 
+(F-B), \;\;\;\;\;\; \tilde{G} \equiv i[\tilde{\phi}_{m},\tilde{\pi}_{m}]
 + \tilde{\Lambda}_{\alpha} \tilde{\Lambda}_{\alpha} + (B-F)   \label{r}
\end{equation}
\begin{equation}
G^{\dagger} = G, \;\;\;\;\;\; \tilde{G}^{\dagger} = \tilde{G}
\end{equation}
\begin{equation}
\tilde{G} = - G = 1 - \mid 0 \rangle \langle 0 \mid 
\end{equation}
\begin{equation}
\tilde{G} \mid 0 \rangle = G \mid 0 \rangle = 0, \;\;\;\;\;\;
\tilde{G} \mid A \rangle = - G \mid A \rangle =   \mid A \rangle . \label{ku}
\end{equation} 
\es 

\vspace{12pt}
\noindent B.  Density maps. For each density relation $R_{rs} = 0$ we obtain a 
pair of reduced equations
\begin{equation}
R_{rs} ( \frac{\rho}{\sqrt N} , \frac{\dot{\rho}}{\sqrt N}  
 ) = 0 \; \; \longrightarrow \; \; R(\rho , \dot{\rho}  
 ) = \tilde{R}(\tilde{\rho} , \dot{\tilde{\rho}} ) = 0 
\label{ka}
\end{equation}
where $\rho$ includes all the canonical variables.
Similarly, if  $R_{rs} \mid 0\dd \rangle = 0$, then $R\mid 0 \rangle = 
\tilde{R} \mid 0 \rangle = 0$.

This map gives us, for example, an untilde and a tilde version
of the reduced equations of motion (see for example Subsec.~2.6).

\vspace{12pt}
\noindent C.  Canonical  maps.  When  $\rho$ and $\sigma$ are 
any of the canonical
variables $\phi,\; \pi \; or \; \Lambda$,  we find that 
\begin{equation}
 [\rho_{rs} , \sigma_{pq} ]_{\mp} = ic \delta_{sp} \delta_{rq}
\; \; \longrightarrow \; \; [\tilde{\rho} , \sigma]_{\mp} =  [\rho , 
\tilde{\sigma}]_{\mp} =  ic  \mid 0 \rangle \langle 0 \mid \label{u}
\end{equation}
which is derived by considering matrix elements such as 
\begin{equation}
\langle \dd 0 \mid [\rho_{rs}, \sigma_{sq}]_{\mp} \mid 0\dd \rangle, \;\;\;\;\;\;
\langle ps,A \mid [\rho_{rs}, \; \sigma_{pq} ]_{\mp} \mid qr,B\rangle .
\end{equation} 
These contributions to the reduced equal-time algebra of the theory
 are free algebraic 
relations because they contain   no relations among the untilde 
operators  or among the tilde operators. See Subsec.~2.5
for further discussion of the equal-time algebra.

\vspace{12pt}
\noindent D.  A density and a trace class operator. The form of relations 
involving the product of a
density and a trace class operator are preserved in reduced space. 
We phrase this in terms of commutators and anticommutators:
\bs \label{s}
\begin{equation}
 [T\dd \; , \;  R_{rs} (  \frac{\rho}{\sqrt N} , \frac{\dot{\rho}}{\sqrt N}
) ]_{\mp} =  S_{rs} (  \frac{\rho}{\sqrt N} , \frac{\dot{\rho}}{\sqrt N} )
\end{equation}
\begin{equation}
\longrightarrow \;\; \left\{  
\begin{array}{l}
[T \; , R(\rho, \dot{\rho})]_{\mp} = S(\rho, \dot{\rho})  \\ 
{[} T \; , \tilde{R} ( \tilde{\rho}, \dot{\tilde{\rho}}) ]_{\mp} 
 = \tilde{S}(\tilde{\rho}, \dot{\tilde{\rho}})
\end{array} \right.
\end{equation}
\es 
where $R_{rs}$ and $S_{rs}$ are general densities.
This map gives us for example the commutator form of the reduced equations 
of motion in terms of the reduced Hamiltonian $H$,
\begin{equation}
\dot{\rho} = i [H,\rho], \;\;\;\;\;\; \dot{\tilde{\rho}} = i 
[H,\tilde{\rho}], \;\;\;\;\;\;\rho = \phi, \pi, \; or \; 
\Lambda 
\end{equation}
which supplements the explicit form of the reduced equations of motion obtained 
from the map (\ref{ka}). Moreover, the reduced images of (\ref{lj}) are 
\begin{equation}
\dot{G} = \dot{\tilde{G}} = [G,H] = [\tilde{G},H] = 0
\end{equation}
where $G$ and $\tilde{G}$ are the reduced gauge generators in (\ref{sy}). 
The map (\ref{s}) also tells us  that  the transformation properties of 
the  reduced operators $\phi, \; \pi \; or \; \Lambda$  under  the reduced 
rotation or supersymmetry generators is unchanged by the reduction 
(see Subsecs.~2.6, 3.2 and 3.3 and Secs.~5 and 6).

\vspace{12pt}
\noindent E.  Two trace class operators. Algebraic relations among trace 
class operators are preserved  in the reduction:
\begin{equation}
 [T_{\textbf{\raisebox{.3ex}{.}} 1} \; , \;
  T_{\textbf{\raisebox{.3ex}{.}} 2}]_{\mp} = 
  T_{\textbf{\raisebox{.3ex}{.}} 3} \; \; \longrightarrow
\; \; [T_{1} \; , \; T_{2} ]_{\mp} = T_{3} . \label{lo}
\end{equation}
This map tells us, for example, that the angular momentum algebra or 
the supersymmetry algebra of the theory is preserved in the reduction 
 (See Subsecs.~2.6, 3.3, 5.3, 5.5 and Sec.~6). In the case of a gauged 
matrix model such as Matrix theory, however, the explicit form of the 
reduced supersymmetry algebra (see Eq.~(\ref{kp})) can be surprisingly 
different from that of the original unreduced supersymmetry algebra.

\subsection{Reduced Equal-Time Algebra}

In this section, we familiarize ourselves with the reduced equal-time  
free algebras of  the general matrix model,  temporarily deferring the 
contributions of any conserved quantity.

The explicit form of  the reduced equal-time algebras follows from  
(\ref{t}) and (\ref{u}):
\bs \label{sc}
\begin{equation}
[\phi_{m} , \tilde{\pi}_{n} ] = [\tilde{\phi}_{m} , \pi_{n} ] =
i \delta_{mn} \mid 0 \rangle \langle 0 \mid \label{v}
\end{equation}
\begin{equation}
[\phi_{m} , \tilde{\phi}_{n} ] = [\pi_{m} , \tilde{\pi}_{n} ] = 0 \label{z}
, \;\;\;\;\;\; 
[\Lambda_{\alpha} , \tilde{\Lambda}_{\beta} ]_{+} = \delta_{\alpha \beta} 
\mid 0 \rangle \langle 0 \mid \label{v2}
\end{equation}
\begin{equation}
[\Lambda_{\alpha} , \tilde{\phi}_{m}] = [\tilde{\Lambda}_{\alpha} , \phi_{m}]
 = [\Lambda_{\alpha} , \tilde{\pi}_{m}] = [\tilde{\Lambda}_{\alpha} , \pi_{m}]
  = 0
\end{equation}
\begin{equation}
 [\phi_{m} , \pi_{m} ] - i \Lambda_{\alpha} \Lambda_{\alpha}  = 
 [\tilde{\phi}_{m} , \tilde{\pi}_{m} ] - i \tilde{\Lambda}_{\alpha}
 \tilde{\Lambda}_{\alpha}
  =  i  [ (B-F-1) + \mid 0 \rangle \langle 0 \mid ]   \label{w} 
\end{equation} 
\begin{equation}
m, n = 1 \ldots B, \;\;\;\;\;\; \alpha, \beta = 1 \ldots f, 
\;\;\;\;\;\; F = \frac{f}{2}
\end{equation}
\es 
where B and f are integers.  The equal-time algebras (\ref{sc}) provide our 
first examples of \textit{symmetric free algebras}, so called because each 
algebra is symmetric under the exchange of tilde and untilde 
operators:
\begin{equation}
\rho \; \longleftrightarrow \; \tilde{\rho}, \;\;\;\;\;\; \rho = 
\phi, \; \pi \; and \; \Lambda . \label{li}
\end{equation}
Historically,  a Euclidean operator  isomorphic to $\tilde{\pi}$ in 
(\ref{v}) 
was first introduced in Ref.~\cite{Haan}, and later as a differential realization in 
Ref.~\cite{Gopa}.

In what follows, we discuss a number of properties of the symmetric 
free algebras (\ref{sc}) in combination with the vacuum relations 
(\ref{x}) and (\ref{y}), which we repeat here for reference
\bs \label{ki}
\begin{equation}
\tilde{\rho} \mid 0 \rangle = \rho \mid 0 \rangle , \;\;\;\;\;\; 
  \langle 0 \mid \tilde{\rho} = \langle 0 \mid \rho
\end{equation}
\begin{equation}
\tilde{R} (\tilde{\rho}) \mid 0 \rangle = R(\rho) \mid 0 \rangle
 , \;\;\;\;\;\;
\langle 0 \mid \tilde{R} (\tilde{\rho}) = \langle 0 \mid R(\rho) . \label{lr}
\end{equation}
\es

\vspace{12pt}
\noindent A.  Consistency check. We note that (\ref{ki}) and the vacuum 
expectation value of  (\ref{w}) imply the relation
\begin{equation}
\langle 0 \mid [\phi_{m} , \tilde{\pi}_{m} ] - \frac{i}{2} [\Lambda_{\alpha}
, \tilde{ \Lambda}_{\alpha}]_{+} \mid 0 \rangle =  i (B-F)
\end{equation}
which is  consistent with  (\ref{v}), (\ref{v2}). \nopagebreak 

\vspace{12pt}
\noindent B.  Tilde operators as right multipliers.  Consider a general state 
(word) formed by the 
action of any number of $\phi$ operators on the vacuum (fermionic
operators can be added as well). The action of another 
$\phi$ on the state is of course the addition of the operator on the left 
of the word. On the other hand,
 (\ref{z}) and (\ref{ki}) tell us that the action of a $\tilde{\phi}$
operator is equivalent to adding a $\phi$ field on the right:
\begin{equation}
\tilde{\phi}_{m} \phi_{m_{1}} \ldots \phi_{m_{n}}\mid 0 \rangle =
\phi_{m_{1}} \ldots \phi_{m_{n}} \phi_{m} \mid 0 \rangle .
\end{equation}
In word notation, the action of $\phi$ and $\tilde{\phi}$ is
\bs \label{kq}
\begin{equation}
\mid w \rangle  = \phi^{w} \mid 0 \rangle \equiv \phi_{m_{1}} \phi_{m_{2}}
\ldots \phi_{m_{n}}
 \mid 0 \rangle,\;\;\;\;\;\; w \equiv m_{1}m_{2} \ldots m_{n}
\end{equation}
\begin{equation}
\phi_{m} \mid w \rangle = \mid mw \rangle , \;\;\;\;\;\; \tilde{\phi}_{m} 
\mid w \rangle = \mid wm \rangle . \end{equation}
\es

\vspace{6pt}
\noindent C.  Cyclicity of  ground state averages. The equal-time 
reduced vacuum 
expectation value of any number of reduced $\phi$'s is cyclically symmetric. 
To see this use (\ref{z}) and (\ref{ki}) to follow the steps:
\begin{eqnarray}
\langle 0 \mid \phi_{m_{1}} \phi_{m_{2}} \ldots \phi_{m_{n}} \mid 0 \rangle
 & = &   \langle 0 \mid \tilde{\phi}_{m_{1}} \phi_{m_{2}}\ldots 
 \phi_{m_{n}} \mid 0 \rangle  \nonumber \\
= \langle 0 \mid \phi_{m_{2}} \ldots \phi_{m_{n}} \tilde{\phi}_{m_{1}}
\mid 0 \rangle & = & 
 \langle 0 \mid \phi_{m_{2}} \ldots \phi_{m_{n}} \phi_{m_{1}} \mid 0 \rangle .
 \label{fp}
\end{eqnarray} 
According to (\ref{i}),  this result is only the image in reduced space of  
the cyclic  property
of the unreduced equal-time traced Wightman functions.\footnote{Related 
identities such as $\langle 0 \!\! \mid \!\! 
[R(\phi),S(\phi)] \!\!\mid \!\!  0 \rangle = 0$ also follow in the same way 
from (\ref{lr}) and (using the results of App.~C) we see that this identity 
is the image 
of $N^{-1} Tr[R( \phi /{\sqrt N}),S(\phi /{\sqrt N})] = 0$ 
in the unreduced theory.}  We 
emphasize however the central role of the tilde operators in 
establishing this property directly in the reduced space.

The same cyclicity is found for the vacuum expectation value of 
many reduced $\pi$'s, as expected, but vacuum expectation values of mixed 
products of $\phi$'s, $\pi$'s and $\Lambda$'s are 
generally not cyclic. Following steps similar to those in (\ref{fp}), 
however, the corrections to cyclicity can always be  
computed directly from the equal-time algebra and (\ref{ki}). 
Here are some simple examples
\bs 
\begin{equation}
\langle 0 \mid \phi_{m} \pi_{n} \mid 0 \rangle = 
\langle 0 \mid \pi_{n} \phi_{m} \mid 0 \rangle + i\delta_{mn} 
\end{equation}
\begin{equation}
\langle 0 \mid \Lambda_{\alpha} \Lambda_{\beta} \mid 0 \rangle = 
 - \langle 0 \mid \Lambda_{\beta} \Lambda_
{\alpha} \mid 0 \rangle + \delta_{\alpha \beta}
\end{equation}
\es 
which the reader is invited to verify.

\vspace{8pt}
\noindent D.  Tilde of general reduced densities.  The $\tilde{R}$  
corresponding to  a general composite 
density $R$ is defined in  Subsec.~2.2 and satisfies (\ref{ki}). The form of 
$\tilde{R}$ is simple when the operators
of the original density commute,  as noted in  (\ref{k}). We give here a 
useful algorithm for the 
form of the general $\tilde{R}$ which nicely packages the results of 
App.~D:
One can compute $\tilde{R}$ from $R$ using the equal-time algebra and 
(\ref{ki}), remembering that
$\tilde{R}$ is a function only of  tilde fields. This means that we 
eliminate any
vacuum projectors $\mid 0 \rangle \langle 0 \mid$ which arise  
by using the identity   $\mid 0 \rangle \langle 0 \mid 0 
\rangle = \mid 0 \rangle$. 
As a simple example, consider 
\begin{equation}
\phi_{m} \pi_{n} \mid 0 \rangle = \phi_{m} \tilde{\pi}_{n} \mid 0 \rangle = 
[\phi_{m}, \tilde{\pi}_{n}] \mid 0 \rangle + \tilde{\pi}_{n} \phi_{m} 
\mid 0 \rangle = (i \delta_{mn}  + \tilde{\pi}_{n} 
\tilde{\phi}_{m}) \mid 0 \rangle
\end{equation}
which tells us that
\begin{equation}
\widetilde{(\phi_{m} \pi_{n})} = \tilde{\pi}_{n} \tilde{\phi}_{m} 
+ i \delta_{mn} .
\end{equation}
Another example is $R = \Lambda_{\alpha} \phi_{m} \Lambda_{\beta} 
\pi_{n}$, for which we \vspace{-3pt} find 
\begin{equation}
\tilde{R}  =  - \tilde{\pi}_{n} \tilde{\Lambda}_{\beta} 
\tilde{\phi}_{m} \tilde{\Lambda}_{\alpha} + i \delta_{mn} \langle 0 
\mid \Lambda_{\beta} \mid 0 \rangle \tilde{\Lambda}_{\alpha} 
+ \delta_{\alpha \beta} \langle 0 \mid \phi_{m} \mid 0 \rangle \tilde
{\pi}_{n} . \vspace{-3pt}
\end{equation}
The form of the reduced 
gauge generator $\tilde{G}$ in (\ref{r}) is also easily computed  
in this way from the form of the reduced gauge generator $G$.

\vspace{12pt}
\noindent E.  An even number of real fermions.  When the number of  
real adjoint  fermions is even, we may introduce complex reduced 
fermions \vspace{-3pt} as
\begin{equation}
\Lambda_{\alpha} = \frac {1}{\sqrt 2} 
\left( \begin{array}{c}
\psi^{\dagger}_{\dot \alpha} + \psi_{\dot \alpha} \\ 
i(\psi^{\dagger}_{\dot \alpha} - \psi_{\dot \alpha}) 
\end{array} \right) ,
\; \; \; \; \; \; \dot{\alpha} = 1 \ldots F , \; \; \; \; \; \; 
F=\frac{f}{2} = \textup{integer}  \vspace{-3pt}
\end{equation}
and similarly for $\tilde{\Lambda}_{\alpha} \rightarrow 
\tilde{\psi}^{\dagger}_{\dot{\alpha}} , \tilde{\psi}_{\dot{\alpha}}$. 
Then  the fermionic part of the reduced equal-time algebra 
\vspace{-3pt} becomes:
\bs \label{tn}
\begin{equation}
[ \psi_{\dot{\alpha}} , \tilde{\psi}^{\dagger}_{\dot{\beta}}]_{+} =
[ \tilde{\psi}_{\dot{\alpha}} , \psi^{\dagger}_{\dot{\beta}}]_{+} = 
\delta_{\dot{\alpha} \dot{\beta}} \mid 0 \rangle \langle 0 \mid
 ,\;\;\;\;\;\; 
[ \psi_{\dot{\alpha}} , \tilde{\psi}_{\dot{\beta}}]_{+} =
[ \psi^{\dagger}_{\dot{\alpha}} , \tilde{\psi}^{\dagger}_{\dot{\beta}}]_{+} = 0
 \label{lh}
\end{equation}
\begin{equation}
 [\phi_{m} , \pi_{m} ] - i[ \psi_{\dot{\alpha}} , \psi^{\dagger}_{\dot{\alpha}}
 ]_{+}  =  [\tilde{\phi}_{m} , \tilde{\pi}_{m} ] - i[ \tilde{\psi}_
 {\dot{\alpha}}  , \tilde{\psi}^{\dagger}_{\dot{\alpha}}
 ]_{+} 
   =  i [ (B-F-1) + \mid 0 \rangle  \langle 0 \mid ] 
\end{equation} 
\begin{equation}
\tilde{\psi}_{\dot{\alpha}} \mid 0 \rangle = \psi_{\dot{\alpha}} \mid 0 
\rangle, 
\;\;\;\;\;\; \tilde{\psi}^{\dagger}_{\dot{\alpha}} \mid 0 \rangle =
\psi^{\dagger}_{\dot{\alpha}} \mid 0 \rangle .
\end{equation} \vspace{-3pt}
\es 
The complex fermionic operators continue to commute with the bosonic tilde 
operators and vice-versa with respect to the tilde. 

This decomposition allows us 
to see many of the properties discussed above for the bosonic operators. 
For example the \vspace{-3pt} relation
\begin{equation}
\tilde{\psi}^{\dagger}_{\dot{\alpha}} \psi^{\dagger}_{\dot{\alpha}_{1}} \ldots 
\psi^{\dagger}_{\dot{\alpha}_{n}} \mid 0 \rangle = (-1)^{n} 
\psi^{\dagger}_{\dot{\alpha}_{1}} \ldots \psi^{\dagger}_{\dot{\alpha}_{n}}
\psi^{\dagger}_{\dot{\alpha}} \mid 0 \rangle \vspace{-3pt}
\end{equation}
shows that $\tilde{\psi}^{\dagger}$ is  a right multiplication operator with 
respect to the daggered fermionic words. Similarly, the tilde of 
the composite fermionic \vspace{-3pt} operators
\begin{equation}
R = \psi^{\dagger}_{\dot{\alpha}_{1}} \ldots  \psi^{\dagger}_{\dot{\alpha}_{n}}, 
\;\;\;\;\;\; \tilde{R} = (-1)^{\frac{1}{2} n(n-1)} \tilde{\psi}^{\dagger}_{\dot
{\alpha}_{n}} \ldots  \tilde{\psi}^{\dagger}_{\dot{\alpha}_{1}} \vspace{-3pt}
\end{equation}
is easily computed from (\ref{ki}) and \nopagebreak (\ref{lh}).

\subsection{Example: General Bosonic System}

As an explicit example, we  collect here the setup for a general 
system of B bosons, starting with the hermitian  \vspace{-3pt} Hamiltonian
\bs \label{ts}
\begin{equation}
H\dd = Tr (\frac{1}{2} \pi^{m} \pi^{m} + N V(\frac{\phi}{\sqrt N})) 
  \label{aa}
\end{equation}
\begin{equation}
Tr \; V(\phi ) = Tr (v^{(0)} + \sum_{n=1}^{\infty} \frac{1}{n} \;
 v^{(n)}_{m_{1} \ldots m_{n}} \; \phi^{m_{1}} \ldots \phi^ {m_{n}} )   \label{ae}
\end{equation}
\begin{equation}
v^{(n)}_{m_{1} m_{2} \ldots m_{n}} =  v^{(n)}_{m_{2} \ldots m_{n} m_{1}} \label{bx}
,\;\;\;\;\;\; (Tr \; V)^{\dagger} = Tr \; V, \;\;\;\;\;\;
v^{(n) \; *}_{m_{1}\ldots m_{n}} = v^{(n)}_{m_{n}\ldots m_{1}}   \label{by}
\end{equation}
\es 
where the numerical coefficients $v^{(n)}$ of  the potential are cyclically 
symmetric in their subscripts.  These coefficients are also  independent of
 N to maintain `t~Hooft scaling at large N. Comparing (\ref{aa})
  with (\ref{ab}) we see that 
$C(N) = N$ for the Hamiltonian, as noted above.

Going over now to the reduced formulation at large N, we record first 
the equal-time free algebra of the system
\bs \label{sz}
\begin{equation}
[\phi_{m} , \tilde{\pi}_{n} ] = [\tilde{\phi}_{m},\pi_{n}] = 
i \delta_{mn} \mid 0 \rangle \langle 0 \mid \label{bg}
\end{equation}
\begin{equation}
[\phi_{m} , \tilde{\phi}_{n} ] = [\tilde{\pi}_{m},\pi_{n}] = 0
\end{equation} 
\begin{equation}
[\phi_{m} , \pi_{m} ] = [\tilde{\phi}_{m},\tilde{\pi}_{m}] = 
  i [B-1 + \mid 0 \rangle \langle 0 \mid ] \label{bg1}
\end{equation}
\begin{equation}
\tilde{\rho} \mid 0 \rangle = \rho \mid 0 \rangle, \;\;\;\;\;\; 
\langle 0 \mid \tilde{\rho} = \langle 0 \mid \rho, \;\;\;\;\;\; 
\rho=\phi \; o r\; \pi
\end{equation}
\es 
and then the  reduced equations of motion
\bs 
\begin{equation}
\dot{\phi}_{m} = i [H,\phi_{m}] = \pi_{m}, \; \; \; \; \; \; 
\dot{\pi}_{m} = i [H, \pi_{m}] = - V'_{m}(\phi) \label{bv}
\end{equation}
\begin{equation}
\dot{\tilde{\phi}}_{m} = i [H,\tilde{\phi}_{m}] = \tilde{\pi}_{m}, \; \; 
\; \; \; \; \dot{\tilde{\pi}}_{m} = i [H, \tilde{\pi}_{m}] = - \tilde{V}'
_{m} (\tilde{\phi})
\end{equation}
\begin{equation}
 V'_{m}(\phi) = \sum_{n=1}^{\infty} v^{(n)}_{m m_{2} \ldots m_{n}} \phi_{m_{2}}
 \ldots \phi_{m_{n}}, \;\;\;\;\;\; 
\tilde{V}'_{m} (\tilde{\phi}) = \sum_{n=1}^{\infty} v^{(n)}_{m m_{2} \ldots 
m_{n}} \tilde{\phi}_{m_{n}} \ldots \tilde{\phi}_{m_{2}}
\end{equation}
\begin{equation}
V'^{\dagger}_{m} = V'_{m}, \;\;\;\;\;\; \tilde{V}'^{\dagger}_{m} =
\tilde{V}'_{m}, \;\;\;\;\;\; \tilde{V}'_{m} \mid 0 \rangle =
V'_{m} \mid 0 \rangle
\end{equation}
\es 
are obtained from the original equations of motion and maps B and D 
of Subsec.~2.4. Here $H$ is the reduced Hamiltonian of the system which 
satisfies 
\bs 
\begin{equation}
E_{0} = \langle \dd 0 \mid H\dd \mid 0\dd \rangle =  \langle 0 \mid H \mid 0 
\rangle = N^{2} \; \langle 0 \mid
\frac{1}{2} \pi_{m} \pi_{m} + V(\phi ) \mid 0 \rangle
\end{equation}
\begin{equation}
(H - E_{0}) \mid 0 \rangle = 0, \;\;\;\;\;\;(H - E_{0}) \mid A \rangle = 
\omega_{A 0} \mid A \rangle 
\end{equation}
\begin{equation}
\omega_{\mu \nu} = E_{\mu} - E_{\nu} = O(N^{0}) , \; \; \; \; \; \; \mu = (0,A) ,
\; \; \; \; \; \; \nu = (0,B)
\end{equation}
\es 
and governs the time dependence of the reduced system \cite{me81, us} 
according to
\bs \label{sd}
\begin{equation}
\rho (t) = e^{iHt} \rho(0) e^{-iHt}, \;\;\;\;\;\;
\tilde{\rho} (t) = e^{iHt} \tilde{\rho}(0) e^{-iHt}  \label{ad}
\end{equation}
\begin{equation}
\rho (t) _{\mu \nu} = e^{i\omega_{\mu \nu} t}\rho (0) _{\mu \nu}, 
\;\;\;\;\;\;  \tilde{\rho} (t) _{\mu \nu} = e^{i\omega_{\mu \nu} t}
\tilde{\rho} (0) _{\mu \nu}   \label{ac}
\end{equation}
\es
where $\rho = \phi \; or \; \pi$.  As noted in the Introduction, the
matrix elements in (\ref{ac}) are 
 the master fields of the theory, and the results (\ref{sd}) hold as
well  in general  matrix  models including fermions. As emphasized in 
Subsec.~2.3, we do not yet know the composite structure of the reduced $H$.

In what follows, we discuss a number of useful aspects of this system.

\vspace{12pt}
\noindent A.  Connection and curvature. Define a connection
on the free Hilbert space as the collection  $J \equiv \{ J_{m}(\phi), 
\tilde{J}_{m}(\tilde{\phi}), \; m=1\ldots B \}$
 where   $J_{m}(\phi)$ is a set of reduced 
densities whose tildes are $\tilde{J}_{m}$. Further  define the curvature 
 of the connection $J$ as 
\begin{equation}
{\cal F}_{mn} (J) \equiv [\pi_{m}, \tilde{J}_{n}] - [\tilde{\pi}_{n},
J_{m}] .
\end{equation}
As an example, use the equations of motion to compute
\begin{equation}
\frac{d}{dt} [\pi_{m} , \tilde{\pi}_{n} ] = 0 \; \; \; \longrightarrow
\; \; \; [\tilde{\pi}_{n} , V'_{m}] - [\pi_{m} , \tilde{V}'_{n} ] = 0 .
\end{equation}
This  shows that $V' \equiv \{ V'_{m}, \tilde{V}'_{m} \}$, which is a natural 
``gradient'' associated to the reduced potential $V$, is a flat 
connection on the free Hilbert space,
\begin{equation}
{\cal F}_{mn} (V') = 0 .
\end{equation}

We have also checked that the notions of flat connection  and cyclic  
coefficients are  equivalent:  Any connection $J$ of the form 
\bs
\begin{equation} 
J_{m}(\phi) = \sum_{n=1}^{\infty} j^{(n)}_{m m_{2} \ldots m_{n}} 
\phi_{m_{2}} \ldots \phi_{m_{n}}, \;\;\;\;\;\; 
\tilde{J}_{m}(\tilde{\phi}) = \sum_{n=1}^{\infty} j^{(n)}_{m m_{2} \ldots m_{n}} 
\tilde{\phi}_{m_{n}} \ldots \tilde{\phi}_{m_{2}} 
\end{equation}
\begin{equation}
j^{(n)}_{m_{1} m_{2} \ldots m_{n}} = j^{(n)}_{m_{2} \ldots m_{n} m_{1}}
\end{equation}
\es 
is flat and we have also solved the flatness condition to prove that 
any flat connection has this form.  The notions of flat connection  
and integrability are  equivalent as well, so that every 
flat connection on the free Hilbert space is associated to a trace 
class generating function of the form (\ref{ae}) in the unreduced large N 
theory.  Flat connections will play an important role in the 
development of Sec.~4.

\vspace{12pt}
\noindent B.  Rotational invariance. We consider the special case when the original 
bosonic theory is rotation invariant  with trace class generators 
$J\dj^{mn}$:
\bs 
\begin{equation}
J\dj^{mn} = Tr(\pi^{[m} \phi^{n]}), \;\;\;\;\;\; m,n = 1 \ldots B \label{z1}
\end{equation} 
\begin{equation}
\dot{J}\dj^{mn} = i [H\dd,J\dj^{mn}] = 0, \;\;\;\;\;\; J\dj^{mn} 
\mid 0\dd \rangle = 0
\end{equation}
\es 
which satisfy the algebra of spin(B). The operators $\phi$ and $\pi$ are 
in the vector representation of spin(B), and (\ref{z1}) specifies the 
constant in (\ref{ab}) as $C(N) = N$.  At large N, these  
generators map onto  reduced generators $J_{mn}$ which satisfy
\bs 
\begin{equation}
\langle 0 \mid J_{mn} \mid 0 \rangle = \langle 0 \mid \pi_{[m}
\phi_{n]} \mid 0 \rangle = 0
\end{equation}
\begin{equation}
\dot{J}_{mn} = i [H,J_{mn}] = 0, \;\;\;\;\;\; J_{mn} \mid 0 \rangle = 0
\end{equation}
\begin{equation}
[J_{mn}, B_{p}] = -i \delta_{p [m} \; B_{n]}, \;\;\;\;\;\; 
B = \phi, \pi, \tilde{\phi} \; or \; \tilde{\pi} \label{be}
\end{equation}
\begin{equation}
[J_{mn}, J_{pq}] = i ( \delta_{q [m} J_{n] p} -  \delta_{p [m} J_{n] q}) 
\end{equation}
\es 
including (according to map E of Subsec.~2.4) the same algebra of spin(B). 
As in the case of the reduced Hamiltonian $H$, we do not yet know the 
composite structure of the reduced generators $J_{mn}$ (see Subsec.~2.3).

\vspace{12pt}
\noindent C.  One hermitian matrix . The case $B=1$ above is called the (Hamiltonian) 
one hermitian matrix model, whose solution \cite{me81, us}  we review here as a 
special case of our general development. 

We have seen above that untilde and tilde operators correspond respectively
to left and right multiplication in the word notation (see Eq.~(\ref{kq})), 
but  left and right multiplication are indistinguishable when $B=1$,  so that
\begin{equation}
B = 1: \;\;\;\tilde{\phi} = \phi, \; \; \; \; \; \; \tilde{\pi} = \pi . 
\label{af}
\end{equation}
The  identification (\ref{af}), which does not hold for higher B (or 
for $B=1$ and $F \neq 0$),  is the 
essential simplification of the 1-matrix model. Then the reduced system 
 reads  simply \cite{me81, us} 
\bs \label{tk}
\begin{equation}
\dot{\phi} = i[H,\phi] = \pi , \; \; \; \; \; \dot{\pi} = i[H,\pi ] =
 -V'(\phi ) 
\end{equation}
\begin{equation}
[\phi , \pi ] = i \mid 0 \rangle \langle 0 \mid \label{ko}
\end{equation}
\begin{equation}
\rho (t) _{\mu \nu} = e^{i\omega_{\mu \nu} t} \rho (0) _{\mu \nu}, 
\;\;\;\;\;\; \rho = \phi \;  or \; \pi \label{kw}
\end{equation}
\es 
where the master fields are given in (\ref{kw}). 

We mention  two early approaches to the solution of this model. 
The case of the oscillator was solved in Ref.~\cite{me81}: 
\bs
\begin{equation}
V = \frac{1}{2} \omega^{2} \phi^{2},\;\;\;\;\;\; 
\phi = \frac{1}{\sqrt {2 \omega}} (e^{i \omega t} a^{\dagger} +
e^{-i \omega t} a ), \;\;\;\;\;\;
\pi = i \sqrt \frac{\omega}{2} (e^{i \omega t} a^{\dagger} -
e^{-i \omega t} a )
\end{equation}
\begin{equation}
(a^{\dagger})_{\mu \nu} = \delta_{\mu , \nu +1}, \; \; \; \; \; \; 
(a)_{\mu \nu} = \delta_{\mu , \nu -1} \label{ag}
\end{equation}
\begin{equation}
 a a^{\dagger} = 1 , \; \; \; \; \; \; a \mid 0 \rangle = 0, \; \; \; 
\; \; \;  a^{\dagger} a = 1 - \mid 0 \rangle \langle 0 \mid .  \label{ah}
\end{equation}
\es 
This solution was originally written with  Kronecker deltas, as   
in (\ref{ag}), but we recognize this today as a realization of  the one 
dimensional  Cuntz algebra  in (\ref{ah}). We will return to this 
approach for many oscillators in Sec.~3.

The general system (\ref{tk}) was solved \cite{us} in the coordinate basis:  
\bs 
\begin{equation}
\phi \mid q \rangle = q \mid q \rangle, \;\;\;\;\;\; 
\pi_{q,q'} = i \frac{{\cal P}}{q-q'} \;  \psi_{0} (q) \psi^{\ast}_{0} (q')
\end{equation}
\begin{equation}
(H - E_{0})_{q, q'} = - \frac{{\cal P}}{(q-q')^{2}} \; \psi_{0} (q) \psi^
{\ast}_{0} (q') +\; \delta (q-q') \int dq'' \frac{{\cal P}}{(q-q'')^{2}}
 \rho (q'')  \label{ai}
\end{equation}
\begin{equation}
\rho (q) = \psi^{\ast}_{0} (q) \psi_{0} (q) = 
\frac{\sqrt 2}{\pi} (\epsilon - V(q))^{\frac{1}{2}}, \;\;\;\;\;\;
\int dq \; \rho (q) = 1
\end{equation}
\begin{equation}
\langle 0 \mid \phi^{n} \mid 0 \rangle = \int dq \; \rho (q) q^{n}
\end{equation}
\es 
where ${{\cal P}}$ is principal value, $\psi_{0}$ is the reduced coordinate space 
ground state wave function, and sub $q,q'$ denotes matrix elements in the 
coordinate basis. See Ref.~\cite{us} for further 
details of this solution, including the ground state energy $E_{0}$ and 
the energies of the dominant adjoint states.  We note in particular 
the explicit construction (\ref{ai}) of the reduced Hamiltonian $H$, whose 
composite structure is seen to be highly nonlocal.  A similarly nonlocal
reduced Hamiltonian  for the (one polygon) unitary matrix model  was obtained 
in Ref.~\cite{me82}. 
This approach is also considered for higher B in Subsec.~3.6.

\section{Bosonic Oscillators}

\subsection{Symmetric Cuntz Algebras}

In this section we consider the special case of B bosonic oscillators
\bs
\begin{equation}
V = \frac{1}{2} \sum_{m=1}^{B} \omega^{2}_{m} \phi_{m} \phi_{m}
\end{equation}
\begin{equation}
\dot{\phi}_{m} = i [H,\phi_{m} ] = \pi_{m}, \; \; \; \; \; \; 
\dot{\pi}_{m} = i [H,\pi_{m} ] = - \omega^{2}_{m} \phi_{m} \label{aj}
\end{equation}
\begin{equation}
\dot{\tilde{\phi}}_{m} = i [H,\tilde{\phi}_{m} ] = \tilde{\pi}_{m}, 
\; \; \; \; \; \; \dot{\tilde{\pi}}_{m} = i [H,\tilde{\pi}_{m} ] 
= - \omega^{2}_{m} \tilde{\phi}_{m} \label{ak}
\end{equation}
\es
in order to understand the relationship between our equal-time 
free algebra (\ref{sz}) and the Cuntz algebra (\ref{km}).

The solution to the reduced equations of motion (\ref{aj}) and (\ref{ak}) is
\bs \label{tr}
\begin{equation}
\phi_{m} = \frac{1}{\sqrt {2 \omega_{m}}} (e^{i \omega_{m} t} a^
{\dagger}_{m} + e^{-i \omega_{m} t} a_{m} ), \;\;\;\;\;\; 
\pi_{m} = i \sqrt \frac{\omega_{m}}{2} (e^{i \omega_{m} t} a^
{\dagger}_{m} - e^{-i \omega_{m} t} a_{m} )
\end{equation}
\begin{equation}
\tilde{\phi}_{m} = \frac{1}{\sqrt {2 \omega_{m}}} (e^{i \omega_{m} t}
\tilde{a}^{\dagger}_{m} + e^{-i \omega_{m} t} \tilde{a}_{m} )
, \;\;\;\;\;\; 
\tilde{\pi}_{m} = i \sqrt \frac{\omega_{m}}{2} (e^{i \omega_{m} t} 
\tilde{a}^{\dagger}_{m} - e^{-i \omega_{m} t} \tilde{a}_{m} )
\end{equation}
\es
and we know that
\bs  \label{al}
\begin{equation}
a_{m} \mid 0 \rangle = \tilde{a}_{m} \mid 0 \rangle =
\langle 0 \mid a^{\dagger}_{m} = \langle 0 \mid \tilde{a}^
{\dagger}_{m} = 0 
\end{equation}
\begin{equation}
\tilde{a}^{\dagger}_{m} \mid 0 \rangle = a^{\dagger}_{m} \mid 0 
\rangle , \;\;\;\;\;\; \langle 0 \mid \tilde{a}_{m} = \langle 0 \mid a_{m} .
\end{equation}
\es
The relations (\ref{al}) follow from the maps of Subsec.~2.4 because the 
corresponding matrix creation and annihilation operators 
$(a^{\dagger})_{rs},\; a_{rs}$ 
are densities.  We can also define time-dependent creation and annihilation 
operators
\bs \label{se}
\begin{equation}
a_{m} (t) = \frac{1}{\sqrt {2 \omega_{m}}} ( \omega_{m} \phi_{m}
 + i \pi_{m} ) = e^{-i \omega_{m} t} a_{m} \label{as}
\end{equation}
\begin{equation}
a^{\dagger}_{m} (t) = \frac{1}{\sqrt {2 \omega_{m}}} ( \omega_{m}
 \phi_{m} - i \pi_{m} ) = e^{i \omega_{m} t} a^{\dagger}_{m} 
 \label{at}
\end{equation}
\es
and similarly for $\tilde{a}_{m}(t), \; \tilde{a}^{\dagger}_{m} (t)$. 
The time-dependent creation/annihilation operators also satisfy 
(\ref{al}) and, similarly, the relations below can be read in terms of 
either the time-independent or the time-dependent operators.

In terms of these operators, the equal-time algebra (\ref{sc}) now 
reads
\bs \label{sf}
\begin{equation}
[ a_{m} , \tilde{a}^{\dagger}_{n} ] =[ \tilde{a}_{m} , a^{\dagger}_{n} ] 
= \delta_{mn} \mid 0 \rangle \langle 0 \mid 
\end{equation}
\begin{equation}
[ a_{m} , \tilde{a}_{n} ] = [ a^{\dagger}_{m} , \tilde{a}^{\dagger}_{n} ]
 = 0 \label{an}
\end{equation}
\begin{equation}
[ a_{m} , a^{\dagger}_{m} ] =[ \tilde{a}_{m} , \tilde{a}^{\dagger}_{m} ] 
= B-1 + \mid 0 \rangle \langle 0 \mid . \label{am}
\end{equation}
\es
We consider next the construction of complete sets of states, using 
(\ref{al}) and (\ref{sf}). Any state involving mixed untilde and tilde 
creation operators on the vacuum can be expressed entirely in terms 
of untilde creation operators, e.g.,
\begin{eqnarray}
&& a^{\dagger}_{m_{1}} \ldots a^{\dagger}_{m_{n}} 
\tilde{a}^{\dagger}_{n_{1}} \ldots \tilde{a}^{\dagger}_{n_{m}}
a^{\dagger}_{p_{1}} \ldots a^{\dagger}_{p_{q}} \mid 0 \rangle 
\nonumber \\
&& = a^{\dagger}_{m_{1}} \ldots a^{\dagger}_{m_{n}} 
 a^{\dagger}_{p_{1}} \ldots a^{\dagger}_{p_{q}}
a^{\dagger}_{n_{m}} \ldots a^{\dagger}_{n_{1}}  \mid 0 \rangle . 
\label{kk}
\end{eqnarray}
Similarly, mixed states involving $a$'s and $a^{\dagger}$'s can be 
expressed entirely in terms of $a^{\dagger}$'s.  To see this follow 
the steps
\begin{eqnarray}
a_{m} a^{\dagger}_{m_{1}} \ldots a^{\dagger}_{m_{n}} \mid 0 \rangle = 
a_{m} \tilde{a}^{\dagger}_{m_{n}} \ldots \tilde{a}^{\dagger}_{m_{1}} 
\mid 0 \rangle &=& [ a_{m} , \tilde{a}^{\dagger}_{m_{n}} \ldots 
\tilde{a}^{\dagger}_{m_{1}} ] \mid 0 \rangle \nonumber \\
 =  \tilde{a}^{\dagger}_{m_{n}} \ldots \tilde{a}^{\dagger}_{m_{2}} 
\delta_{m , m_{1}} \mid 0 \rangle &=& \delta_{m , m_{1}} a^{\dagger}_
{m_{2}} \ldots a^{\dagger}_{m_{n}} \mid 0 \rangle  \label{ap}
\end{eqnarray} 
where we have used the relation
\begin{equation}
[a_{m} , \tilde{a}^{\dagger}_{n}] \tilde{a}^{\dagger}_{p} = 
\delta_{mn} \mid 0 \rangle \langle 0 \mid \tilde{a}^{\dagger}_{p} = 0 
\label{kl}
\end{equation}
which follows from the equal-time algebra and (\ref{al}).  It follows 
from (\ref{kk}) and (\ref{ap}) that the $a^{\dagger}$ states are complete.

The relations 
 (3.6~-~3.8) are also true, 
however, under exchange of tilde and untilde labels, so that the 
$\tilde{a}^{\dagger}$ states are also complete
\bs
\begin{equation}
\{ a^{\dagger}_{m_{1}} \ldots a^{\dagger}_{m_{n}} \mid 0 \rangle \}  = 
\{ \tilde{a}^{\dagger}_{m_{1}} \ldots \tilde{a}^{\dagger}_{m_{n}} 
\mid 0 \rangle \}  =
\textup{complete set of states}
\end{equation}
\begin{equation}
\textbf{1}  =  \sum_{n=0}^{\infty} a^{\dagger}_{m_{1}} \ldots  a^
{\dagger}_{m_{n}}  \mid 0 \rangle \langle 0 \mid a_{m_{n}} \ldots 
a_{m_{1}}
 =  \sum_{n=0}^{\infty} \tilde{a}^{\dagger}_{m_{1}} \ldots  \tilde{a}^
{\dagger}_{m_{n}}  \mid 0 \rangle \langle 0 \mid \tilde{a}_{m_{n}} 
\ldots \tilde{a}_{m_{1}}
\end{equation} 
\begin{equation}
\tilde{a}^{\dagger}_{m_{1}} \ldots  \tilde{a}^{\dagger}_{m_{n}}  \mid 0 \rangle = 
a^{\dagger}_{m_{n}} \ldots  a^{\dagger}_{m_{1}}  \mid 0  \rangle . \label{ao}
\end{equation}
\es
Indeed, the tilde states can be rewritten in terms of the untilde 
states\footnote{For $B=1$, 
Eq.(\ref{ao}) implies that $\tilde{a}^{\dagger} = a^{\dagger}$
 and hence $\tilde{a} = a$, in accord with (\ref{af}).}, 
 as shown explicitly in (\ref{ao}).

Since (\ref{ap}) and its tilde $\leftrightarrow$ untilde version 
are true on complete sets of states, we have established the full 
equal-time algebra of the reduced creation/annihilation operators 
\bs \label{sg}
\begin{equation}
a_{m} \; a^{\dagger}_{n} = \tilde{a}_{m} \; \tilde{a}^{\dagger}_{n}
 = \delta_{mn}, \;\;\;\;\;\; m,n = 1 \ldots B  \label{aq}
\end{equation}
\begin{equation}
a^{\dagger}_{m} \; a_{m} = \tilde{a}^{\dagger}_{m} \; \tilde{a}_{m}
 = 1 - \mid 0 \rangle \langle 0 \mid  \label{ar}
\end{equation}
\begin{equation}
[a_{m}, \tilde{a}^{\dagger}_{n}] = [\tilde{a}_{m}, a^{\dagger}_{n}] = 
\delta_{mn} \mid 0 \rangle \langle 0 \mid  \label{ay}
, \;\;\;\;\;\; 
[a_{m}, \tilde{a}_{n}] = [a^{\dagger}_{m}, \tilde{a}^{\dagger}_{n}] = 0
 \label{az}
\end{equation}
\begin{equation}
a_{m} \mid 0 \rangle = \tilde{a}_{m} \mid 0 \rangle =
\langle 0 \mid a^{\dagger}_{m} = \langle 0 \mid \tilde{a}^
{\dagger}_{m} = 0 
\end{equation}
\begin{equation}
\tilde{a}^{\dagger}_{m} \mid 0 \rangle = a^{\dagger}_{m} \mid 0 \rangle, 
\;\;\;\;\;\; \langle 0 \mid \tilde{a}_{m} = \langle 0 \mid a_{m} . \label{az1}
\end{equation} 
\es
In particular, the  argument (\ref{ap}) on all $a^{\dagger}$ states gives the 
ordinary Cuntz relation (see (\ref{km})) in (\ref{aq}), and the 
tilde $\leftrightarrow$ untilde version of (\ref{ap}) gives the tilde 
Cuntz relation in (\ref{aq}).  Then (\ref{ar}) follows from (\ref{aq}) 
and (\ref{am}). In what follows, these algebras will be called 
\textit{symmetric Cuntz algebras}: Each 
algebra is symmetric under the interchange of tilde and untilde operators, 
and contains two Cuntz subalgebras (untilde and tilde).

\subsection{Reduced Hamiltonian}

Using (\ref{tr}) and the symmetric Cuntz algebra (\ref{sg}), it is 
straightforward to compute the large N ground state energy for the oscillators 
\begin{equation}
E_{0} =  \; \langle 0 \mid H \mid 0 \rangle = 
\frac{N^{2}}{2} \; \langle 0 \mid \sum_{m} ( \pi_{m} \pi_{m} + \omega^{2}
_{m }\phi_{m} \phi_{m}) \mid 0 \rangle = \frac{N^{2}}{2} \; \sum_{m}
 \omega_{m}
\end{equation}
where $H$ is the reduced Hamiltonian.  The reduced Hamiltonian also 
appears in the reduced equations of motion (\ref{aj}), (\ref{ak}) and, 
using these relations, it is not 
difficult to construct the reduced Hamiltonian explicitly in this case.

The reduced Hamiltonian has many equivalent forms, beginning with\footnote
{The $B=1$ form of (\ref{au}) was given in footnote 8 of 
Ref.~\cite{us} and a number operator of this type (with $\omega = 1$) 
was later considered for all B in Ref.~\cite{Gopa}.}
\bs \label{tj}
\begin{eqnarray}
H - E_{0}  =  \sum_{n=0}^{\infty} \;
a^{\dagger}_{m_{1}} \ldots a^{\dagger}_{m_{n}} \; (a^{\dagger} \omega 
a) \; a_{m_{n}} \ldots a_{m_{1}} \label{au} \\
 =  \sum_{n=0}^{\infty} \; \tilde{a}^{\dagger}_{m_{1}} \ldots \tilde{a}
^{\dagger}_{m_{n}} \; (\tilde{a}^{\dagger} \omega \tilde{a}) \; 
\tilde{a}_{m_{n}} \ldots \tilde{a}_{m_{1}} \label{av}
\end{eqnarray}
\begin{equation}
 (a^{\dagger} \omega a) = \sum_{m} a^{\dagger}_{m} \omega_{m} a_{m} .
\end{equation}
\es
These forms can be used to check the commutators
\bs \label{sh}
\begin{equation}
[H, \; a^{\dagger}_{m} ] = \omega_{m}  a^{\dagger}_{m} , \; \; \; \; 
\; \; [H, \; a_{m} ] = - \omega_{m}  a_{m} \label{aw}
\end{equation}
\begin{equation}
[H, \; \tilde{a}^{\dagger}_{m} ] = \omega_{m}  \tilde{a}^{\dagger}_{m} , 
\; \; \; \; \; \; [H, \; \tilde{a}_{m} ] = - \omega_{m}  \tilde{a}_{m} 
\label{ax}
\end{equation}
\es
which guarantee the correct equations of motion.  Here is a roadmap 
for checking these commutators, all four of which are true for both 
forms of $H$ in (\ref{tj}):  Consider first the untilde form of $H$ in (\ref{au}). 
In this case  
the commutators in (\ref{aw}) are easily checked by writing out each 
term,  using (\ref{aq}) and subtracting.  The commutators in 
(\ref{ax}) must be computed directly using the mixed commutators 
(\ref{ay}), and these come out as
\begin{equation}
[H, \; \tilde{a}^{\dagger}_{m}] = \omega_{m} \tilde{a}^{\dagger}_{m} 
 (\sum_{n=0}^{\infty} \; a^{\dagger}_{m_{1}} \ldots a^{\dagger}
_{m_{n}}\mid 0 \rangle \langle 0 \mid   a_{m_{n}} \ldots a_{m_{1}})
 =  \omega_{m} \tilde{a}^{\dagger}_{m} .
\end{equation}
For the tilde form of $H$ in (\ref{av}), the two types of computation 
above are reversed but the same results are obtained. The results in 
(\ref{sh}) 
also show that the two forms of $H$ in (\ref{tj}) are equal: The 
difference $\Delta$ of the two forms is zero because $\Delta$ 
annihilates the vacuum and commutes with all the operators of the 
theory, so that $\Delta = 0$ on any state.

Using (\ref{se}), we see that the reduced Hamiltonian $H$ in (\ref{tj}) is a
 highly nonlocal operator (see (\ref{tv})), and we will see this 
 nonlocality quite generally below for the reduced trace class operators
  of the various theories. This is the price one must pay in using 
  free algebras (which are not local commutators) to solve reduced 
  algebraic relations such as (\ref{sh}).
  
   There are other 
equivalent forms of $H$ which show its spectrum, e.g. 
\bs
\begin{equation}
H - E_{0} = \sum_{n=0}^{\infty} \;
a^{\dagger}_{m_{1}} \ldots a^{\dagger}_{m_{n}}\mid 0 \rangle \; E(m_{1}
\ldots m_{n} ) \; \langle 0 \mid   a_{m_{n}} \ldots a_{m_{1}} \label{ba}
\end{equation} 
\begin{equation}
(H-E_{0}) a^{\dagger}_{m_{1}} \ldots a^{\dagger}_{m_{n}}\mid 0 \rangle =
 E(m_{1} \ldots m_{n}) a^{\dagger}_{m_{1}} \ldots a^{\dagger}_{m_{n}}\mid 0 
\rangle
\end{equation}
\begin{equation}
 E(m_{1} \ldots m_{n}) = \sum_{i=1}^{n}  \omega_{m_{i}} 
\end{equation}
\es
and another form of $H$ is (\ref{ba}) with all operators tilded.

\subsection{Isotropic Oscillators and Angular Momentum}

We consider next the special case of $B$ isotropic oscillators 
$(\omega_{m} = \omega)$,
\begin{equation}
H - E_{0} = \omega \sum_{n=0}^{\infty} a^{\dagger}_{m_{1}} 
\ldots a^{\dagger}_{m_{n}} \; (a^{\dagger} 1 a) \; a_{m_{n}} \ldots a_{m_{1}}
\end{equation}
for which we are also able to find the explicit nonlocal structure of the 
reduced spin(B) generators discussed in Subsec.~2.6.  One form of the 
generators is
\bs
\begin{equation}
J_{mn} =  \sum_{n=0}^{\infty} a^{\dagger}_{m_{1}} \ldots a^{\dagger}_{m_{n}}
 \; (a^{\dagger} L_{mn} a) \; a_{m_{n}} \ldots a_{m_{1}} \label{bd}
\end{equation}
\begin{equation}
(a^{\dagger} L_{mn} a) = a^{\dagger}_{k} (L_{mn})_{kl} a_{l} =
i a^{\dagger}_{[m} a_{n]}, \;\;\;\;\;\; 
(L_{mn})_{kl} = i (\delta_{mk} \delta_{nl} - \delta_{nk} \delta_{ml} )
\end{equation}
\begin{equation}
J_{mn} \mid 0 \rangle = 0 , \; \; \; \; \; \; \dot{J}_{mn} = i [H, \; 
J_{mn} ] = 0
\end{equation}
\begin{equation}
[J_{mn}, \; a^{\dagger}_{k} ] = a^{\dagger}_{l} (L_{mn})_{lk}, \; \; 
\; \; \; \; [J_{mn}, \; a_{k} ] = a_{l} (L_{mn})_{lk} \label{bb}
\end{equation}
\begin{equation}
[J_{mn}, \; \tilde{a}^{\dagger}_{k} ] = \tilde{a}^{\dagger}_{l} (L_{mn})_{lk}, 
\; \; \; \; \; \; [J_{mn}, \; \tilde{a}_{k} ] = \tilde{a}_{l} (L_{mn})_{lk}
 \label{bc}
\end{equation}
\begin{equation}
[J_{mn}, J_{pq}] = i( \delta_{q [m} J_{n] p} -  \delta_{p [m} J_{n] q}) . 
\end{equation}
\es
The commutators in (\ref{bb}) and (\ref{bc}) are equivalent to (\ref{be}) and tell 
us that the 
reduced fields transform in the vector representation of spin(B). 
Another form of the reduced generators is obtained by replacing all 
the operators in (\ref{bd}) by tilde operators, as discussed above for $H$.

\subsection{Algebraic Identities}

We note here some generalizations of the algebraic identities 
 above. \nopagebreak 

The reduced Hamiltonian (\ref{au}) and the reduced angular momentum operators 
(\ref{bd}) are special cases of the  family  of nonlocal operators  
\bs
\begin{eqnarray}
{\cal M} (M) & = & \sum_{n=0}^{\infty} a^{\dagger}_{m_{1}} \ldots a^{\dagger}
_{m_{n}} \; (a^{\dagger} M a) \; a_{m_{n}} \ldots a_{m_{1}} \\
& = & \sum_{n=0}^{\infty} \tilde{a}^{\dagger}_{m_{1}} \ldots 
\tilde{a}^{\dagger}_{m_{n}} \; (\tilde{a}^{\dagger} M \tilde{a}) \; 
\tilde{a}_{m_{n}} \ldots \tilde{a}_{m_{1}}
\end{eqnarray}
\begin{equation}
(a^{\dagger} M a) = a^{\dagger}_{m} M_{mn} a_{n} 
\end{equation}
\es
where $M$ is any constant matrix. For each such $M$, we find that
\bs 
\begin{equation}
[ {\cal M}  (M) , \; a^{\dagger}_{m} ] = a^{\dagger}_{n} M_{nm}, \; \; 
\; \; \; \; [ {\cal M} (M) , \; a_{m} ] = - M_{mn} a_{n} \label{sss}
\end{equation}
\begin{equation}
[ {\cal M}  (M) , \; \tilde{a}^{\dagger}_{m} ] = \tilde{a}^{\dagger}_{n} M_{nm},
 \; \; \; \; \; \; [ {\cal M} (M) , \; \tilde{a}_{m} ] = - M_{mn} 
 \tilde{a}_{n}
\end{equation}
\es
and when $M$ and $N$ are any two constant matrices we also find
\begin{equation}
[ {\cal M} (M) , {\cal M} (N) ] =  {\cal M} ([M,N]) 
\end{equation}
so that the algebra of the ${\cal M}$'s is faithful to the algebra of the 
matrices. These identities include the nontrivial commutators among 
$H$ and $J_{mn}$ above, and allow, for example, the construction 
of general (reduced) Lie algebras when the reduced fields are in any matrix 
representation of the algebra.

Using the same constant matrices, we also consider a second family of nonlocal
operators
\begin{equation}
{\cal M}^{\diamond} (M) = \sum_{n=0}^{\infty} a^{\dagger}_{m_{1}} \ldots 
a^{\dagger}_{m_{n}} \; (a^{\dagger}_{m} \mid 0 \rangle  M_{mn} 
\langle 0 \mid a_{n}) \; a_{m_{n}} \ldots a_{m_{1}}
\end{equation}
which satisfy
\bs
\begin{equation}
[{\cal M}^{\diamond} (M), \; a^{\dagger}_{m}] = a^{\dagger}_{n} \mid 
0 \rangle \langle 0 \mid M_{nm}, \;\;\;\;\;\; 
[{\cal M}^{\diamond} (M), \; a_{m}] =  - M_{mn} \mid 0 \rangle \langle 
0 \mid a_{n}
\end{equation}
\begin{equation}
{\cal M}^{\diamond} (M) {\cal M}^{\diamond} (N) = {\cal M}^
{\diamond} (MN) . \label{bf}
\end{equation}
\es
We see in (\ref{bf}) that products of the ${\cal M}^{\diamond}$ operators follow 
the matrix products; and moreover we find that 
the  ${\cal M}^{\diamond}$ operators transform as 
\begin{equation}
[{\cal M} (M), \; {\cal M}^{\diamond} (N) ] = {\cal M}^{\diamond} ([M,N])
\end{equation}
so they form a representation of the 
algebra of ${\cal M}$ operators above. \nopagebreak 

\subsection{Large N Density-Trace Identifications}

In this section, we use the examples above to point out  a new phenomenon 
at large N which we call large N density-trace identification. This 
phenomenon 
involves an unexpected relation between trace class operators (such as the 
Hamiltonian, the angular momenta and the supercharges) and their densities 
at large N, and the phenomenon  constructs 
 new nonlocal  densities in the original unreduced  theory, which 
 are generically conserved only at large N. 

We have seen that the reduced conserved trace class operators $T$ 
of the theory have a highly nonlocal composite structure, although they 
are the images of local conserved trace class 
operators $T\dd = C(N) Tr (t)$ in the original unreduced large N theory. 
Given the composite structure of any such reduced operator $T$, however,  it 
is not difficult to work backward to construct a new  
nonlocal density class operator $D_{rs}$ which also corresponds at 
large N (via the density maps of Sec.~2) to $T$ in the reduced theory. 
It follows that $D_{rs}$ is 
itself conserved at large N, at least in the large N Hilbert space 
defined by (\ref{b}). Pictorially,  we find the 2 to 1 map
\bs
\begin{equation}
\left. \begin{array}{c} T\dd\;(local) \\ D_{rs}\;(nonlocal) \end{array}
\right\rangle \begin{array}{c} \; \\ {\longrightarrow} \\ ^{N} \end{array} 
T \; ( nonlocal )
\end{equation} 
\begin{equation}
\dot{T}\dd = \dot{T} = 0, \;\;\;\;\;\; \dot{D}_{rs} 
\;\stackrel{_{\textstyle =}}{_{_{N}}} \;  0
\end{equation}
\es
in which  both the conserved local trace class operator $T\dd$  and the 
large N-conserved nonlocal density  $D_{rs}$ correspond to the same 
conserved reduced 
operator $T$ at large N. As we will see in the examples, the new 
nonlocal density $D_{rs}$ can be understood as a  nonlocally dressed form of 
the local  density $t_{rs}$  of the original trace class operator $T\dd$. 

To illustrate this field identification phenomenon most simply, we consider  
the reduced Hamiltonian (\ref{au}) and reduced angular momentum generators 
(\ref{bd})  of  the $B$ isotropic oscillators at unit frequency 
($\omega_{m} = \omega = 1$) 
\bs \label{tv}
\begin{eqnarray}
H' \equiv H - E_{0} = \sum_{n=0}^{\infty} \;
\frac{1}{2^{n+1}} (\phi - i\pi)_{m_{1}} \ldots (\phi - i\pi)_{m_{n}}
\nonumber \\
\times(\pi_{m} \pi_{m} + \phi_{m} \phi_{m} + i[\phi_{m}, \; \pi_{m}] )
 (\phi + i\pi)_{m_{n}} \ldots (\phi + i\pi)_{m_{1}}
\end{eqnarray} 
\begin{eqnarray}
J_{mn} = \sum_{n=0}^{\infty} \;
\frac{1}{2^{n+1}} (\phi - i\pi)_{m_{1}} \ldots (\phi - i\pi)_{m_{n}}
\nonumber \\
\times (i (\phi - i \pi)_{[m} (\phi + i \pi )_{n]})
 (\phi + i\pi)_{m_{n}} \ldots (\phi + i\pi)_{m_{1}}
\end{eqnarray} 
\begin{equation}
\dot{H} = \dot{J}_{mn} = 0, \;\;\;\;\;\;
H' \mid 0 \rangle = J_{mn} \mid 0 \rangle = 0
\end {equation}
\es
where we have  used (\ref{se}) to reexpress $H$ and $J_{mn}$ in terms of the 
time dependent  reduced fields $\phi(t)$ and $\pi(t)$.  These forms 
are easily pulled back to new unreduced densities $H_{rs}$ and 
$(J_{mn})_{rs}$ 
\bs \label{tt}
\begin{eqnarray}
&&(H)_{rs}\equiv   \sum_{n=0}^{\infty} \;
\frac{1}{2^{n}} [ (\phi - i\pi)^{m_{1}} \ldots (\phi - i\pi)^{m_{n}}
\nonumber  \\
&& \;\;\;\; \times (h^{(0)}) (\phi + i\pi)^{m_{n}} \ldots (\phi + i\pi)^{m_{1}}]_{rs}
= ( h^{(0)})_{rs} + \ldots
\end{eqnarray}
\begin{equation}
 (h^{(0)})_{rs} = \frac{1}{2} (\pi^{m} \pi^{m} + \phi^{m} \phi^{m} + i[\phi^{m}, 
 \; \pi^{m}] )_{rs}
\end{equation} 
\begin{eqnarray}
&&(J_{mn})_{rs} \equiv \sum_{n=0}^{\infty} \;
\frac{1}{2^{n}}[ (\phi - i\pi)^{m_{1}} \ldots (\phi - i\pi)^{m_{n}} 
\nonumber \\
&& \;\;\;\; \times (j^{(0)}_{mn}) \;  (\phi + i\pi)^{m_{n}} \ldots (\phi + i\pi)^{m_{1}}]_{rs}
= (j^{(0)}_{mn})_{rs} + \ldots
\end{eqnarray} 

\begin{equation}
(j^{(0)}_{mn})_{rs} = \frac{i}{2} (\phi^{[m} \phi^{n]} + \pi^{[m} \pi^{n]} 
+ i (\phi^{[m} \pi^{n]} - \pi^{[m} \phi^{n]} ) )_{rs}
\end{equation}

\begin{equation}
 (\dot{H})_{rs} = (\dot{j}_{mn})_{rs} = 0, \;\;\;\;\;\;
H_{rs} \mid 0\dd \rangle = (J_{mn})_{rs} \mid 0\dd \rangle = 0
\end{equation}
\es
which also correspond at large N (via the density maps of Sec.~2.4) 
to the same reduced operators $H'$ and $J_{mn}$. 
Our construction guarantees that these new densities are conserved at 
large N, since they map to the conserved reduced trace class 
operators in this limit. (Oscillator examples are special in that the 
new densities are conserved at all N.)

To see that these new densities are nonlocally dressed forms of the original
energy and angular momentum densities of the theory, note that the first 
terms of the new densities satisfy
\bs
\begin{equation}
Tr( h^{(0)} ) = \frac{1}{2} Tr ( \pi^{m} \pi^{m} + 
\phi^{m} \phi^{m} ) - E_{0} = H\dd - E_{0}
\end{equation}
\begin{equation}
Tr (j^{(0)}_{mn}) = Tr (\pi^{[m} \phi^{n]} ) = J\dj^{mn}
\end{equation}
\es
where $H\dd$ and $J\dj^{mn}$ are the original Hamiltonian and angular 
momenta.

Finding new conserved nonlocal  quantities in oscillator theories is never 
surprising, but these new densities are important quantities in the large N 
theory since they map onto the important reduced trace class operators.
 Moreover, this 
field identification phenomenon is apparently \textit{universal} for each  
conserved trace class operator in any large N theory,   given the 
explicit composite structure of the reduced trace class operator. 
For more general matrix models, one expects that these  new nonlocal densities 
are generically conserved only  at large N, and only in the large N 
Hilbert space defined by (\ref{b}).  We will return to this  
phenomenon for oscillator supercharges and supercharge densities in Subsec.~5.3 
(see also Subsec.~4.6).

\subsection{Coordinate Bases and the Rank of the Equal-Time Algebras}

The oscillators also allow us to make some useful comments about the rank 
of the equal-time algebras and the corresponding coordinate bases.

For $B=1$, the rank of the equal-time algebra (\ref{ko})  is 1, and we may 
construct the coordinate eigenstates explicitly for the oscillator:
\bs
\begin{equation}
B=1, \; \; \; \; \; \; \phi \mid q \rangle = q \mid q \rangle
\end{equation}
\begin{equation}
\mid q \rangle = C_{1} \sum_{m=0}^{\infty} \frac{\sin ((m+1) \theta_{q} )} {\sqrt 
{\sin \theta_{q}}} (a^{\dagger})^{m} \mid 0 \rangle 
\end{equation}
\begin{equation}
{\sqrt 2} \cos \theta_{q} = q , \; \; \; \; \; {\textbf 1} = \int dq \; 
\mid q \rangle \langle q \mid .
\end{equation}
\es
The coordinate eigenstates are complete in this case, and similarly 
complete coordinate bases  
 provide the starting point for the solution 
 \cite{us} of the general 1-matrix model. 

For $B\geq 2$,  we find that the rank of 
the equal-time algebra 
(2.51a-c) is 2. Choosing the 
commuting set as $\phi_{1}$ and $\tilde{\phi}_{2}$, we may again construct the 
coordinate eigenstates explicitly for the oscillators:
\bs
\begin{equation}
B \geq 2 ,\; \; \; \; \; \;  [\phi_{1}, \; \tilde{\phi}_{2} ] = 0
\end{equation}
\begin{equation}
\phi_{1} \mid x y \rangle = x \mid x y \rangle , \; \; \; \; \; \; 
\tilde{\phi}_{2} \mid x y \rangle = y \mid x y \rangle
\end{equation}
\begin{equation}
\mid x y \rangle = C_{2} \sum_{m,n=0}^{\infty}  \frac{\sin ((m+1) \theta_{x} )} 
{\sqrt {\sin \theta_{x}}} \;  \frac{\sin ((n+1) \theta_{y} )} {\sqrt 
{\sin \theta_{y}}} \; (a^{\dagger}_{1})^{m} \; 
(a^{\dagger}_{2})^{n} \mid 0 \rangle 
\end{equation}
\begin{equation}
{\textbf 1} = \int dx \; dy \; \mid x y \rangle \langle x y \mid + 
\; \Delta  .
\end{equation}
\es
In this case, the coordinate eigenstates are explicitly not complete,
since they have no overlap with more complicated words such as  
$ a^{\dagger}_{1} a^{\dagger}_{2} a^{\dagger}_{1} \mid 0 \rangle$. It follows 
that the coordinate-basis approach of Ref.~\cite{us} cannot be extended 
to matrix models with $B \geq 2$.

\section{Interacting Symmetric Cuntz Algebras}

The symmetric Cuntz algebras of Sec.~3 arose in the context 
of large N oscillators, which may be considered to be free theories. In 
this section, we find the generalization of these algebras for arbitrary 
interactions, which we call interacting symmetric Cuntz algebras.  The 
final form of these algebras, and their associated new large 
N-conserved quantities, are found in Eq.~(\ref{sl}) and Subsec.~4.5 
respectively.

\subsection{Generalized Creation and Annihilation Operators}

We begin with the invariant, real and nodeless ground state wave function 
of the general bosonic system
\begin{equation}
\psi_{0} (\phi ) = \langle \phi \mid 0\dd \rangle, \;\;\;\;\;\; \phi = 
\{ \phi^{m}_{rs} \} 
\end{equation}
in the coordinate basis of the unreduced theory. The explicit 
form of the ground state will not be required in this construction. 
Operating with the matrix momenta defines a set of matrix-valued 
functions $F^{m}(\phi)$,
\begin{equation}
i \pi^{m}_{rs} \psi_{0} (\phi ) = \frac{\partial}{\partial \phi^{m}_{sr}}
\; \psi_{0} (\phi ) = - F^{m}_{rs} (\phi )  \; \psi_{0} (\phi ) 
, \;\;\;\;\;\; 
( F^{m}_{rs}(\phi ))^{\dagger} =  F^{m}_{sr}(\phi )
\end{equation}
which lead us to generalized matrix creation and annihilation 
operators
\bs \label{tl}
\begin{equation}
A^{m}_{rs} \equiv \frac{1}{\sqrt 2} (  F^{m}_{rs}(\phi ) \; + \; i 
\pi^{m}_{rs})
,\; \; \; \; \; \; (A^{m \dagger})_{rs} \equiv \frac{1}{\sqrt 2}
 (  F^{m}_{rs}(\phi ) \; - \; i \pi^{m}_{rs}) 
\end{equation}
\begin{equation}
 A^{m}_{rs}\;  \psi_{0} (\phi ) = 0 
,\; \; \; \; \; \; \psi_{0} (\phi ) \; (A^{m \dagger})_{rs} = 0  \label{bs}
\end{equation}
\es
for any interaction.  A useful property of this system is 
\bs
\begin{equation}  \nopagebreak 
0 = [A^{m}_{pq}, \; A^{n}_{rs}]\psi_{0} (\phi ) = \frac{i}{2} 
([\pi^{m}_{pq},\; F^{n}_{rs}]
\; -\; [\pi^{n}_{rs},\; F^{m}_{pq}] ) \; \psi_{0} (\phi ) \label{bi}
\end{equation}
\begin{equation}
[\pi^{m}_{pq},\; F^{n}_{rs}(\phi )] \; +\; [ F^{m}_{pq}(\phi ), 
\pi^{n}_{rs}] = 0 \label{bh}
\end{equation}
\es
where (\ref{bh}), which may be considered as the ground state 
integrability condition, 
follows from (\ref{bi}) because the ground state is nodeless. Further discussion 
of these operators in the unreduced theory is found in App.~E, where it 
is also shown that $F$, $A$ and $A^{\dagger}$  may be considered as densities 
at large N. Here we go directly to the reduced theory at large N.

Following the line of the canonical maps in Subsec.~2.4, we find first that
\bs
\begin{equation}
[\tilde{\pi}_{m},F_{n}(\phi )]-[\pi_{n},\tilde{F}_{m}(\tilde{\phi} 
)]=0 \label{bj}
\end{equation}
\begin{equation} 
F_{m}^{\dagger}(\phi )  =  F_{m}(\phi ), \;\;\;\;\;\;
\tilde{F}^{\dagger}_{m}(\tilde{\phi}) = \tilde{F}_{m}(\tilde{\phi}), 
\;\;\;\;\;\;
\tilde{F}_{m}(\tilde{\phi} ) \mid 0 \rangle  =  F_{m}(\phi ) 
\mid 0 \rangle \label{bl}
\end{equation} 
\es
where the reduced operators $F_{m}$, $\tilde{F}_{m}$ are the images of 
$F^{m}_{rs}$. The result (\ref{bj}), which is the image of (\ref{bh}), 
tells us that the pair $F = \{F_{m}, \tilde{F}_{m}\}$ comprises a flat 
connection, ${\cal F}_{mn}(F) = 0$,  on the reduced Hilbert space. It follows 
that  $F_{m}$ and $\tilde{F}_{m}$ have the form (see Subsec.~2.6)
\bs
\begin{equation}
F_{m}(\phi ) = \sum_{n=1}^{\infty} \; f^{(n)}_{m m_{2} \ldots m_{n}} \phi_{m_{2}}
\ldots \phi_{m_{n}}, \;\;\;\;\;\; 
\tilde{F}_{m}(\tilde{\phi}) = \sum_{n=1}^{\infty} \; f^{(n)}_{m m_{2} \ldots m_{n}} 
\tilde{\phi}_{m_{n}}\ldots \tilde{\phi}_{m_{2}}
\end{equation}
\begin{equation}
f^{(n)}_{m_{1} \ldots m_{n}} = f^{(n)}_{m_{2} \ldots m_{n}m_{1}} \label{cb}
, \;\;\;\;\;\; 
f^{(n)\;*}_{m_{1} \ldots m_{n}} = f^{(n)}_{m_{n} \ldots m_{1}} \label{bk}
\end{equation}
\es
where the as yet undetermined 
coefficients $f$ are cyclically symmetric in their lower indices. 
The reality condition in (\ref{bk}) follows from (\ref{bl}). 

We continue with the reduced creation and annihilation operators\footnote
{Using (\ref{ki}), the relation $A_{m}\!\mid 0 \rangle~=~0$ can be 
written as 
 $(\tilde{\pi}_{m} - iF_{m}(\phi))\!\mid 0 \rangle~=~0$. According to the 
remark below (\ref{li}), this is the Hamiltonian 
analogue of Haan's Euclidean equation of motion \cite{Haan, Gopa}.}
\bs \label{ta}
\begin{equation}
A_{m} = \frac{1}{\sqrt 2} (F_{m}+i\pi_{m}), \; \; \; \; \; \; 
A^{\dagger}_{m}= \frac{1}{\sqrt 2} (F_{m}-i\pi_{m}) \label{kj}
\end{equation}
\begin{equation}
\tilde{A}_{m} = \frac{1}{\sqrt 2} (\tilde{F}_{m}+i\tilde{\pi}_{m}), \; \; \; \; \; \; 
\tilde{A}^{\dagger}_{m}= \frac{1}{\sqrt 2} (\tilde{F}_{m}-i\tilde{\pi}_{m}) 
\end{equation}
\begin{equation}
A_{m}\mid 0 \rangle = \tilde{A}_{m}\mid 0 \rangle = \langle 0 \mid 
A^{\dagger}_{m} =  \langle 0 \mid \tilde{A}^{\dagger}_{m} = 0 \label{bu}
\end{equation}
\begin{equation}
\tilde{A}^{\dagger}_{m}\mid 0 \rangle = A^{\dagger}_{m}\mid 0 
\rangle, \; \; \; \; \; \; 
 \langle 0 \mid \tilde{A}_{m} =  \langle 0 \mid A_{m} \label{bo}
\end{equation}
\es
which are the images of the matrix creation and annihilation operators 
in (\ref{tl}). 
The state $\mid 0 \rangle$ is the reduced ground state of the interacting system. 
The equal-time algebra of these operators 
\bs \label{tb}
\begin{equation}
[A_{m},\tilde{A}_{n}] = [A^{\dagger}_{m},\tilde{A}^{\dagger}_{n}] = 0 
\label{bp1}
\end{equation}
\begin{equation}
[A_{m},\tilde{A}^{\dagger}_{n}] = [\tilde{A}_{n},A^{\dagger}_{m}] = 
i[\tilde{\pi}_{n},F_{m}] = i[\pi_{m},\tilde{F}_{n}] \label{bp}
\end{equation}
\es
follows directly from the equal-time algebra (\ref{sz}) and the flatness 
condition (\ref{bj}). The relations (\ref{bu}) and (\ref{bp1}) tell us that 
mixed words involving both $A^{\dagger}$'s and $\tilde{A}^{\dagger}$'s 
on the vacuum can be rewritten in terms of only  $A^{\dagger}$'s or 
only $\tilde{A}^{\dagger}$'s. We will argue in Subsec.~4.4 that both 
sets of states are complete, at least for potentials in some 
neighborhood of the oscillator potential. 

The equal-time algebra (\ref{sz})  also allows us to compute
\bs
\begin{equation}
[\tilde{\phi}_{m},A_{n}] = [\phi_{m},\tilde{A}_{n}] = -\frac{1}{\sqrt 
2} \; \delta_{mn} \mid 0 \rangle \langle 0 \mid 
\end{equation}
\begin{equation}
[\tilde{\phi}_{m},A^{\dagger}_{n}] = [\phi_{m},\tilde{A}^{\dagger}_{n}] =
 \frac{1}{\sqrt 2} \; \delta_{mn} \mid 0 \rangle \langle 0 \mid 
\end{equation}
\begin{equation}
[\tilde{\phi}_{l},A_{m}A^{\dagger}_{n}] = [\phi_{l},\tilde{A}_{m} 
\tilde{A}^{\dagger}_{n}] = 0 . \label{bm}
\end{equation}
\es
Equation (\ref{bm}) strongly suggests\footnote{One may conjecture 
that any reduced operator $X$ which obeys $[\tilde{\phi}_{m},X]=0, \forall m$ 
may be expressed as $X=X(\phi)$.} that the products $A A^{\dagger}$ 
and $\tilde{A} \tilde{A}^{\dagger}$ are equal to functions
of the reduced operators $\phi$ and $\tilde{\phi}$ respectively, and 
this intuition is confirmed in App.~E.
We will call these unknown functions $C$ and $\tilde{D}$:
\bs \label{tc}
\begin{equation}
A_{m}A^{\dagger}_{n} = C_{mn}(\phi), \; \; \; \; \; \; \tilde{A}_{m} 
\tilde{A}^{\dagger}_{n} = \tilde{D}_{mn}(\tilde{\phi}) \label{bn}
\end{equation}
\begin{equation}
C^{\dagger}_{mn} = C_{nm}, \;\;\;\;\;\; \tilde{D}^{\dagger}_{mn}
 = \tilde{D}_{nm} . \label{kv}
\end{equation}
\es
The relations (\ref{bn}) comprise two copies of a generalized free algebra 
for arbitrary interaction.

The functions $C$ and $\tilde{D}$ are closely related.  To see this,  
we first evaluate the functions on the ground state using (\ref{ta}), 
(\ref{tb}) and (\ref{tc}):
\begin{equation}
C_{mn}(\phi) \mid 0 \rangle = \tilde{D}_{nm}(\tilde{\phi})
 \mid 0 \rangle = i[\tilde{\pi}_{n},F_{m}]\mid 0 \rangle
= i[\pi_{m},\tilde{F}_{n}]\mid 0 \rangle . \label{bq}
\end{equation}
The first relation in (\ref{bq}) is easily solved as
\bs
\begin{equation}
C_{mn}(\phi) = \sum_{q=2}^{\infty} \; C^{(q)}_{mnm_{3} \ldots m_{q}} 
\; \phi_{m_{3}} \ldots \phi_{m_{q}}
\end{equation}
\begin{equation}
\tilde{D}_{mn}(\tilde{\phi}) = \sum_{q=2}^{\infty} \; C^{(q)}_{nmm_{3}
 \ldots m_{q}} \; \tilde{\phi}_{m_{q}} \ldots \tilde{\phi}_{m_{3}}
  = \tilde{C}_{nm}(\tilde{\phi})
\end{equation}
\begin{equation}
C_{mn}(\phi) = D_{nm}(\phi) \label{fv}
\end{equation}
\begin{equation}
C^{(q) \;*}_{mnm_{3} \ldots m_{q}} = C^{(q)}_{nmm_{q} \ldots m_{3}} \label{br}
\end{equation}
\es
where the coefficients $C^{(q)}$ are so far undetermined and 
the reality condition in (\ref{br}) follows from (\ref{kv}). 
The other relations 
in (\ref{bq}) will be helpful in computing the explicit forms of $C$ 
and $D$ below.

We turn now to some explicit computations involving the new operators, 
returning to formal developments, including completeness, in Subsec.~4.4.

\subsection{Starting from $F_{m}(\phi)$}

Given the reduced operators $F_{m}$ one can in principle compute 
the potential  of the system, as well as the commutators (\ref{bp}) and the 
functions $C$ and $D$ which enter into the generalized free 
 algebras 
 (\ref{tc}). 

We begin with the relations
\bs
\begin{equation}
0 = \dot{A}_{m} \mid 0 \rangle = (\dot{F}_{m} - i V_{m}') \mid 0 \rangle 
\label{bt}
\end{equation}
\begin{equation}
F_{m} = \sum_{n=1}^{\infty} \; f^{(n)}_{m m_{2} \ldots m_{n}} \phi_{m_{2}}
\ldots \phi_{m_{n}} \label{bz}
, \;\;\;\;\;\; 
V_{m}' = \sum_{n=1}^{\infty} \; v^{(n)}_{m m_{2} \ldots m_{n}} \phi_{m_{2}}
\ldots \phi_{m_{n}} \label{ca}
\end{equation}
\es
where (\ref{bt}) follows from (\ref{bu}) by the equations of motion 
(\ref{bv}). 
Consider the relation obtained by substitution of the forms (\ref{bz}) 
into (\ref{bt}), remembering that $\dot{\phi} = \pi$.
Using the equal-time algebra (\ref{sz}), it is not difficult to work out a  
sequence of relations  beginning with 
\bs
\begin{equation}
\pi_{m} \mid 0 \rangle =i F_{m}(\phi) \mid 0 \rangle
\end{equation}
\begin{equation}
\pi_{m} \phi_{n} \mid 0 \rangle = [\pi_{m},\tilde{\phi}_{n}] 
\mid 0 \rangle + i\tilde{\phi}_{n} F_{m}(\phi) \mid 0 \rangle = 
i (- \delta_{mn} +  F_{m}(\phi) \phi_{n}) \mid 0 \rangle
\end{equation}
\es
which can be used to eliminate all $\pi$ operators and rewrite the
 relation (\ref{bt}) in terms of the $\phi$ operators alone. The coefficient 
 of each  $\phi$ monomial must vanish separately,  allowing us
to compute $V$ in terms of $F$. We record the results of this computation 
including the $f$ coefficients through n=3:
\bs
\begin{equation}
v^{(1)}_{m} = f^{(2)}_{mn} f^{(1)}_{n} - f^{(3)}_{mnn} + \ldots
\end{equation}
\begin{equation}
v^{(2)}_{mn} = f^{(2)}_{mp} f^{(2)}_{pn} +(f^{(3)}_{mpn} +f^{(3)}_{npm})
f^{(1)}_{p} + \ldots
\end{equation}
\begin{equation}
v^{(3)}_{mnp} = f^{(2)}_{mq}f^{(3)}_{qnp} + f^{(2)}_{pq} f^{(3)}_{qmn}
+ f^{(2)}_{nq} f^{(3)}_{qpm} + \ldots
\end{equation}
\begin{equation}
v^{(4)}_{mnpq} = f^{(3)}_{mnr} f^{(3)}_{rpq} + f^{(3)}_{qmr} 
f^{(3)}_{rnp} + \ldots
\end{equation}
\begin{equation}
v^{(5)}_{mnpqr} = \ldots
\end{equation}
\es
where the dots indicate the contributions of $f^{(n)}, n\geq 4$. The symmetries
 of the $v$ coefficients in (\ref{bx})  are guaranteed by the symmetries 
 of the $f$ coefficients in (\ref{cb}).
This means that $V' = \{V'_{m},  \tilde{V}'_{m}\}$ is a flat connection, as 
it should be, 
when $F = \{F_{m},\;\tilde{F}_{m}\}$ is a flat connection.  Although we will 
not present 
the proof here, we have checked that  this statement is 
true to all orders in the expansions (\ref{bz}). 

We turn next to the evaluation of the functions $C$ and $\tilde{D}$ in the 
generalized free algebras (\ref{bn}). First evaluate the commutators 
(\ref{bp})
\begin{eqnarray}
[A_{m}, \; \tilde{A}^{\dagger}_{n}] = [\tilde{A}_{n}, \; A^{\dagger}_{m}]
= i[\tilde{\pi}_{n}, \; F_{m}]
= f^{(2)}_{mn} \mid 0 \rangle\langle 0 \mid  \nonumber \\ 
+ f^{(3)}_{mnp} \mid 0 \rangle\langle 0 \mid \phi_{p} +  
f^{(3)}_{mpn} \phi_{p} \mid 0 \rangle\langle 0 \mid + \ldots \label{cc}
\end{eqnarray} 
through this order in the $f$ coefficients, using (\ref{v}) and  (\ref{bz}). 
For  $C$ and $\tilde{D}$, we use (\ref{bq}), (\ref{fv}) and  (\ref{cc}) to 
evaluate
\bs
\begin{equation}
C_{mn}(\phi) \mid 0 \rangle = i [\tilde{\pi}_{n}, \; F_{m}] \mid 0 
\rangle \label{cd}
\end{equation}
\begin{equation}
C_{mn}(\phi) = f^{(2)}_{mn} + f^{(3)}_{mnp} \langle 0 \mid \phi_{p}
\mid 0 \rangle + f^{(3)}_{mpn} \phi_{p} + \ldots \label{ce}
\end{equation}
\begin{equation}
\tilde{D}_{mn}(\tilde{\phi}) = f^{(2)}_{nm} + f^{(3)}_{nmp} \langle 0
 \mid \phi_{p}\mid 0 \rangle + f^{(3)}_{npm} \tilde{\phi}_{p} + \ldots  \;\; 
\end{equation}
\es
where the result (\ref{ce}) is obtained from (\ref{cd}) by eliminating vacuum 
projectors to obtain a function of $\phi$ only on the vacuum. Another 
form of $\tilde{D}$ is found in Eq.~(\ref{lg}).

As a simple check on the results above, we note  the special case of 
the anharmonic oscillators
\bs \label{dm}
\begin{equation}
F_{m} = \omega_{m} \phi_{m}, \;\;\;\;\;\; V'_{m} = \omega^{2}_{m} \phi_{m},
\;\;\;\;\;\; C_{mn} = \omega_{m} \delta_{mn} 
\end{equation}
\begin{equation}
\tilde{F}_{m} = \omega_{m} \tilde{\phi}_{m}, \;\;\;\;\;\; \tilde{V}'_{m} =
 \omega^{2}_{m} \tilde{\phi}_{m},
\;\;\;\;\;\; D_{mn} = \omega_{m} \delta_{mn} 
\end{equation}
\es
for which  the rescaled  operators
\begin{equation}
\{ a_{m}, a^{\dagger}_{m}, \tilde{a}_{m}, \tilde{a}^{\dagger}_{m} \} \equiv 
\{ A_{m}, A^{\dagger}_{m}, \tilde{A}_{m}, \tilde{A}^{\dagger}_{m} 
\}/{\sqrt \omega_{m}} \label{dn}
\end{equation}
are seen to satisfy the symmetric Cuntz algebra (\ref{sg}).

\subsection{Basis-independent Analysis of the 1-Matrix Model}

We have noted in Subsec.~3.6 that the coordinate-basis approach of 
Ref.~\cite{us} cannot be extended to the case of many matrices. Here, we develop 
a basis-independent approach to the general 1-matrix model which 
constructs the generalized creation and annihilation operators 
$A^{\dagger}, A$ as well as the exact form  of $C(\phi)$ in their  
interacting Cuntz algebra
\begin{equation}
A A^{\dagger} = C(\phi) .
\end{equation}
This approach is in principle extendable to many matrices, although 
we will confine ourselves here to preliminary remarks in this direction.  
(In this subsection only, we use boldface {\boldmath $\pi$} for the momentum 
operators, to distinguish them from the \nopagebreak number $\pi$.)
 
We begin with the relations
\bs
\begin{equation}
\textup{\boldmath $\pi$} \mid 0 \rangle = iF\mid 0 \rangle ,\; \; \; \; \; \; 
\langle 0 \mid \textup{\boldmath $\pi$} = -i \langle 0 \mid F \label{cg}
\end{equation}
\begin{equation}
[i\textup{\boldmath $\pi$}, \; \frac{1}{z-\phi} ] = \frac{1}{z-\phi}\mid 0 
\rangle \langle 
0 \mid  \frac{1}{z-\phi}, \; \; \; \; \; \; Im \; z > 0 \label{ch}
\end{equation}
\begin{equation}
\langle 0 \mid [F(\phi) , \; \frac{1}{z-\phi} ]_{+} \mid 0 \rangle =
\langle 0 \mid  \frac{1}{z-\phi} \mid 0 \rangle ^{2} \label{cf}
\end{equation}
\es
where (\ref{cf}) follows from the vacuum properties (\ref{cg}) and the identity  
(\ref{ch}). We intend to let the complex variable $z$ approach the real axis  
$z \rightarrow q+i\epsilon$, where we will  need the following facts (${\cal P}$ 
is principal value)
\bs
\begin{equation}
\frac{1}{q - \phi +i\epsilon} = \frac{{\cal P}}{q -\phi} - i\pi \delta(q-\phi)
\end{equation}
\begin{equation}
\frac{{\cal P}}{q-a} \; \frac{{\cal P}}{q-b} = \frac{{\cal P}}{a-b}  
(\frac{{\cal P}}{q-a} - \frac{{\cal P}}{q-b} ) + \pi^{2} \delta (q-a) 
\; \delta (q-b) . \label{cn}
\end{equation}
\es
We also define the ground state density function 
$\rho (q)$ and the function $F(q)$
\bs
\begin{equation}
\rho(q) \equiv \langle 0 \mid \delta(q-\phi)\mid 0 \rangle \geq  0 , \; 
\; \; \; \; \; \int \; dq \; \rho(q) = 1 \label{co}
\end{equation}
\begin{equation}
F(q) \equiv \rho(q)^{-1} \langle 0 \mid F(\phi) \delta(q-\phi) \mid 0 \rangle 
\end{equation}
\es
where the latter is just $F(\phi)$ with $\phi$ replaced \nopagebreak  by  $q$. 

Letting $z$ approach the real axis, we find 
\bs
\begin{equation}
F(q) = \langle 0 \mid  \frac{{\cal P}}{q -\phi} \mid 0 \rangle = 
\int \; dq'\;  \frac{{\cal P}}{q -q'} \; \rho(q') \label{ci}
\end{equation}
\begin{equation}
\langle 0 \mid F(\phi) \; \frac{{\cal P}}{q-\phi} \mid 0 \rangle =
\frac{1}{2} F^{2}(q) - \frac{\pi^{2}}{2} \rho^{2}(q)
\end{equation}
\es
from the imaginary and real parts respectively of (\ref{cf}). The result in 
(\ref{ci}) gives $F(q)$ and the generalized creation and annihilation operators 
\begin{equation}
A = \frac{1}{\sqrt 2}( \int  dq \; \frac{{\cal P}}{\phi - q} \; \rho (q)
+ i \textup{\boldmath $\pi$}), \;\;\;\;\;\; A^{\dagger} = \frac{1}{\sqrt 2}
 (\int  dq \; 
\frac{{\cal P}}{\phi - q} \; \rho (q) - i \textup{\boldmath $\pi$})
\end{equation}
in terms of the ground state density $\rho$.

The result in (\ref{ci}) can also be used to compute $C(\phi)$ in terms of $\rho$ :
\bs
\begin{equation}
F(\phi) = Re \int \; dq \frac{1}{\phi - q -i\epsilon} \rho(q) \label{ck}
 , \;\;\;\;\;\;  C(\phi) \mid 0 \rangle = i 
 [\textup{\boldmath $\pi$}, F(\phi)] \mid 0 \rangle \label{cm}
\end{equation}
\begin{equation}
 i [\textup{\boldmath $\pi$}, F(\phi)] \mid 0 \rangle = - Re \int \; dq 
 \frac{1}{\phi - q -i
 \epsilon} \rho(q) \mid 0 \rangle \langle 0 \mid \frac{1}{\phi - q -i
 \epsilon} \mid 0 \rangle \label{cj}
\end{equation}
\begin{equation}
C(\phi) = - \int \; dq \; dq' \; \rho(q) \; \rho(q') \;  \frac{{\cal P}}
{q-q'}\;  \frac{{\cal P}}{q-\phi} + \pi^{2} \rho^{2} (\phi) . \label{cl}
\end{equation}
\es 
Here (\ref{cj}) follows from (\ref{ck}) and  (\ref{ch}),  while 
(\ref{cl}) follows from (\ref{ck}) by  rearranging  (\ref{cj}) into a 
function of $\phi$ on the vacuum.  A final  form for $C(\phi)$
\begin{equation}
C(\phi) = \frac{1}{2} (F^{2}(\phi) + \pi^{2} \rho^{2} (\phi)) \ge 0 \label{dg}
\end{equation}
is obtained by symmetrizing the double integral in (\ref{cl}) and using 
(\ref{cn}).

This completes the first stage of the analysis, in which we have 
expressed the new operators $F$, $A$, $A^{\dagger}$ and $C(\phi)$ in terms of the ground 
state density $\rho$.

In the second stage, we evaluate the ground state density $\rho$ in 
terms of the reduced potential $V$ of the system. We  begin this 
stage with  the  identity
\begin{eqnarray}
\langle 0 \mid \frac{1}{z-\phi} \; V'(\phi) \mid 0 \rangle  &=&  -
\langle 0 \mid \frac{1}{z-\phi} \; \dot{\textup{\boldmath $\pi$}} \mid 0 \rangle 
\nonumber \\
= \langle 0 \mid (\frac{d}{dt} \frac{1}{z-\phi}) \; \textup{\boldmath $\pi$} 
\mid 0 \rangle 
 &=& \langle 0 \mid \frac{1}{z-\phi} \; \textup{\boldmath $\pi$} \; 
 \frac{1}{z-\phi} \; \textup{\boldmath $\pi$}
 \mid 0 \rangle  
\nonumber \\
= \langle 0 \mid F \;\frac{1}{(z-\phi)^{2}} \;F \mid 0 \rangle  &-& 
\langle 0 \mid \frac{1}{z-\phi} \mid 0 \rangle \langle 0 \mid \frac{1}
{(z-\phi)^{2}} F \mid 0 \rangle \label{cp}
\end{eqnarray} \nopagebreak 
where we have used the fact that $\langle 0 \mid  \dot{A} \mid 0 \rangle = 0$ 
for any $A$.  Letting $z$ approach the real axis, taking the imaginary part and 
using $(z-\phi)^{-2} = - \partial_{z} (z-\phi)^{-1}$ along with previous 
formulas then allows us to compute $\rho$ as a function of $V$
\begin{equation}
V'(q) = -\pi^{2} \rho(q) \rho '(q) \; \;\longrightarrow \; \; \rho(q) = 
\frac{1}{\pi} \sqrt{2(\epsilon - V(q))}
\end{equation}
where the constant $\epsilon$ is determined by the normalization 
condition in (\ref{co}).  The ground state 
density  $\rho$ is  the same function introduced in 
Ref.~\cite{us}.

Although a complete discussion is beyond the scope of this paper, 
the basis-independent analysis above can be extended to many matrices, 
beginning with the extension of (\ref{ch}),
\begin{eqnarray}
B \geq 2: \;\;\;\;\;\;i[\tilde{\textup{\boldmath $\pi$}}_{k} , \; \frac{1}
{z_{m_{1}} - \phi_{m_{1}}}
 \ldots  \frac{1}{z_{m_{n}} - \phi_{m_{n}}}] = \sum_{i=1}^{n} \delta_{k,m_{i}}
\nonumber \\
\times \frac{1}{z_{m_{1}} - \phi_{m_{1}}} \ldots  \frac{1}{z_{m_{i}}
- \phi_{m_{i}}}\mid 0 \rangle \langle 0 \mid \frac{1}{z_{m_{i}} - 
\phi_{m_{i}}} \ldots  \frac{1}{z_{m_{n}}- \phi_{m_{n}}}
\end{eqnarray} 
and the corresponding extension of (\ref{cp}).

\subsection{``Ordinary'' Cuntz Algebras and Completeness in 
Interacting Theories}

In Subsecs.~4.1-4.3, we have constructed generalized creation and 
annihilation operators which satisfy generalized free algebras in 
interacting theories, but we have not yet discussed completeness for 
these operators.  Here we note first that ordinary Cuntz algebras 
can, under certain technical assumptions, be constructed in the 
interacting theories as well, and this will help us understand 
completeness in the case of the generalized operators.

The form of the generalized free algebra in (\ref{bn}) guarantees that 
$C_{mn}$ and $\tilde{D}_{mn}$ are non-negative operators.  The 
results of Subsec.~4.2 show that $C_{mn}$ and $\tilde{D}_{mn}$ are in 
fact positive operators at least where the potential of the 
interacting theory is in some (say perturbative) neighborhood of the 
oscillator potential.  Moreover (\ref{dg}) shows that $C$ is positive for 
almost all 1-matrix models.  The following discussion is limited 
to the broad class of theories for which these operators are strictly 
positive
\begin{equation}
C(\phi), \; \tilde{D}(\tilde{\phi}) > 0 \label{kx}
\end{equation}
although we do not yet have a complete characterization of these 
theories in terms of the potential.

For this class of theories, we can construct the ``ordinary'' Cuntz 
operators
\bs \label{si}
\begin{equation}
a_{m} \equiv (C^{-\frac{1}{2}})_{mn} A_{n} , \; \; \; \; \; \; 
a^{\dagger}_{m} \equiv  A^{\dagger}_{n}(C^{-\frac{1}{2}})_{nm} \label{di}
\end{equation}
\begin{equation}
\tilde{a}_{m} \equiv (\tilde{D}^{-\frac{1}{2}})_{mn} \tilde{A}_{n} , 
\; \; \; \; \; \; \tilde{a}^{\dagger}_{m} \equiv \tilde{A}^{\dagger}_{n}
(\tilde{D}^{-\frac{1}{2}})_{nm} \label{dj}
\end{equation}
\es
which satisfy the symmetric free algebra
\bs \label{tm}
\begin{equation}
a_{m} a^{\dagger}_{n} = \tilde{a}_{m} \tilde{a}^{\dagger}_{n} = 
\delta_{mn}
\end{equation}
\begin{equation}
a_{m} \mid 0 \rangle = \tilde{a}_{m} \mid 0 \rangle =
\langle 0 \mid a^{\dagger}_{m} = \langle 0 \mid \tilde{a}^{\dagger}_{m} = 0
\end{equation}
\es
as a consequence of (\ref{ta}) and (\ref{tc}). Using (\ref{bo}), (\ref{bp1}) and 
(\ref{si}), other relations, such as the mixed commutators 
$[a_{m}, \tilde{a}_{n}]$, $[a^{\dagger}_{m},\tilde{a}^{\dagger}_{n}]$, 
can also be computed in terms of $C$, $\tilde{D}$ and the Cuntz 
operators. 

The states formed by the $a^{\dagger}$'s or the $\tilde{a}^{\dagger}$'s 
on the vacuum should be complete at least in some neighborhood of the 
oscillator potential, and the expected completeness relations for the 
Cuntz operators
\begin{equation}
a^{\dagger}_{m} a_{m} = \tilde{a}^{\dagger}_{m} \tilde{a}_{m} = 
 1 - \mid 0 \rangle \langle 0 \mid \label{dk}
\end{equation}
follow with (\ref{tm}) by operating on the complete set of $a^{\dagger}$ 
or $\tilde{a}^{\dagger}$ states.  

The results 
 (4.32 - 4.34) complete the construction of 
a symmetric pair of Cuntz algebras in the interacting theories.  Our 
construction is in agreement with the complementary discussion 
of Ref.~\cite{qdef}, which assumed the existence of the (untilde) Cuntz algebra 
for potentials in a perturbative neighborhood of the oscillator. 

Returning to the generalized creation and annihilation operators, we 
can now show that the completeness of the $a^{\dagger}$ or 
$\tilde{a}^{\dagger}$ states is equivalent to the completeness of the 
generalized  $A^{\dagger}$ or $\tilde{A}^{\dagger}$ states:  The 
corresponding completeness relations for the generalized creation and 
annihilation operators
\begin{equation}
A^{\dagger}_{m} (C^{-1})_{mn} A_{n} = \tilde{A}^{\dagger}_{m} 
(\tilde{D}^{-1})_{mn} \tilde{A}_{n} = 1 - \mid 0 \rangle \langle 0 \mid 
\label{dl}
\end{equation}
follow immediately from (\ref{dk}) and (\ref{bn}).  These relations are also 
obtained by studying the action of the left hand sides of (\ref{dl}) on the
 $A^{\dagger}$ or $\tilde{A}^{\dagger}$ states, implying the 
 completeness of these sets of states as well.  The argument can 
  easily be run 
 backward, so that all four types ($a^{\dagger}, \; \tilde{a}^{\dagger}
 , \; A^{\dagger}, \; \tilde{A}^{\dagger}$) of completeness 
   are equivalent.
 
For reference we collect here the final form of our \textit{generalized or 
interacting symmetric Cuntz algebras}
\bs \label{sl}
\begin{equation}
A_{m}A^{\dagger}_{n} = C_{mn}(\phi), \; \; \; \; \; \; \tilde{A}_{m} 
\tilde{A}^{\dagger}_{n} = \tilde{D}_{mn}(\tilde{\phi}) 
\end{equation}
\begin{equation}
D_{mn}(\phi) = C_{nm}(\phi)
\end{equation}
\begin{equation}
[A_{m},\tilde{A}_{n}] = [A^{\dagger}_{m},\tilde{A}^{\dagger}_{n}] = 0 
\end{equation}
\begin{equation}
[A_{m},\tilde{A}^{\dagger}_{n}] = [\tilde{A}_{n},A^{\dagger}_{m}] = 
i[\tilde{\pi}_{n},F_{m}] = i[\pi_{m},\tilde{F}_{n}] 
\end{equation}
\begin{equation}
A^{\dagger}_{m} (C^{-1})_{mn} A_{n} = \tilde{A}^{\dagger}_{m} 
(\tilde{D}^{-1})_{mn} \tilde{A}_{n} = 1 - \mid 0 \rangle \langle 0 \mid
\end{equation}
\begin{equation}
A_{m}\mid 0 \rangle = \tilde{A}_{m}\mid 0 \rangle = \langle 0 \mid 
A^{\dagger}_{m} =  \langle 0 \mid \tilde{A}^{\dagger}_{m} = 0 
\end{equation}
\begin{equation}
\tilde{A}^{\dagger}_{m}\mid 0 \rangle = A^{\dagger}_{m}\mid 0 
\rangle, \; \; \; \; \; \; 
 \langle 0 \mid \tilde{A}_{m} =  \langle 0 \mid A_{m} 
\end{equation}
\es
which includes (\ref{dl}), the results of Subsec.~4.1 and assumes (\ref{kx}).  
For the special case of 
the oscillators (see (\ref{dm}) and (\ref{dn})), these generalized 
free algebras reduce to the symmetric Cuntz algebra (\ref{sg}).

The interacting symmetric Cuntz algebras (\ref{sl}) are a central 
result of this paper.  We turn now to two applications of these 
algebras.

\subsection{New Local Conserved Quantities at Large N}

Our first application is simple but quite remarkable. The interacting symmetric
 Cuntz algebras (\ref{sl}) and their associated 
``ordinary'' Cuntz algebras 
 (4.32 - 4.34) imply  \textit{new local conserved  quantities at large N}:
\bs \label{tu} 
\begin{equation}
\dot{{\cal J}}_{mn} = \dot{\cal J} = 0, \;\;\;\;\;\; m, n = 1 \ldots B
\end{equation} 
\begin{eqnarray}
{\cal J}_{mn} &\equiv&  a_{m} a^{\dagger}_{n} = (C^{-\frac{1}{2}})_{mp}
A_{p} A^{\dagger}_{q} (C^{-\frac{1}{2}})_{qn} \label{lp}\\ 
&=&  \frac{1}{2}[(C^{-\frac{1}{2}}(\phi))_{mp}(F_{p}(\phi) + i \pi_{p})
(F_{q}(\phi) - i \pi_{q})(C^{-\frac{1}{2}}(\phi))_{qn}] \nonumber
\end{eqnarray}
\begin{eqnarray} 
 {\cal J} &\equiv&  a^{\dagger}_{m} a_{m} = 
A^{\dagger}_{m} (C^{-1})_{mn} A_{n} \label{lq} \\ 
 &=&  \frac{1}{2}[(F_{m}(\phi) - i \pi_{m})
(C^{-1}(\phi))_{mn}(F_{n}(\phi) + i \pi_{n})] \nonumber 
\end{eqnarray}
\begin{equation}
{\cal J} \mid 0 \rangle = 0, \;\;\;\;\;\; {\cal J}^{\dagger} = {\cal J} 
\end{equation}
\es 
for all bosonic theories with $C, D > 0$. 
Similarly conserved operators $\tilde{\cal J}_{mn}$ and $\tilde{\cal 
J}$ are constructed by replacing $C \rightarrow \tilde{D}$ and each 
of the other operators by their tilde form.  \nopagebreak 

The conservation of ${\cal J}_{mn}$ in (\ref{tu}) understates the 
information we have because we also know that
\begin{equation}
{\cal J}_{mn} = \delta_{mn} \;\; \longleftrightarrow \;\; (A_{m} A^{\dagger}_{n} 
- C_{mn} (\phi)) = 0 
\end{equation}
which is properly interpreted as a set of $B^{2}$ local constraints at large N. 

For the 1-matrix model, the results of Subsec.~4.3, 
\bs
\begin{equation}
F(\phi) = \int dq \frac{\cal P}{\phi - q} \rho(q), \;\;\;\;\;\; 
C(\phi) = \frac{1}{2}(F^{2}(\phi) + \pi^{2} \rho^{2}(\phi)) 
\end{equation}
\begin{equation}
\rho(\phi) = \frac{1}{\pi} \sqrt{2(\epsilon - V(\phi))}, \;\;\;\;\;\; 
\int dq \rho (q) = 1 
\end{equation}
\es
give these new large N-conserved quantities in closed form, and it is 
possible in principle to evaluate (\ref{lp}) and (\ref{lq}) for higher B 
to any desired 
order in the coefficients $v^{(n)}$ of the potential (see 
Eq.~(\ref{ts})). Explicit forms of these reduced quantities 
can be pulled back (as in Subsec.~3.5) into hidden local (second order in 
momenta) but nonpolynomial (in coordinates)
unreduced densities $({\cal J}_{mn})_{rs}$ and  $({\cal J})_{rs}$ 
which  are conserved only at large N, and only in the large N Hilbert 
space of (\ref{b}).

These new large N-conserved quantities are another central result of 
this paper, since they apparently realize an old dream of hidden local conserved 
quantities in quantum field theory.  

\subsection{General Reduced Hamiltonian}

The interacting symmetric Cuntz algebras (\ref{sl}) and their associated 
``ordinary'' Cuntz algebras 
 (4.32 - 4.34) also allow us 
in principle to construct the general reduced Hamiltonian for bosonic 
systems with $C, D > 0$.

Following Subsecs.~3.2 and 3.4, we consider a fitting procedure based on 
a family of nonlocal reduced Hamiltonians,
\bs \label{to}
\begin{equation}
H' = H - E_{0} = \sum_{n=0}^{\infty} \; a^{\dagger}_{m_{1}} 
\ldots a^{\dagger}_{m_{n}}
\; h(a,a^{\dagger}) \;  a_{m_{n}} \ldots a_{m_{1}} \label{le}
\end{equation}
\begin{equation}
H'\mid 0 \rangle = h(a,a^{\dagger}) \mid 0 \rangle 
\end{equation}
\begin{equation}
\dot{a}^{\dagger}_{m} = i[H',\; a^{\dagger}_{m}] = ih(a,a^{\dagger}) 
a^{\dagger}_{m} \label{la}
, \;\;\;\;\;\; 
\dot{a}_{m} = i[H',\; a_{m}] = -i a_{m} h(a,a^{\dagger}) \label{lb}
\end{equation}
\begin{equation}
\pi_{m} = \frac{i}{\sqrt 2} (a^{\dagger}_{n} (C^{\frac{1}{2}})_{nm}
- (C^{\frac{1}{2}})_{mn} a_{n})
\end{equation}
\es
where the arbitrary hermitian operator $h(a,a^{\dagger})$ is to be determined in 
terms of the potential.  According to the Cuntz algebra, a formal 
solution of (\ref{la}) is
\bs
\begin{equation}
h = h^{\dagger} = i a^{\dagger}_{m} \dot{a}_{m} = -i 
\dot{a}^{\dagger}_{m} a_{m} \label{lc}
\end{equation}
\begin{equation}
H' \mid 0 \rangle = h \mid 0 \rangle = 0
\end{equation}
\es
and we can make contact with the theory in question by using (\ref{si}) to 
reexpress this system in terms of the interacting Cuntz operators.  This gives 
in particular the useful form of $h$:
\begin{eqnarray}
 h &=& i(A^{\dagger}C^{-\frac{1}{2}})_{m} \frac{d}{dt} (C^{-\frac{1}{2}} A)_{m}
  \label{ld} \\ \nonumber
 &=& \frac{i}{2} (F_{m} -i \pi_{m})[(C^{-\frac{1}{2}} \frac{d}{dt} 
C^{-\frac{1}{2}})_{mn}(F_{n} + i \pi_{n}) + (C^{-1})_{mn}
(\frac{d}{dt} F_{n} -i V'_{n})].
\end{eqnarray}
Using the data of Subsec.~4.3 it is straightforward to evaluate (\ref{ld}) 
and the reduced Hamiltonian $H'$ of the general 1-matrix model in 
closed form.  More generally, it is possible in principle to evaluate 
the reduced Hamiltonian (\ref{le}) to any  order in the coefficients
$v^{(n)}$ of the potential (see Eq.~(\ref{ts})). Explicit forms of the 
general reduced Hamiltonian can be pulled back (see Subsec.~3.5) into new large 
N-conserved unreduced 
densities $H_{rs}$, which also correspond at large N (by the density 
maps of Sec.~2) to the same reduced $H'$.

\section{Bose, Fermi and SUSY Oscillators}

\subsection{Symmetric Bose/Fermi/Cuntz Algebras}

We turn now to study the reduced equal-time algebra of  a set of B real 
bosonic and F complex fermionic oscillators. Drawing on the discussion 
of Subsec.~2.5 and Sec.~3, our goal in this subsection is to find the 
Bose/Fermi generalization of the symmetric Cuntz algebras (\ref{sg}) 
discussed above. 
  
For the reduced oscillators we may assume that
\bs
\begin{equation}
a_{m} \mid 0 \rangle = \tilde{a}_{m} \mid 0 \rangle = 0, \;\;\;\;\;\;
\tilde{a}^{\dagger}_{m} \mid 0 \rangle =  a^{\dagger}_{m} \mid 0 \rangle
\end{equation}
\begin{equation}
\psi_{\dot \alpha} \mid 0 \rangle = \tilde{\psi}_{\dot \alpha} \mid 0 
\rangle = 0, \;\;\;\;\;\; \tilde{\psi}^{\dagger}_{\dot \alpha} \mid 
0 \rangle  = \psi^{\dagger}_{\dot \alpha} \mid 0 \rangle
\end{equation}
\begin{equation}
m=1\ldots B, \; \; \; \; \; \; \dot \alpha = 1\ldots F, \;\;\;\;\;\;
F = \frac{f}{2} = \textup{integer}
\end{equation}
\es
and the equal-time algebra (see (\ref{tn})) takes the form
\bs
\begin{equation}
[ a_{m} , \tilde{a}^{\dagger}_{n} ] =[ \tilde{a}_{m} , a^{\dagger}_{n} ] 
= \delta_{mn} \mid 0 \rangle \langle 0 \mid 
\end{equation}
\begin{equation}
[ \psi_{\dot{\alpha}} , \tilde{\psi}^{\dagger}_{\dot{\beta}}]_{+} =
[ \tilde{\psi}_{\dot{\alpha}} , \psi^{\dagger}_{\dot{\beta}}]_{+} = 
\delta_{\dot{\alpha} \dot{\beta}} \mid 0 \rangle \langle 0 \mid
\end{equation}
\begin{equation}
[ a_{m} , \tilde{a}_{n} ] = [ a^{\dagger}_{m} , \tilde{a}^{\dagger}_{n} ]
 = 0 , \;\;\;\;\;\; 
[ \psi_{\dot{\alpha}} , \tilde{\psi}_{\dot{\beta}}]_{+} =
[ \psi^{\dagger}_{\dot{\alpha}} , \tilde{\psi}^{\dagger}_{\dot{\beta}}]_{+} = 0 
\end{equation}
\begin{equation}
[a_{m},a^{\dagger}_{m}] - [\psi_{\dot \alpha}, \;  \psi^{\dagger}_{\dot 
\alpha}]_{+} = [\tilde{a}_{m},\tilde{a}^{\dagger}_{m}] - [\tilde{\psi}
_{\dot \alpha}, \;  \tilde{\psi}^{\dagger}_{\dot 
\alpha}]_{+} = B-F-1+\mid 0 \rangle \langle 0 \mid . \label{cu}
\end{equation}
\es
The bosonic and fermionic operators also commute with each other when 
only one is tilded. Following Sec.~3, we
see that the set of all untilded words is complete
\begin{equation}
\{ \psi^{\dagger}_{{\dot\alpha}_{1}}\ldots a^{\dagger}_{m_{1}}
\ldots \psi^{\dagger}_{{\dot\alpha}_{n}}\ldots a^{\dagger}_{m_{p}}
\mid 0 \rangle\} = \textup{complete}
\end{equation}
as well as the set of all tilded words, and indeed that each word of 
one set can be reexpressed as a word in the other set.

Studying the action of the annihilation operators on the complete 
sets of states (see Eq.~(\ref{ap})) one finds the \textit{symmetric 
Bose/Fermi/Cuntz algebra}
\bs \label{sj}
\begin{equation}
a_{m} a^{\dagger}_{n} = \tilde{a}_{m} \tilde{a}^{\dagger}_{n} =\delta_{mn}
 \label{cq}
 , \;\;\;\;\;\; 
\psi_{\dot \alpha} \psi^{\dagger}_{\dot \beta} = \tilde{\psi}_{\dot \alpha} 
\tilde{\psi}^{\dagger}_{\dot \beta} =\delta_{{\dot \alpha},{\dot 
\beta}} \label{cr}
\end{equation}
\begin{equation}
a_{m}\psi^{\dagger}_{\dot \alpha} = \psi_{\dot \alpha} a^{\dagger}_{m} =
\tilde{a}_{m}\tilde{\psi}^{\dagger}_{\dot \alpha} = \tilde{\psi}_{\dot  
\alpha} \tilde{a}^{\dagger}_{m} = 0 \label{cs}
\end{equation}
\begin{equation}
a^{\dagger}_{m} a_{m} + \psi^{\dagger}_{\dot \alpha} \psi_{\dot \alpha} = 
\tilde{a}^{\dagger}_{m} \tilde{a}_{m} + \tilde{\psi}^{\dagger}_{\dot \alpha}
 \tilde{\psi}_{\dot \alpha} = 1-\mid 0 \rangle \langle 0 \mid  \label{ct}
\end{equation}
\begin{equation}
[a_{m},\; \tilde{a}^{\dagger}_{n}] = [\tilde{a}_{m},\; a^{\dagger}_{n}] =
 \delta_{mn} \mid 0 \rangle \langle 0 \mid \label{da}
\end{equation}
\begin{equation}
[\psi_{{\dot \alpha}},\; \tilde{\psi}^{\dagger}_{{\dot \beta}}]_{+} = [\tilde
{\psi}_{{\dot \alpha}},\; \psi^{\dagger}_{{\dot \beta}}]_{+} = \delta_
{\dot{\alpha} \dot{\beta}} \mid 0 \rangle \langle 0 \mid \label{db}
\end{equation}
\begin{equation}
[ a_{m} , \tilde{a}_{n} ] = [ a^{\dagger}_{m} , \tilde{a}^{\dagger}_{n} ]
 = [ \psi_{\dot{\alpha}} , \tilde{\psi}_{\dot{\beta}}]_{+} =
[ \psi^{\dagger}_{\dot{\alpha}} , \tilde{\psi}^{\dagger}_{\dot{\beta}}]_{+} = 0
 \label{dp}
\end{equation}
\begin{equation}
[a_{m},\;\tilde{\psi}_{\dot \alpha}] = [a_{m},\;\tilde{\psi}^{\dagger}
_{\dot \alpha}] = [\tilde{a}_{m},\;\psi_{\dot \alpha}] = [\tilde{a}_{m},\;
\psi^{\dagger}_{\dot \alpha}] = 0
\end{equation}
\begin{equation}
[a^{\dagger}_{m},\;\tilde{\psi}_{\dot \alpha}] = [a^{\dagger}_{m},\;\tilde
{\psi}^{\dagger}_{\dot \alpha}] = [\tilde{a}^{\dagger}_{m},\;\psi_{\dot 
\alpha}] = [\tilde{a}^{\dagger}_{m},\;\psi^{\dagger}_{\dot \alpha}] = 
0  \label{cv}
\end{equation}
\begin{equation}
a_{m} \mid 0 \rangle = 0, \;\;\;\; \tilde{a}^{\dagger}_{m} \mid 0 \rangle =
a^{\dagger}_{m} \mid 0 \rangle, \;\;\;\;
\psi_{\dot \alpha} \mid 0 \rangle = 0, \;\;\;\;
 \tilde{\psi}^{\dagger}_{\dot \alpha} \mid 
0 \rangle  = \psi^{\dagger}_{\dot \alpha} \mid 0 \rangle  \label{ct1}
\end{equation}
\begin{equation}
m,n = 1 \ldots B, \;\;\;\;\;\; \dot{\alpha}, \dot{\beta} = 1 \ldots F
\end{equation}
\es
where B and F are integers.
In particular, the relations (\ref{cq}),  (\ref{cs})
 come from the analysis 
of the annihilation operators, while the completeness relations (\ref{ct}) 
are the result of using these three relations in (\ref{cu}).

The symmetric Bose/Fermi/Cuntz algebra (\ref{sj}) is symmetric under 
interchange of tilde and untilde operators. It may also be considered 
as a family of algebras which interpolates from the symmetric 
Bose/Cuntz algebra (\ref{sg}) at $F=0$ to a symmetric 
Fermi/Cuntz algebra at $B=0$. In the special case of Fadeev-Popov 
ghosts, a subalgebra similar to the untilde part of  (\ref{cq}), (\ref{cs}) 
 was written down in Ref.~\cite{qdef}. \nopagebreak

\subsection{Cuntz Superalgebras}

A striking feature of the symmetric Bose/Fermi/Cuntz algebra (\ref{sj}) 
is that it contains an important free subalgebra which shows a 
Bose-Fermi equivalence. To highlight 
this fact, it is convenient to introduce the oscillator superfields  
\begin{eqnarray}
A_{M} = \left( \begin{array}{c}
a_{m} \\ \psi_{\dot \alpha} \end{array} \right), \; \; \; \; \;  
A^{\dagger}_{M} = \left( \begin{array}{c}
a^{\dagger}_{m} \\ \psi^{\dagger}_{\dot \alpha} \end{array} \right), & &
    \tilde{A}_{M} = \left( \begin{array}{c}
\tilde{a}_{m} \\ \tilde{\psi}_{\dot \alpha} \end{array} \right), \; \; \; 
\; \;  \tilde{A}^{\dagger}_{M} = \left( \begin{array}{c}
\tilde{a}^{\dagger}_{m} \\ \tilde{\psi}^{\dagger}_{\dot \alpha} 
\end{array} \right) \nonumber \\
M & = &1 \ldots (B+F) \label{cz}
\end{eqnarray}
in terms of which the free subalgebra 
(5.4a-c)  and (\ref{ct1})  takes
 the Bose-Fermi equivalent form
\bs \label{sk}
\begin{equation}
A_{M} A^{\dagger}_{N} = \tilde{A}_{M} \tilde{A}^{\dagger}_{N} = 
\delta_{MN} \label{cw}
\end{equation}
\begin{equation}
A^{\dagger}_{M} A_{M} = \tilde{A}^{\dagger}_{M} \tilde{A}_{M} =
 1-\mid 0 \rangle \langle 0 \mid \label{cx}
\end{equation}
\begin{equation}
A_{M} \mid 0 \rangle = \tilde{A}_{M} \mid 0 \rangle = 0, \;\;\;\;\;\;
A^{\dagger}_{M} \mid 0 \rangle = \tilde{A}^{\dagger}_{M} \mid 0 \rangle .
\label{cy}
\end{equation}
\es
This algebra is a subalgebra of a symmetric Cuntz algebra, but since 
it contains both bose and fermi oscillators, we will refer to it  as a 
\textit{symmetric Cuntz superalgebra}.

It is well known that the Cuntz algebras, being free algebras, dictate 
classical or Boltzmann statistics for the states. The states of the 
symmetric Bose/Fermi/Cuntz algebra (\ref{sj}) are formed 
by the application of any number of 
$A^{\dagger}$'s (or $\tilde{A}^{\dagger}$'s) on the vacuum
\begin{equation}
\{ A^{\dagger}_{M_{1}}\ldots A^{\dagger}_{M_{n}} \mid 0 \rangle\} = 
\{ \tilde{A}^{\dagger}_{M_{1}}\ldots \tilde{A}^{\dagger}_{M_{n}} \mid 0 \rangle\}
 =  \textup{complete}
\end{equation}
so the free superalgebra (\ref{sk}) dictates the same 
Boltzmann statistics for the large N fermions and bosons.  In particular, 
the Pauli Principle is lost for large N fermions.
The Bose-Fermi equivalence is also related to the fact 
that the identification $\tilde{\phi} = \phi,\; \tilde{\pi} = \pi$ 
(see Subsec.~2.6) is lost for $B=1$ 
when $F \neq 0$,  just as it is when $F=0$ and $B \geq 2$.

Finally, we emphasize that the Bose-Fermi equivalence seen in the 
free subalgebra 
(5.4a-c), (\ref{ct1}) is not sustained in  the full 
 symmetric Bose/Fermi/ Cuntz algebra (\ref{sj}), where
  the mixed relations 
(5.4d-f)  distinguish bosons from  fermions. 
 See Subsec.~5.5 for the superfield form of the full algebra. 

\subsection{One SUSY Oscillator}

As a simple example, we consider the large N formulation and solution of the 
($\omega = 1$)  supersymmetric 1-matrix oscillator, with supercharges
$Q\dd$ and $\bar{Q}\dd$. The unreduced form of this system is
\bs \label{tw}
\begin{equation}
Q\dd = Tr(\psi \pi^{+}), \; \; \; \; \; \; \bar{Q}\dd = Tr(\psi^{\dagger} 
\pi^{-}) , \; \; \; \; \; \; \pi^{\pm}_{rs} = (\pi \pm i \phi)_{rs} \label{dq}
\end{equation}
\begin{equation}
[\phi_{rs},\pi_{pq}] = i\delta_{sp} \delta_{rq}, \;\;\;\;\;\; 
[\psi_{rs}, \psi^{\dagger}_{pq}]_{+} = \delta_{sp} \delta_{rq}
\end{equation}
\begin{equation}
Q\dd^{2} = \bar{Q}\dd^{2} = 0, \;\;\;\;\;\; [Q\dd, \bar{Q}\dd]_{+} = 2 
H\dd
\end{equation}
\begin{equation}
H\dd = \frac{1}{2} Tr(\pi^{2} + \phi^{2} + [\psi^{\dagger},\psi]) \label{dr}
\end{equation}
\begin{equation}
Q\dd \mid 0\dd \rangle = \bar{Q}\dd \mid 0\dd \rangle = H\dd \mid 0\dd \rangle = 0 
\end{equation}
\es
together with implied relations such as
\begin{equation}
\dot{Q}\dd = \dot{\bar{Q}}\dd = \dot{H}\dd = [Q\dd,H\dd] = 
[\bar{Q}\dd,H\dd] = 0
\end{equation}
and the algebra of the supercharges with the  fields.

For the reduced formulation we will temporarily employ a mixed notation, 
including sometimes the component fields and sometimes the 
\nopagebreak superfield 
notation (\ref{cz}) with $M=1,2$:
\bs
\begin{equation}
A_{M} = \left( \begin{array}{c}
a \\ \psi \end{array} \right), \; \; \; \; \;  
A^{\dagger}_{M} = \left( \begin{array}{c}
a^{\dagger} \\ \psi^{\dagger} \end{array} \right)  ,
 \; \;  \tilde{A}_{M} = \left( \begin{array}{c}
\tilde{a} \\ \tilde{\psi} \end{array} \right), \; \; \; \; \;  
\tilde{A}^{\dagger}_{M} = \left( \begin{array}{c}
\tilde{a}^{\dagger} \\ \tilde{\psi}^{\dagger} \end{array} \right) 
\end{equation}
\begin{equation}
a = \frac{i}{\sqrt 2} \pi_{-} , \;\;\;\;\;\; a^{\dagger} =
 - \frac{i}{\sqrt 2} \pi_{+}, \;\;\;\;\;\; \pi_{\pm} = \pi \pm i \phi 
\end{equation}
\es
where the superfields satisfy the Cuntz superalgebra (\ref{sk}) 
with  $B+F=2$. The remaining components of the reduced equal-time algebra are
\bs \label{ti}
\begin{equation}
[a,\tilde{a}^{\dagger}] = [\tilde{a},a^{\dagger}] = [\psi, 
\tilde{\psi}^{\dagger}]_{+} = [\tilde{\psi},\psi^{\dagger}]_{+} = \; 
\mid 0 \rangle \langle 0 \mid \label{ks}
\end{equation}
\begin{equation}
[a,\tilde{a}] = [a^{\dagger},\tilde{a}^{\dagger}] = [\psi, 
\tilde{\psi}]_{+} = [\psi^{\dagger},\tilde{\psi}^{\dagger}]_{+} = 0 
\label{kt}
\end{equation}
\begin{equation}
[a,\tilde{\psi}] = [a,\tilde{\psi}^{\dagger}] = [\tilde{a}, 
\psi] = [\tilde{a},\psi^{\dagger}] = 0
\end{equation}
\begin{equation}
[a^{\dagger},\tilde{\psi}] = [a^{\dagger},\tilde{\psi}^{\dagger}] = [\tilde{a}
^{\dagger}, \psi] = [\tilde{a}^{\dagger},\psi^{\dagger}] = 0 
\end{equation}
\es
which can also be written in terms of the 
superfields (see Subsec.~5.5).

The reduced equations of motion, involving the reduced Hamiltonian $H$, are
\bs \label{tf}
\begin{equation}
\dot{A}_{M} = i[H,A_{M}] = -i A_{M}, \;\;\;\;\;\;
\dot{A}^{\dagger}_{M} = i[H,A^{\dagger}_{M}] = i A^{\dagger}_{M}
\end{equation}
\begin{equation}
\dot{\tilde{A}}_{M} = i[H,\tilde{A}_{M}] = -i\tilde{A}_{M}, 
\;\;\;\;\;\; \dot{\tilde{A}}^{\dagger}_{M} = i[H,\tilde{A}^
{\dagger}_{M}] = i \tilde{A}^{\dagger}_{M} \label{hj}
\end{equation}
\es
and the ground state energy of 
the system is evaluated with (\ref{cy}) as
\begin{eqnarray}
E_{0} = \langle 0 \mid H \mid 0 \rangle & = & \frac{N^{2}}{2} \langle 0 \mid 
[a^{\dagger} ,a]_{+} + [\psi^{\dagger},\psi]  \mid 0 \rangle 
\nonumber \\ & = & 
\frac{N^{2}}{2} \langle 0 \mid [a ,\tilde{a}^{\dagger}] - [\psi,
\tilde{\psi}^{\dagger}]_{+}  \mid 0 \rangle = 0 .
\end{eqnarray}
The properties of the reduced supercharges $Q$ and 
$\bar{Q}$ also follow from the maps of Sec.~2: 
\bs \label{tg}
\begin{equation}
Q^{2} = \bar{Q}^{2} = 0,\;\;\;\;\;\;[\bar{Q},Q]_{+} = 2H
\end{equation}
\begin{equation}
\dot{Q} = \dot{\bar{Q}} = \dot{H} = [Q,H] = [\bar{Q},H] = 0
\end{equation}
\begin{equation}
Q\mid 0 \rangle = \bar{Q}\mid 0 \rangle = H\mid 0 \rangle = 0
\end{equation}
\begin{equation}
\langle 0 \mid Q\mid 0 \rangle = \langle 0 \mid i {\sqrt 2} \psi 
a^{\dagger}\mid 0 \rangle = 0, \;\;\;\;\;\;
\langle 0 \mid \bar{Q} \mid 0 \rangle = \langle 0 \mid -i {\sqrt 2} 
\psi^{\dagger} a \mid 0 \rangle = 0
\end{equation}
\es
\bs \label{th}
\begin{equation}
[Q,a] = -i{ \sqrt 2} \psi, \;\;\;\;\;\; [\bar{Q},a^{\dagger}] =
-i {\sqrt 2} \psi^{\dagger}
\end{equation}
\begin{equation}
[Q,\psi^{\dagger}]_{+} = i{ \sqrt 2}a^{\dagger}, \;\;\;\;\;\; 
[\bar{Q},\psi]_{+} = -i{ \sqrt 2}a 
\end{equation}
\begin{equation}
[Q,a^{\dagger}] = [\bar{Q},a] = [Q,\psi]_{+} = 
[\bar{Q},\psi^{\dagger}]_{+} = 0
\end{equation}
\es

\bs \label{sm}
\begin{equation}
[Q,\tilde{a}] = -i{ \sqrt 2} \tilde{\psi}, \;\;\;\;\;\; 
[\bar{Q},\tilde{a}^{\dagger}] = -i {\sqrt 2} \tilde{\psi}^{\dagger}
\end{equation}
\begin{equation}
[Q,\tilde{\psi}^{\dagger}]_{+} = i{ \sqrt 2}\tilde{a}^{\dagger}, \;\;\;\;\;\; 
[\bar{Q},\tilde{\psi}]_{+} = -i{ \sqrt 2}\tilde{a} 
\end{equation}
\begin{equation}
[Q,\tilde{a}^{\dagger}] = [\bar{Q},\tilde{a}] = [Q,\tilde{\psi}]_{+} = 
[\bar{Q},\tilde{\psi}^{\dagger}]_{+} = 0 .
\end{equation}
\es
Owing to the opacity phenomenon for trace class operators (see 
Subsec.~2.3), we do not yet know the composite form of the reduced 
supercharges and the reduced Hamiltonian.

Drawing on experience in earlier sections, we have solved the relations
 (\ref{tf}) and 
  (5.14 - 5.16) to obtain the explicit form
  of the reduced
supercharges and Hamiltonian. The results can be expressed entirely 
in terms of the superfields
\bs
\begin{eqnarray}
Q & = & i \sum_{n=0}^{\infty} \; (A^{\dagger}\tau_{3})_{M_{1}} \ldots 
(A^{\dagger}\tau_{3})_{M_{n}}(A^{\dagger} {\sqrt 2}\tau_{+}A)
A_{M_{n}} \ldots A_{M_{1}} \label{ha} \\
& = &  i \sum_{n=0}^{\infty} \; (\tilde{A}^{\dagger}\tau_{3})_{M_{1}} \ldots 
(\tilde{A}^{\dagger}\tau_{3})_{M_{n}}(\tilde{A}^{\dagger} {\sqrt 2}\tau_{+}
\tilde{A}) \tilde{A}_{M_{n}} \ldots \tilde{A}_{M_{1}} \label{hg}
\end{eqnarray}
\begin{eqnarray}
\bar{Q} & = &- i \sum_{n=0}^{\infty} \; (A^{\dagger}\tau_{3})_{M_{1}} \ldots 
(A^{\dagger}\tau_{3})_{M_{n}}(A^{\dagger} {\sqrt 2}\tau_{-}A)
A_{M_{n}} \ldots A_{M_{1}} \label{hb} \\
 &= & - i \sum_{n=0}^{\infty} \; (\tilde{A}^{\dagger}\tau_{3})_{M_{1}} \ldots 
(\tilde{A}^{\dagger}\tau_{3})_{M_{n}}(\tilde{A}^{\dagger} {\sqrt 
2}\tau_{-} \tilde{A}) \tilde{A}_{M_{n}} \ldots \tilde{A}_{M_{1}} \label{hh}
\end{eqnarray}
\begin{eqnarray}
H & = &  \sum_{n=0}^{\infty} \; A^{\dagger}_{M_{1}} \ldots 
A^{\dagger}_{M_{n}}(A^{\dagger}A) A_{M_{n}} \ldots A_{M_{1}} \label{hc} \\
& = &  \sum_{n=0}^{\infty} \; \tilde{A}^{\dagger}_{M_{1}} \ldots 
\tilde{A}^{\dagger}_{M_{n}}(\tilde{A}^{\dagger}\tilde{A}) \tilde{A}_{M_{n}}
 \ldots \tilde{A}_{M_{1}} \label{hi}
\end{eqnarray}
\begin{equation}
\tau_{+} = \left( \begin{array}{c} 0\;\; 1 \\ 0\;\;0 \end{array} \right)
, \;\;\;\;\;\; \tau_{-} = \left( \begin{array}{c} 0\;\; 0 \\ 1\;\;0 
\end{array} \right) , \;\;\;\;\;\; \tau_{3} = \left( \begin{array}{c}
 1\;\;\;\;\; 0 \\ 0 \;-1 \end{array} \right)
\end{equation}
\es
where the Pauli matrices $\tau$ operate in the reduced two 
dimensional superspace $(\tau A)_{M} = \tau_{MN} A_{N}$.

The reduced supercharges and Hamiltonian can be understood
as nonlocally dressed forms of their ``zeroth order'' factors
\bs
\begin{equation}
Q = i A^{\dagger}{\sqrt 2} \tau_{+}A + \ldots = i{\sqrt 2} a^{\dagger} \psi
 + \ldots
\end{equation}
\begin{equation}
\bar{Q} = - i A^{\dagger}{\sqrt 2} \tau_{-}A + \ldots = -i{\sqrt 2} \psi^
{\dagger} a + \ldots
\end{equation}
\begin{equation}
H = A^{\dagger} A + \ldots = a^{\dagger} a + \psi^{\dagger} \psi + \ldots
\end{equation}
\es
which closely resemble the unreduced supercharge and energy densities 
in (\ref{dq}) and (\ref{dr}). \nopagebreak 

Indeed, following the discussion of large N field identification 
in Subsec.~3.5, we may construct the new nonlocal unreduced densities
$Q_{rs}$, $\bar{Q}_{rs}$ and $H_{rs}$,
\bs \label{tq}
\begin{equation}
\dot{Q}_{rs} = \dot{\bar{Q}}_{rs} = \dot{H}_{rs} = 0, \;\;\;\;\;\; r,s = 1 \ldots N
\end{equation}
\begin{equation}
Q_{rs} \mid 0\dd \rangle = \bar{Q}_{rs} \mid 0\dd \rangle = H_{rs} \mid 0\dd 
\rangle = 0
\end{equation}
\es
which also correspond at large N to the same reduced supercharges and 
Hamiltonian\footnote{It is again an oscillator artifact that the 
relations (\ref{tq}) are true at finite N.}.  These densities have exactly the 
forms given in (\ref{ha}),
(\ref{hb}) and
(\ref{hc}), now with all operators interpreted as matrix-valued operators
\begin{equation}
(A_{M})_{rs} = \left( \begin{array}{c} \frac{i}{\sqrt 2} \pi^{-}_{rs}
\\ \psi_{rs} \end{array} \right) , \;\;\;\;\;\;
(A^{\dagger}_{M})_{rs} = \left( \begin{array}{c} - \frac{i}{\sqrt 2} 
\pi^{+}_{rs} \\ \psi^{\dagger}_{rs} \end{array} \right)
\end{equation}
and all products as matrix products. The operators $\pi^{\pm}_{rs}$ are defined
 in (\ref{dq}). As an example, the first two terms of the new supercharge 
density   are 
\bs
\begin{eqnarray}
Q_{rs} & = & i \sum_{n=0}^{\infty}  \{(A^{\dagger}\tau_{3})_{M_{1}} \ldots 
(A^{\dagger}\tau_{3})_{M_{n}}(A^{\dagger} {\sqrt 2}\tau_{+}A)
A_{M_{n}} \ldots A_{M_{1}} \}_{rs} \\
& = & i (A^{\dagger}_{M})_{rt} {\sqrt 2}(\tau_{+})_{MN}(A_{N})_{ts}
 \nonumber \\ & + &
i((A^{\dagger}\tau_{3})_{M})_{rt} (A^{\dagger}{\sqrt 2}\tau_{+}A)
_{tu} (A_{M})_{us} + \ldots \;\; . 
\end{eqnarray}
\es 
As in Sec.~3.5 we find that the new conserved densities are dressed nonlocal
forms
\bs
\begin{equation}
Q_{rs} = \pi^{+}_{rt} \psi_{ts} + \ldots, \;\;\;\;\;\; 
\bar{Q}_{rs} = \psi^{\dagger}_{rt} \pi^{-}_{ts} + \ldots
\end{equation}
\begin{equation}
H_{rs} = \frac{1}{2} \pi^{+}_{rt} \pi^{-}_{ts} + \psi^{\dagger}_{rt}\psi_{ts}
 + \ldots
\end{equation}
\begin{equation}
Q_{rr} = Q. + \ldots, \;\;\;\;\;\; \bar{Q}_{rr} = \bar{Q}. + \ldots,
\;\;\;\;\;\;H_{rr} = H. + \ldots
\end{equation}
\es
of the original unreduced supercharge and energy densities.

\subsection{Bosonic Construction of Supersymmetry}

We note here that the Bose-Fermi equivalence of the free superalgebra 
(\ref{sk})  allows a purely bosonic construction of supersymmetry
at large N.

Suppose that the unreduced operator $\psi_{rs} \rightarrow b_{rs}$ above was 
a complex boson. 
This does not change the free algebra (\ref{sk}) of the reduced operators 
\begin{equation}
A = \left( \begin{array}{c} a \\ b \end{array} \right), \;\;\;\;\;\;
A^{\dagger} = \left( \begin{array}{c} a^{\dagger} \\ b^{\dagger} 
\end{array} \right)
\end{equation}
so we retain that part of our
algebraic construction\footnote{One can alternately start with the tilde forms
of $Q$, $\bar{Q}$ in (\ref{hg}) and (\ref{hh}), obtaining the SUSY algebra 
(\ref{sn}) and the tilde relations (\ref{hj}), (\ref{sm}) and (\ref{hi}).} 
\bs
\begin{equation}
Q  =  i \sum_{n=0}^{\infty} \; (A^{\dagger}\tau_{3})_{M_{1}} \ldots 
(A^{\dagger}\tau_{3})_{M_{n}}(A^{\dagger} {\sqrt 2}\tau_{+}A)
A_{M_{n}} \ldots A_{M_{1}} \label{hd}
\end{equation}
\begin{equation}
\bar{Q}  = - i \sum_{n=0}^{\infty} \; (A^{\dagger}\tau_{3})_{M_{1}} \ldots 
(A^{\dagger}\tau_{3})_{M_{n}}(A^{\dagger} {\sqrt 2}\tau_{-}A)
A_{M_{n}} \ldots A_{M_{1}} \label{he}
\end{equation}
\begin{equation}
H  =   \sum_{n=0}^{\infty} \; A^{\dagger}_{M_{1}} \ldots 
A^{\dagger}_{M_{n}}(A^{\dagger}A) A_{M_{n}} \ldots A_{M_{1}} \label{hf}
\end{equation}
\begin{equation}
\dot{A}_{M} = i[H,A_{M}] = -i A_{M}, \;\;\;\;\;\;
\dot{A}^{\dagger}_{M} = i[H,A^{\dagger}_{M}] = i A^{\dagger}_{M}
\end{equation}
\es

\bs \label{sn}
\begin{equation}
Q^{2} = \bar{Q}^{2} = 0,\;\;\;\;\;\;[\bar{Q},Q]_{+} = 2H
\end{equation}
\begin{equation}
\dot{Q} = \dot{\bar{Q}} = \dot{H} = [Q,H] = [\bar{Q},H] = 0
\end{equation}
\begin{equation}
Q\mid 0 \rangle = \bar{Q}\mid 0 \rangle = H\mid 0 \rangle = 0
\end{equation}
\es

\bs \label{sp}
\begin{equation}
[Q,a] = -i{ \sqrt 2} b, \;\;\;\;\;\; [\bar{Q},a^{\dagger}] =
-i {\sqrt 2} b^{\dagger}
\end{equation}
\begin{equation}
[Q,b^{\dagger}]_{+} = i{ \sqrt 2}a^{\dagger}, \;\;\;\;\;\; 
[\bar{Q},b]_{+} = -i{ \sqrt 2}a 
\end{equation}
\begin{equation}
[Q,a^{\dagger}] = [\bar{Q},a] = [Q,b]_{+} = 
[\bar{Q},b^{\dagger}]_{+} = 0
\end{equation}
\es
which follows from the free algebra alone. 

The nonlocal forms 
 (5.24a-c) provide a purely 
bosonic construction of 
supersymmetry at large N, but the construction is apparently not 
equivalent to two unreduced  bosonic oscillators, or indeed to any local
unreduced theory: In the first place, the anticommutators persist for
$b$'s in (\ref{sp}). Moreover, the mixed relations in (\ref{ks}),  
 (\ref{kt}) for $b$ now
involve commutators, which loses the tilde forms of $Q$, $\bar{Q}$, 
$H$ in (\ref{hg}), (\ref{hh}) and (\ref{hi}) and causes for example a nonlocal
 equation of motion for $\tilde{b}$.

\subsection{Higher Supersymmetry}

In this section, we generalize the construction of Subsec.~5.3 to higher
supersymmetry. We will work directly in reduced space, starting 
with nothing but the Cuntz superalgebra 
\bs \label{sq}
\begin{equation}
A_{M} A^{\dagger}_{N} = \delta_{MN}, \;\;\;\;\;\; A_{M} \mid 0 
\rangle = \langle 0 \mid A^{\dagger}_{M} = 0 \label{ho}
\end{equation}
\begin{equation}
A^{\dagger}_{M} A_{M} = 1 - \mid0\rangle\langle0\mid , \;\;\;\;\;\;
M,N = 1 \ldots (B+F)
\end{equation}
\es
where B and F are presently undetermined.  Using only 
(\ref{ho}), we consider the operators
\begin{equation}
Q_{i} \equiv \sum_{n=0}^{\infty} \; (A^{\dagger} \gamma)_{M_{1}} \ldots 
(A^{\dagger}\gamma)_{M_{n}}(A^{\dagger} \Gamma_{i} A)
A_{M_{n}} \ldots A_{M_{1}}
\end{equation}
which are functions of arbitrary matrices $\gamma$ and $\Gamma_{i}$. 
It is straightforward to see that these operators satisfy
\bs 
\begin{equation}
Q_{i} \mid 0 \rangle = 0
\end{equation}
\begin{equation}
Q_{i} A^{\dagger}_{M} - (A^{\dagger} \gamma )_{M} Q_{i} = 
(A^{\dagger} \Gamma_{i} )_{M} \label{hk}
, \;\;\;\;\;\; 
A_{M} Q_{i} - Q_{i} (\gamma A)_{M} = (\Gamma_{i} A)_{M}. \label{hl}
\end{equation}
\es
Now choose the matrices to satisfy
\begin{equation}
\gamma^{2} = 1,\;\;\;\;\;\;[\gamma , \Gamma_{i}]_{+}
= 0 , \;\;\; \forall i 
\end{equation}
so that the relations 
\begin{equation}  \label{so}
Q_{i} A^{\dagger}_{M} = (A^{\dagger} \Gamma_{i} )_{M} + 
(A^{\dagger} \gamma )_{M} Q_{i}  
, \;\;\;\;\;\; 
Q_{i} A_{M} = - (\gamma \Gamma_{i} A)_{M} + (\gamma A)_{M} Q_{i}
\end{equation}
follow from (\ref{hk}). These relations can be used in $Q_{i} Q_{j}$ 
to push $Q_{i}$ through all the $A$'s and $A^{\dagger}$'s in $Q_{j}$, 
with the result
\begin{equation}
[Q_{i}, \; Q_{j}]_{+} = \sum_{n=0}^{\infty} \; A^{\dagger}_{M_{1}} \ldots 
A^{\dagger}_{M_{n}}(A^{\dagger} [\Gamma_{i},\; \Gamma_{j}]_{+} A)
A_{M_{n}} \ldots A_{M_{1}}. \label{ds}
\end{equation}
To close the algebra (\ref{ds}), we now choose the matrices $\Gamma_{i}$ 
to be a Dirac representation of a Clifford algebra in 2d  Euclidean dimensions
\begin{equation}
[\Gamma_{i},\; \Gamma_{j}]_{+} = 2 \delta_{ij}, \;\;\;\;\;\; 
(\Gamma_{i})^{\dagger} = \Gamma_{i}, \;\;\;\;\;\; i = 1 \ldots 2d
\end{equation}
where $\gamma =$ ``$\gamma_{5}$'' $= \prod_{i} \Gamma_{i}$ and 
size$(\Gamma_{i},\; \gamma) = 2^{d}$.

We have therefore constructed a reduced system with $n=2d$ 
supersymmetries
\bs
\begin{equation}
[Q_{i}, \; Q_{j}]_{+} = 2 \delta_{ij} H, \;\;\;\;\;\; i,j = 1 \ldots 2d
\end{equation}
\begin{equation}
H =  \sum_{n=0}^{\infty} \; A^{\dagger}_{M_{1}} \ldots 
A^{\dagger}_{M_{n}}(A^{\dagger}  A) A_{M_{n}} \ldots A_{M_{1}}
\end{equation}
\begin{equation}
\dot{A}_{M} = i[H,A_{M}] = -i A_{M}, \;\;\;\;\;\;
\dot{A}^{\dagger}_{M} = i[H,A^{\dagger}_{M}] = i A^{\dagger}_{M}
\end{equation}
\begin{equation}
Q_{i} \mid 0 \rangle = H \mid 0 \rangle = 0
\end{equation}
\es
which includes the construction of Subsec.~5.3 as the special case with $d=1$ 
and size$(\Gamma_{i},\; \gamma) = 2$.

In a Weyl representation (with blocks of size $2^{d-1}$)
\begin{equation}
\gamma = \left( \begin{array}{c} {\textbf 1}\;\;\;\; 0 \\
0\; -{\textbf 1} \end{array} \right), \;\;\;\;\;\;
\Gamma_{i} = \left( \begin{array}{c} 0\;\;\;\; \gamma_{i} \\
\gamma^{\dagger}_{i}\;\;\;\; 0 \end{array} \right)
\end{equation}
we may identify $B=F=2^{d-1}$ bosons and fermions in the superfields as
\begin{equation}
A_{M} = \left( \begin{array} {c} a_{m} \\ \psi_{\dot{\alpha}}
\end{array} \right), \;\;\;\;\;\; A^{\dagger}_{M} = \left( \begin{array}
 {c} a^{\dagger}_{m} \\ \psi^{\dagger}_{\dot{\alpha}} \end{array} 
 \right) ,\;\;\;\;\;\;
 m,\;\dot{\alpha} = 1\ldots2^{d-1}, \;\;\;\;\;\; M=1 \ldots  2^{d}. 
 \label{hn}
\end{equation}
This assignment follows because the relations (\ref{so}) read
\bs
\begin{equation}
[Q_{i},\;a^{\dagger}_{m}] = \psi^{\dagger}_{\dot{\alpha}} 
(\gamma^{\dagger}_{i})_{\dot{\alpha} m}, \;\;\;\;\;\;
[Q_{i},\; \psi^{\dagger}_{\dot{\alpha}}]_{+} = a^{\dagger}_{m} 
(\gamma_{i})_{m \dot{\alpha}}
\end{equation}
\begin{equation}
[Q_{i},\;a_{m}] = - (\gamma_{i})_{m \dot{\alpha}}\psi_{\dot{\alpha}} 
, \;\;\;\;\;\;
[Q_{i},\; \psi_{\dot{\alpha}}]_{+} = (\gamma^{\dagger}_{i})_
{\dot{\alpha} m} a_{m} 
\end{equation}
\es
under the identification (\ref{hn}).  Similarly, one has for the reduced 
supercharges and Hamiltonian
\bs
\begin{equation}
Q_{i} = A^{\dagger} \Gamma_{i} A + \ldots = a^{\dagger}_{m}(\gamma_{i})
_{m \dot{\alpha}} \psi_{\dot{\alpha}} + \psi^{\dagger}_{\dot{\alpha}}
(\gamma^{\dagger}_{i})_{\dot{\alpha} m} a_{m} + \ldots
\end{equation}
\begin{equation}
H = A^{\dagger}A + \ldots = a^{\dagger}_{m} a_{m} + \psi^{\dagger}_{
\dot{\alpha}} \psi_{\dot{\alpha}} + \ldots 
\end{equation}
\es
and these operators can be pulled back straightforwardly into 
unreduced supercharge and energy densities $(Q_{i})_{rs},\; H_{rs}$ 
\begin{equation}
(\dot{Q}_{i})_{rs} = \dot{H}_{rs} = 0, \;\;\;\;\;\; 
(Q_{i})_{rs} \mid 0\dd \rangle = H_{rs} \mid 0\dd \rangle = 0
\end{equation}
which are nonlocally dressed forms of the original densities of the 
theory (see Subsecs.~3.5 and 5.3).

In this case, we have also the reduced generators of spin(2d)
\bs
\begin{equation}
J_{ij} = \sum_{n=0}^{\infty} \; A^{\dagger}_{M_{1}} \ldots 
A^{\dagger}_{M_{n}}( -\frac{i}{4} A^{\dagger} [\Gamma_{i},\; \Gamma_{j}]
  A) A_{M_{n}} \ldots A_{M_{1}} \label{hm}
\end{equation}
\begin{equation}
[H, \; J_{ij}] = 0 , \;\;\;\;\;\; J_{ij} \mid 0 \rangle = 0 .
\end{equation}
\es
Using (\ref{hm}), we have checked that the reduced fields $a_{m}$ and 
$\psi_{\dot{\alpha}}$
both transform as Weyl spinors of spin(2d), while the reduced superfield
$A_{M}$ transforms as a Dirac spinor. The reduced supercharges themselves
transform in the vector representation of spin(2d).

In the discussion above, we used only the free superalgebra (\ref{ho})
of the untilde superfields to study only the untilde forms of the reduced
operators and fields\footnote{Up to this point, the construction 
above can also be interpreted, as in Subsec.~5.4, as a bosonic 
construction of $n=2d$ reduced supersymmetries.}. With the inclusion of the 
full $B=F=2^{d-1}$ Bose/Fermi/Cuntz
algebra (\ref{sj}), one finds also that all the equations of this section
hold when every field is also tilded (including the tilde form of
the equation of motion and the tilde forms of $Q_{i}$ and $H$). 
The form of the corresponding unreduced generators $Q\dqi$ and $H\dd$, as a 
function of the unreduced SUSY oscillators, is left as an exercise for the reader.

In this connection we note that when  $B=F=2^{d-1}$ the full symmetric 
Bose/Fermi/Cuntz algebra (\ref{sj}) can be written entirely in terms of the 
reduced superfields as
\bs \label{sr}
\begin{equation}
A_{M} A^{\dagger}_{N} = \tilde{A}_{M} \tilde{A}^{\dagger}_{N} = 
\delta_{MN} , \;\;\;\;\;\; M,N = 1 \ldots 2^{d} \label{hp}
\end{equation}
\begin{equation}
A^{\dagger}_{M} A_{M} = \tilde{A}^{\dagger}_{M} \tilde{A}_{M} =
 1-\mid 0 \rangle \langle 0 \mid \label{hq}
\end{equation}
\begin{equation}
A_{M} \mid 0 \rangle = \tilde{A}_{M} \mid 0 \rangle = 0, \;\;\;\;\;\;
A^{\dagger}_{M} \mid 0 \rangle = \tilde{A}^{\dagger}_{M} \mid 0 \rangle 
\label{hr}
\end{equation}

\begin{equation}
A_{M} \tilde{A}^{\dagger}_{N} = \tilde{A}^{\dagger}_{N'} R_{MN',M'N} 
A_{M'} + \delta_{MN} \mid 0 \rangle \langle 0 \mid \label{kb}
\end{equation}
\begin{equation}
A_{M} \tilde{A}_{N} = R_{MN,M'N'} \tilde{A}_{N'} A_{M'}, \;\;\;\;\;\;
 A^{\dagger}_{M} \tilde{A}^{\dagger}_{N}  = \tilde{A}^{\dagger}_{N'}
 A^{\dagger}_{M'}R_{M'N',MN} \label{kc}
\end{equation}

\begin{equation}
R_{MN,M'N'} = (\gamma_{+})_{MM'} \delta_{NN'} +  (\gamma_{-})_{MM'}
\gamma_{NN'} = \delta_{MM'} (\gamma_{+})_{NN'} +
 (\gamma)_{MM'} (\gamma_{-})_{NN'} \label{kd}
\end{equation}
\begin{equation}
\gamma_{\pm} = \frac{1}{2} (1 \pm \gamma) . \label{lk}
\end{equation}
\es
Here 
 (5.41a-c) is the symmetric free superalgebra (\ref{sk}), 
which collects  
 (5.4a-c) and (\ref{ct1}), and the matrix $R$ in 
  (5.41d-f) correctly distinguishes between 
fermions and bosons in the mixed relations 
 (5.4d-h). The matrices $\gamma_{\pm}$ in (\ref{lk}) 
are the natural 
projection operators onto $B=F=2^{d-1}$ bosons and fermions, but the 
full algebra (\ref{sj}) for unrestricted B and F can always be put in superfield 
form by choosing the $R$ matrix as
\begin{eqnarray}
R_{MN,M'N'} & = & (P_{b})_{MM'} (P_{b})_{NN'} +  
(P_{b})_{MM'} (P_{f})_{NN'} \nonumber \\ & + & (P_{f})_{MM'} (P_{b})_{NN'} - 
(P_{f})_{MM'} (P_{f})_{NN'} 
\end{eqnarray}
where $P_{b}$ and $ P_{f}$ are projectors onto the bosons and fermions. 

Finally, we remark on a more general algebraic identity, with 
arbitrary matrix $\gamma$ and operator $q$, 
\bs
\begin{equation}
Q = Q(\gamma; q) \equiv \sum_{n=0}^{\infty} (A^{\dagger} \gamma )_{M_{1}}\ldots 
(A^{\dagger} \gamma )_{M_{n}}\; q \; A_{M_{n}}\ldots A_{M_{1}} 
\end{equation}
\begin{equation}
q = q(A,A^{\dagger}),\;\;\;\;\;\; 
Q(\gamma; q) \mid 0 \rangle = q \mid 0 \rangle 
\end{equation}
\begin{equation}
(A^{\dagger} \gamma )_{M} Q = (Q-q) A^{\dagger}_{M}, \;\;\;\;\;\;
Q (\gamma A)_{M} = A_{M} (Q-q) \label{ll}
\end{equation}
\begin{equation}
Q_{1} Q_{2} = Q(\gamma_{1} \gamma_{2}; \; Q_{1} q_{2} + q_{1} Q_{2} 
-q_{1} q_{2} ) \label{lm}
\end{equation}
\es 
where $Q_{1} = Q(\gamma_{1}; q_{1}),\;\; Q_{2} = Q(\gamma_{2}; q_{2})$.
The relations above follow from the free superalgebra 
 (\ref{ho}) , and this
 result includes the identity used in the construction above.

\section{Matrix Theory}

In this section we  give the large N formulation of the n=16
supersymmetric gauge quantum mechanics \cite{Claud}, now called Matrix 
theory \cite{Bank}, in the temporal gauge (see Sec.~2).
We will take the matrix fields of the theory to be traceless,
so that we may follow the lore \cite{Witt2, Bank} in assuming that 
the ground state is
unique and hence supersymmetric and rotationally invariant. 
Since it costs no further 
effort, we will include in our treatment the entire minimal sequence 
\cite{Claud, Flume, Baake} of matrix models  with 16, 8, 4 and 2 supersymmetries 
(which is obtained by dimensional reduction of pure super Yang-Mills 
theories in 10, 6, 4 and 3 spacetime dimensions.)

The hermitian field form of Matrix theory was given for
any gauge symmetry in Ref.~\cite{Claud}. Our notation for the 
traceless field formulation is changed only slightly from that of 
 \vspace{-4pt} Sec.~2,
\bs
\begin{equation}
(T_{a})_{rs}\;(T_{a})_{uv} = P_{rs,uv},\;\;\;\;\;\; Tr(T_{a} T_{b}) = 
\delta_{ab}, \;\;\;\;\;\; Tr(T_{a}) = 0
\end{equation}
\begin{equation}
\rho_{rs} = \rho_{a} (T_{a})_{rs}, \;\;\;\;\;\;Tr(\rho) = 0, \;\;\;\;\;\; 
\rho = \phi, \; \pi, \; or \; \Lambda
\end{equation}
\begin{equation}
[\phi^{m}_{rs},\;\pi^{n}_{uv}] = i \delta^{mn} 
\;P_{rs,uv},\;\;\;\;\;\;
[(\Lambda_{\alpha})_{rs},\; (\Lambda_{\beta})_{uv}]_{+} = \delta_{
\alpha \beta} \; P_{rs,uv}
\end{equation}
\begin{equation}
  r,s = 1 \ldots N, \;\;\;\;\;\; a,b =1 \ldots N^{2} -1
\end{equation}
\begin{equation}
m = 1 \ldots B,\;\;\;\; \alpha = 1 \ldots f,\;\;\;\; B = F+1,
\;\;\;\; F = \frac{f}{2}, \;\;\;\; f = 16, 8, 4, 2 \label{hs}
\end{equation}
\es
where the projector $P$ is defined in (\ref{a}). The numbers of real 
fermions f in (\ref{hs}) are also the numbers of real supersymmetries in
the minimal sequence.

The matrix field form of the theory is defined 
\vspace{-3pt} by
\bs
\begin{equation}
Q\dqa = Tr((\Gamma^{m}\;\Lambda )_{\alpha} \pi^{m} + g 
(\Sigma^{mn} \; \Lambda)_{\alpha} \; [\phi^{m},\phi^{n}])
\end{equation}
\begin{equation}
[Q\dqa , Q\dqb ]_{+} = 2 \delta_{\alpha \beta} H\dd + 2g
\;\Gamma^{m}_{\alpha \beta} Tr(\phi^{m} \; G) \label{kg}
\end{equation}
\begin{equation}
H\dd = \frac{1}{2} Tr(\pi^{m} \pi^{m} - \frac{g^{2}}{2} \;[\phi^{m},\phi^{n}]
 [\phi^{m},\phi^{n}] + g [\Lambda_{\alpha},\; \Lambda_{\beta}]_{+}
 \Gamma^{m}_{\alpha \beta} \; \phi^{m})
\end{equation}
\begin{equation}
G_{rs} = -i [\phi^{m},\; \pi^{m}]_{rs} - 
(\Lambda_{\alpha}\Lambda_{\alpha})_{rs} - (N - \frac{1}{N}) \delta_{rs} 
 \label{ht}
\end{equation}
\begin{equation}
J\dj^{mn} = Tr (\pi^{[m}\;\phi^{n]} - \frac{1}{2} \Lambda \Sigma^{mn} \label{hu}
\Lambda )
\end{equation}
\begin{equation}
[\Gamma^{m},\Gamma^{n}]_{+} = 2 \delta^{mn},\;\;\;\;\;\; \Sigma^{mn} 
= - \frac{i}{4} [\Gamma^{m},\Gamma^{n}]
\end{equation}
\es
where $g$ is the coupling constant and the matrices $\Gamma^{m}$ are real,
symmetric and traceless. At large N, the SU(N) gauge generators in (\ref{ht}) 
are in agreement with the SU(N) generators  (\ref{st}) at $B = F + 1$.
 The generators of spin(B), 
with B = 9,5,3 and 2, are given in (\ref{hu}). Derived relations 
\vspace{-2pt} include 
\bs
\begin{equation}
[Q\dqa , H\dd ] = ig Tr(\Lambda_{\alpha} G) \label{kh} 
, \;\;\;\;\;\;  
H\dd = \frac{1}{f} \sum_{\alpha = 1}^{f} \; Q\dqa^{2}
\end{equation}
\begin{equation}
[J\dj^{mn} ,H\dd] = 0, \;\;\;\;\;\; [J\dj^{mn} , 
Q\dqa ] =
 (\Sigma^{mn} Q\dd)_{\alpha}
\end{equation}
\begin{equation}
[J\dj^{mn}, J\dj^{pq}] = i( \delta^{q [m} J\dj^{n] p}
 -  \delta^{p [m} J\dj^{n] q})
\end{equation}
\es
and the algebra of $Q\dqa$, $H\dd$ and $J\dj^{mn}$ with the fields.

As noted above, we  assume here that the gauge invariant ground
\vspace{-2pt} state
\begin{equation}
G_{rs} \mid 0\dd \rangle = 0
\end{equation}
of the theory is unique\footnote{To study the possibility of degenerate 
vacua in large N theories with fermions, see the discussion around 
Eq.~(\ref{ss}).}. This means that
the ground state is also supersymmetric and rotationally 
\vspace{-2pt} invariant
\begin{equation}
Q\dqa \mid 0\dd \rangle = H\dd \mid 0\dd \rangle = 
J\dj^{mn} \mid 0\dd 
\rangle = 0
\end{equation}
and that the large N completeness relation has the same 
\vspace{-2pt} form
\begin{equation}
\textbf{1}\dd \;\stackrel{_{\textstyle =}}{_{_{N}}} \;\; \mid 0\dd \rangle \langle  
\dd 0 \mid + \sum_{rs,A} \mid rs,A \rangle \langle  rs,A \mid  
\end{equation}
which we have been studying throughout this paper.

For the reduced large N theory, we maintain `t~Hooft scaling by 
 \vspace{-2pt} choosing
\begin{equation}
g = \lambda / {\sqrt N}
\end{equation}
where the rescaled coupling $\lambda$ is independent of N. The definitions of 
reduced matrix 
elements and operators in Sec.~2 are unchanged, except for
the vacuum expectation values (see \ref{d}), which now 
\vspace{-2pt} read
\begin{equation}
\langle 0 \mid \phi_{m} \mid 0 \rangle = \langle 0 \mid \pi_{m} 
\mid 0 \rangle = \langle 0 \mid \Lambda_{\alpha} \mid 0 \rangle = 0
\end{equation}
for the reduced fields, because the unreduced fields are traceless. Then, 
except for a few corrections which are negligible at large N, we find that
no other changes are necessary, and we may take over all
the reduced results derived  in previous sections.

The equal-time free algebra of reduced Matrix theory is a copy of 
(\ref{sc}) and \vspace{-3pt} (\ref{x})
\bs \label{sx}
\begin{equation}
[\phi_{m} , \tilde{\pi}_{n} ] = [\tilde{\phi}_{m} , \pi_{n} ] =
i \delta_{mn} \mid 0 \rangle \langle 0 \mid \label{hv}
\end{equation}
\begin{equation}
[\phi_{m} , \tilde{\phi}_{n} ] = [\pi_{m} , \tilde{\pi}_{n} ] = 0
\end{equation}
\begin{equation}
[\Lambda_{\alpha} , \tilde{\Lambda}_{\beta} ]_{+} = \delta_{\alpha \beta} 
\mid 0 \rangle \langle 0 \mid 
\end{equation}
\begin{equation}
[\Lambda_{\alpha} , \tilde{\phi}_{m}] = [\tilde{\Lambda}_{\alpha} , \phi_{m}]
 = [\Lambda_{\alpha} , \tilde{\pi}_{m}] = [\tilde{\Lambda}_{\alpha} , \pi_{m}]
  = 0
\end{equation}
\begin{equation}
 [\phi_{m} , \pi_{m} ] - i \Lambda_{\alpha} \Lambda_{\alpha}  = 
 [\tilde{\phi}_{m} , \tilde{\pi}_{m} ] - i \tilde{\Lambda}_{\alpha}
 \tilde{\Lambda}_{\alpha}  =  i   \mid 0 \rangle \langle 0 \mid  \label{hw}   
\end{equation} 
\begin{equation}
(\tilde{\phi}_{m} - \phi_{m} ) \mid 0 \rangle = 
(\tilde{\pi}_{m} - \pi_{m} ) \mid 0 \rangle = 
(\tilde{\Lambda}_{\alpha} - \Lambda_{\alpha} ) \mid 0 \rangle = 0 \label{hx}
\end{equation}
\begin{equation}
(\dot{\tilde{\phi}}_{m} - \dot{\phi}_{m} ) \mid 0 \rangle = 
(\dot{\tilde{\pi}}_{m} - \dot{\pi}_{m} ) \mid 0 \rangle = 
(\dot{\tilde{\Lambda}}_{\alpha} - \dot{\Lambda}_{\alpha} ) \mid 0 \rangle = 0
 \label{hx1}
\end{equation}
\es 
with $B=F+1$. Recall   that (\ref{hw}) summarizes the action of the
reduced gauge generators $G$, $\tilde{G}$ on the reduced states 
(see Eq.~(\ref{sy})).

The reduced equations of motion 
\vspace{-3pt} are
\bs \label{su}
\begin{equation}
\dot{\phi}_{m} = i [H,\phi_{m}] = \pi_{m}
\end{equation}
\begin{equation}
\dot{\pi}_{m} = i [H,\pi_{m}] = \lambda^{2} [\phi_{n}, [\phi_{m}, 
\phi_{n}]] - \lambda \Lambda_{\alpha} \Gamma^{m}_{\alpha \beta} 
\Lambda_{\beta}
\end{equation}
\begin{equation}
\dot{\Lambda}_{\alpha} = i [H,\Lambda_{\alpha}] = i \lambda \Gamma^{m}
_{\alpha \beta} [\phi_{m}, \Lambda_{\beta}]
\end{equation}
\vspace{-2pt} 
\begin{equation}
\dot{\tilde{\phi}}_{m} = i [H,\tilde{\phi}_{m}] = \tilde{\pi}_{m}
\end{equation}
\begin{equation}
\dot{\tilde{\pi}}_{m} = i [H,\tilde{\pi}_{m}] = \lambda^{2} [\tilde{\phi}
_{n}, [\tilde{\phi}_{m}, \tilde{\phi}_{n}]] + \lambda \tilde{\Lambda}
_{\alpha} \Gamma^{m}_{\alpha \beta} \tilde{\Lambda}_{\beta}
\end{equation}
\begin{equation}
\dot{\tilde{\Lambda}}_{\alpha} = i [H,\tilde{\Lambda}_{\alpha}] = - i \lambda 
\Gamma^{m}_{\alpha \beta} [\tilde{\phi}_{m}, \tilde{\Lambda}_{\beta}] 
\end{equation}
\es
where $H$ is the reduced Hamiltonian. 
Notice the sign changes in the tilde equations of motion. These follow
 from the unreduced equations of motion, and one can check with
 (6.9a-f) that these signs are consistent with (\ref{hx1}).

Supersymmetry and rotational invariance tell us that the
corresponding reduced operators annihilate the ground 
\vspace{-2pt} state
\begin{equation}
Q_{\alpha} \mid 0 \rangle = H \mid 0 \rangle = J_{mn} \mid 0 \rangle = 0
\end{equation}
and therefore 
\nopagebreak \vspace{-2pt} that
\bs
\begin{equation}
0 = \langle 0 \mid Q_{\alpha} \mid 0 \rangle = \langle 0 \mid 
(\Gamma^{m}\;\Lambda )_{\alpha} \pi_{m} + \lambda  
(\Sigma^{mn} \; \Lambda)_{\alpha} \; [\phi_{m},\phi_{n}] \mid 0 \rangle 
\end{equation}
\begin{equation}
0 = \langle 0 \mid H \mid 0 \rangle = \langle 0 \mid  \pi_{m} \pi_{m}
 - \frac{\lambda^{2}}{2} \;[\phi_{m},\phi_{n}]
 [\phi_{m},\phi_{n}] + \lambda [\Lambda_{\alpha},\; \Lambda_{\beta}]_{+}
 \Gamma^{m}_{\alpha \beta} \; \phi_{m} \mid 0 \rangle 
\end{equation}
\begin{equation}
0 = \langle 0 \mid J_{mn} \mid 0 \rangle = \langle 0 \mid \pi_{[m}\;\phi_{n]} 
- \frac{1}{2} \Lambda \Sigma^{mn} \Lambda  \mid 0 \rangle 
\end{equation}
\es
where the explicit forms of the unreduced operators have been used
to evaluate the last form in each relation.

For the reduced angular momenta, the maps of Sec.~2 
tell us also \vspace{-4pt} that  
\bs
\begin{equation}
\dot{J}_{mn} = i[H,J_{mn}] = 0
\end{equation}
\begin{equation}
[J_{mn}, b_{p}] = -i \delta_{p [m} b_{n]},\;\;\;\;\;\; b = \phi, \;  \pi, 
\; \tilde{\phi}, \; or \; \tilde{\pi} 
\end{equation}
\begin{equation}
[J_{mn},f_{\alpha}] = (\Sigma^{mn} f)_{\alpha}, \;\;\;\;\;\;
f = \Lambda, \; \tilde{\Lambda}\; or\; Q
\end{equation}
\begin{equation}
[J_{mn}, J_{pq}] = i ( \delta_{q [m} J_{n] p} -  \delta_{p [m} J_{n] q}) 
\end{equation}
\es
so that the rotational properties of the reduced fields are the same as
in the unreduced theory.

For the reduced supercharges, we \vspace{-4pt} find 
\bs
\begin{equation}
[Q_{\alpha}, Q_{\beta}]_{+} = 2(\delta_{\alpha \beta} H + \lambda 
\Gamma^{m}_{\alpha \beta} (\tilde{\phi}_{m} - \phi_{m} )) \label{kp}
\end{equation}
\begin{equation}
\dot{Q}_{\alpha} = i [H, Q_{\alpha}] = \lambda (\tilde{\Lambda}_{\alpha} 
- \Lambda_{\alpha}) \label{ke}
, \;\;\;\;\;\; 
H = \frac{1}{f} \sum_{\alpha = 1}^{f} \; Q^{2}_{\alpha} \label{kf}
\end{equation}
\es
\vspace{-4pt} 
\bs \label{sv}
\begin{equation}
[Q_{\alpha}, \phi_{m}] = -i (\Gamma^{m} \Lambda )_{\alpha}
\end{equation}
\begin{equation}
[Q_{\alpha}, \pi_{m}] = 2i \lambda [\phi_{n}, (\Sigma^{mn} \Lambda) 
 _{\alpha}]
\end{equation}
\begin{equation}
[Q_{\alpha}, \Lambda_{\beta}]_{+} = \Gamma^{m}_{\alpha \beta} \pi_{m} 
+ \lambda \Sigma^{mn}_{\alpha \beta} [\phi_{m}, \phi_{n}]
\end{equation}
\es
\vspace{-4pt} 
\bs \label{sw}
\begin{equation}
[Q_{\alpha}, \tilde{\phi}_{m}] = -i (\Gamma^{m} \tilde{\Lambda} )_{\alpha}
\end{equation}
\begin{equation}
[Q_{\alpha}, \tilde{\pi}_{m}] = -2i \lambda [\tilde{\phi}_{n}, (\Sigma^{mn}
 \tilde{\Lambda})_{\alpha}]
\end{equation}
\begin{equation}
[Q_{\alpha}, \tilde{\Lambda}_{\beta}]_{+} = \Gamma^{m}_{\alpha \beta} 
\tilde{\pi}_{m} - \lambda \Sigma^{mn}_{\alpha \beta} [\tilde{\phi}_{m},
\tilde{\phi}_{n}] .
\end{equation}
\es
Note in particular the unexpected form of the extra term on the right side 
of the reduced supersymmetry algebra (\ref{kp}) and the term on 
the right side of $\dot{Q}_{\alpha}$ in (\ref{ke}), both of which are proportional to the difference 
of a reduced operator and its tilde.  These terms are the reduced analogues 
of the \nopagebreak gauge terms in (\ref{kg}) and (\ref{kh}) respectively.
 Using (\ref{hx}), we see that these terms are consistent 
 \vspace{-4pt} with
\begin{equation}
Q_{\alpha} \mid 0 \rangle = \dot{Q}_{\alpha} \mid 0 \rangle = 0
\end{equation}
as they should be. Although we have derived this system from the 
unreduced relations and the maps of Sec.~2, we have checked
for example that the reduced equations of motion (\ref{su}) also
follow from the form of the reduced Hamiltonian in (\ref{kf}) and the algebra
(\ref{sv}) and (\ref{sw}) of the reduced supercharges with the reduced fields.

A next step toward the solution of  Matrix theory
would be to find explicit nonlocal forms of the
reduced operators $Q_{\alpha}$, $H$ and $J_{mn}$, as we have done for simpler
systems in previous sections. Towards this, it will be helpful to
look for interacting bosonic Cuntz operators, possibly of the 
\vspace{-3pt} form
\begin{equation}
A_{m} A^{\dagger}_{n} = C_{mn} (\phi, \Lambda), \;\;\;\;\;\;
A_{m} \mid 0 \rangle = 0, \;\;\;\;\;\; A_{m} = \frac{1}{\sqrt 2} 
(F_{m}(\phi, \Lambda) + i \pi_{m})
\end{equation}
following the line of our construction for bosonic theories in Sec.~4
and App.~E.  Owing to the complexity of the Matrix theory ground state 
\cite{Halp2, Konec}, however, we have not yet been able to prove the existence 
of such operators in this case. We note, however, that if such bosonic operators
exist, then one also obtains a set of generalized fermionic
 creation and annihilation \vspace{-3pt} operators 
\begin{equation}
A_{m \alpha} = i[Q_{\alpha}, A_{m}], \;\;\;\;\;\; A_{m \alpha} \mid 
0 \rangle = 0
\end{equation}
whose local composite form can be evaluated with (\ref{sv}). These two
types of operators correspond to $A \sim a$ and $A_{\dot{\alpha}} \sim
\psi_{\dot{\alpha}}$  in the simpler supersymmetric models above. 

Further study of these generalized free algebras is particularly 
important for Matrix theory, where the associated new large 
N-conserved quantities (local and nonlocal) may be  related to the question of
 hidden 11-dimensional symmetry~\cite{Bank}.

\vspace{1.2cm}

\noindent {\bf ACKNOWLEDGEMENTS}

For helpful discussions, we thank J. de Boer, M. Douglas, D. Evans, 
J.~Greensite, V. Jones, M. Rieffel, C. Schweigert, I. Singer, 
L. Susskind, W.~Taylor, G.~Veneziano and   D. Voiculescu.

     The work of  M.~B.~H. was supported in part by the Director, Office of 
Energy Research, Office of Basic Energy Sciences, of the U.S. 
\nopagebreak  Department 
of Energy under Contract DE-AC03-76F00098 and in part by the National 
Science Foundation under grant PHY95-14797.

\vspace {1cm}

Note added.  After submission of this manuscript, Ref.~\cite{OWG} was called to 
our attention:  In an application to ``infinite statistics'', this 
reference contains another independent discovery of the Cuntz algebra 
in physics and gives our Eq.~(\ref{sss}).

\vskip 1.0cm
\setcounter{equation}{0}
\def\theequation{A.\arabic{equation}}
\boldmath
\noindent{\bf Appendix A: Time Reversal and the Tilde Operators}
\unboldmath
\vskip 0.5cm

For time-reversal invariant theories, the reduced tilde fields of the text 
can be related to the time-reversed form of the reduced untilde fields. 
 This is simplest to see for bosonic theories, as follows.  

In the coordinate representation, where $\phi^{m}_{a}(t=0)$ is real, the usual 
antiunitary  time-reversal operator $\Theta$ gives
\bs
\begin{equation}
\langle \Theta \alpha \mid \phi^{m}_{a}(t) \mid \Theta \beta \rangle = 
\langle \beta \mid \phi^{m}_{a}(-t) \mid \alpha \rangle 
\end{equation}
\begin{equation}
\langle \Theta \alpha \mid \pi^{m}_{a}(t) \mid \Theta \beta \rangle = 
 - \langle \beta \mid \pi^{m}_{a}(-t) \mid \alpha \rangle 
\end{equation}
\es
for general time-independent states $\alpha, \beta$ in the unreduced theory.
Because the Hamiltonian and the gauge generators obey 
\begin{equation}
\Theta \; H\dd = H\dd\; \Theta, \;\;\;\;\;\; \Theta \; G_{rs} = - G_{sr}\;\Theta 
\end{equation}
we can choose a basis of the unreduced singlet and adjoint states 
such that
\begin{equation}
\Theta \mid 0\dd \rangle = \; \mid 0\dd \rangle, \;\;\;\;\;\; \Theta \mid rs,A 
\rangle = \; \mid sr, A \rangle .
\end{equation}
Then the definitions (\ref{sa}) give the relations between the tilde and 
untilde reduced fields:
\begin{equation}
(\tilde{\phi}_{m}(t))_{\mu \nu} = (\phi_{m}(-t))_{\nu \mu}, \;\;\;\;\;\; 
(\tilde{\pi}_{m}(t))_{\mu \nu} = - (\pi_{m}(-t))_{\nu \mu}
\end{equation}
where $\mu = (0,A)$ and $\nu = (0,B)$ label the reduced matrix 
elements of the reduced fields (see, for example, (\ref{td}), 
(\ref{kn}) and (\ref{sd})).

\vskip 1.0cm
\setcounter{equation}{0}
\def\theequation{B.\arabic{equation}}
\boldmath
\noindent{\bf Appendix B: More on Many-time Wightman Functions}
\unboldmath
\vskip 0.5cm

For globally invariant theories, the set of all invariant Wightman
functions is the set of averages of all fully-contracted products, each 
contraction being a summation of a left matrix  index with a right matrix 
index. We find that the equal-time traces
\begin{equation}
\langle \dd 0 \mid Tr( \frac{\rho_{1}(t)}{\sqrt N} \ldots  \frac{\rho_{n}
(t)}{\sqrt N} ) \mid 0\dd \rangle = N \langle 0 \mid \rho_{1}(t) 
\ldots \rho_{n}(t) \mid 0 \rangle, \;\;\;\; \rho = \phi,\; \pi\; or
 \; \Lambda
\end{equation}
are the most general invariant equal-time Wightman functions, but the
many-time traces
\begin{equation}
\langle \dd 0 \mid Tr( \frac{\rho_{1}(t_{1})}{\sqrt N} \ldots  \frac{\rho_{n}
(t_{n})}{\sqrt N} ) \mid 0\dd \rangle = N \langle 0 \mid \rho_{1}(t_{1}) 
\ldots \rho_{n}(t_{n}) \mid 0 \rangle
\end{equation}
are not the most general invariant Wightman functions.

In particular, there are also a large number of invariant ``twisted
traces'', such as
\begin{equation}
\langle \dd 0 \mid \rho_{rs}(t_{1}) \; \rho_{tr}(t_{2}) \; \rho_{st}(t_{3}) 
\mid 0\dd \rangle = N^{\frac{5}{2}} \langle 0 \mid \rho(t_{1}) \; 
\tilde{\rho}(t_{2}) \; \rho(t_{3}) \mid 0 \rangle
\end{equation}               
which can be expressed in terms of our reduced operators $\rho$ and 
$\tilde{\rho}$ (subscripts are suppressed here for simplicity). Moreover, 
there are an even larger number of invariant twisted traces, such as
\begin{equation}
\langle \dd 0 \mid \rho_{rs}(t_{1}) \; \rho_{tu}(t_{2}) \; \rho_{st}(t_{3}) \; 
\rho_{ur}(t_{4}) \mid 0\dd \rangle 
\end{equation}
which cannot be expressed in terms of our reduced operators.

A rule for deciding whether any such invariant product 
will have its average expressible in terms of our reduced operators is as 
follows. Draw a set of index lines joining each pair of 
contracted indices in the ordered invariant operator product. Then 
draw a vertical line between each neighboring pair of operators in 
the product and count how many index lines each vertical line will cut. This 
operation represents inserting a complete set of intermediate states in the 
channel defined by each vertical line.  If no more than two index
lines are cut in a given channel, then this channel is saturated by singlet 
and adjoint states, and the large N completeness relation (\ref{b}) may
be inserted in this channel. If no more than two index lines are cut in
any channel of the twisted trace, then the corresponding average can
be expressed in terms of our reduced operators. However, if more than
two index lines are cut in any channel, then higher representations are
necessary to saturate that channel, and the average cannot be written
in terms of our reduced operators.

\vskip 1.0cm
\setcounter{equation}{0}
\def\theequation{C.\arabic{equation}}
\boldmath
\noindent{\bf Appendix C: More on Trace Class Operators}
\unboldmath
\vskip 0.5cm

In this appendix we prove a theorem about any bosonic (including even
fermion number) trace class operator $T\dd$ and its corresponding
reduced operator $T$. In particular, we will show that, to leading order
at large N in the large N Hilbert space (\ref{b}),  both  $T\dd$ and $T$  
are proportional to the unit operator in their respective spaces.  

We begin in the unreduced theory, where the large N completeness 
statement (\ref{b}) tells us that
\begin{equation}
T\dd \mid 0\dd \rangle = \mid 0\dd \rangle \langle \dd 0 \mid T\dd \mid 0\dd 
\rangle \times (1+O(N^{-1})) . \label{fb}
\end{equation}
We want to extend this by  examining  the action of $T\dd$ on adjoint states. 
For this, we assume that all the operators and states of interest can be 
built from products of canonical variables. For example,
\bs 
\begin{equation}
T\dd  =  Tr(t^{(w)}), \;\;\;\;\;\; t^{(w)}_{rs} = (\chi^{i_{1}} \chi^{i_{2}}
\ldots \chi^{i_{k}})_{rs}  
\end{equation}
\begin{equation}
 \chi^{i}_{rs}  =  
(\frac{\phi^{m}}{\sqrt N})_{rs}\;, \;\; (\frac{\pi^{m}}{\sqrt N})_{rs} 
\;\; or \;\;(\frac{\Lambda_{\alpha}}{\sqrt N})_{rs} , \;\;\;\;\;\; 
\textup{word} \;\; w = i_{1} i_{2} \ldots i_{k}
\end{equation} 
\es
where $t$ is $O(1)$ and $T\dd$ is $O(N)$. A basis for the  adjoint 
states is  
\begin{equation}
 \{ \mid rs, u \rangle = t^{(u)}_{rs} \mid 0\dd \rangle \}
\end{equation}
where the basis vectors are labeled by the set of all words $u$. Finally,
the energy eigenstates of the text may be expressed as an expansion 
in this \vspace{-4pt} basis
\begin{equation}
\mid rs,A \rangle = \sum_{u} K_{u}(A) \mid rs,u \rangle .  
\end{equation}

Now  consider 
\begin{eqnarray}
& & T\dd \mid rs,u \rangle = \; t^{(u)}_{rs}\; T\dd \mid 0\dd \rangle \; + \;
 [T\dd\; ,\; t^{(u)}_{rs}]  \mid 0\dd \rangle \nonumber \\ & & = \mid rs,u 
 \rangle \langle \dd 0 \mid T\dd \mid 0\dd \rangle (1 + O(N^{-1})) \; + \; 
 \textup{Leftover}  \label{fa}
\end{eqnarray} 
where the ``Leftover'' terms come from the commutator. The leading term 
in (\ref{fa}) is O(N), and we will show that the Leftovers  are 
$O(N^{-1})$, so that
\begin{equation}
 T\dd \mid rs,u \rangle = \mid rs,u \rangle \langle \dd 0 \mid T\dd \mid 0\dd \rangle
 \times (1 + O(N^{-1}) + O(N^{-2})). \label{fc} 
\end{equation}
To see this,  we need the canonical commutation relations
\begin{equation}
[\chi^{i}_{pq},\chi^{j}_{rs}]_{\mp} = \frac{1}{N}\; c_{ij}\;
 \delta_{ps} \delta_{qr}, \;\;\;\;\;\; \label{fh}
c_{ij} = \left\{ \begin{array} {l}
\delta_{\alpha \beta} \;\;\;\;\;\;\; \textup{if} \;\; i,j \;\; 
\textup{denote} \;\; 
\Lambda_{\alpha},\;\Lambda_{\beta} \\
\pm i \delta_{mn}\;\; \textup{if} \;\; i,j \;\; \textup{denote} \;\; 
\phi^{m},\; \pi^{n} \\
0 \;\;\;\;\;\;\;\;\;\;\;\; \textup{otherwise}. \end{array} \right. 
\nopagebreak 
\end{equation}
This allows us to check that
\begin{equation}
 [T\dd\; ,\; t^{(u)}_{rs}] = \frac{1}{N} \sum_{v} t^{(v)}_{rs} \sim 
 O(N^{-1})
\end{equation} 
where the words $v$ are made up from various parts 
of the words $w$ and $u$.

Collecting (\ref{fb}) and (\ref{fc}), we see that the leading term of
 $T\dd$ is proportional to the unreduced unit operator $\textbf{1}\dd$
\begin{equation}
 T\dd  = \textbf{1}\dd\; \langle \dd 0 \mid T\dd \mid 0\dd \rangle \times 
 (1 + O(N^{-1}) + O(N^{-2})).  \label{fd} 
\end{equation}
Similarly, we find for the reduced operator
\begin{equation}
T = \textbf{1} \; \langle 0 \mid T\mid 0 \rangle (1 + O(N^{-1})
 + O(N^{-2})) \label{fe}
\end{equation}
where $\textbf{1}$ is the reduced unit operator. In the case when 
$T\dd$ contains an odd number 
of fermions, a generalization of this result can be obtained involving
the phases $(-1)^{F}$. We emphasize that the result (\ref{fd}) and (\ref{fe}) is 
independent of the scale of $T\dd$ and $T$, since a factor like $C(N)$ in 
(\ref{ab})
can be included if desired on both sides of the result.

As an example, consider the  Hamiltonian $H\dd$, for which  
\begin{equation}
H\dd = E_{0}\; \textbf{1}\dd + H'\dd, \;\;\;\;\;\; E_{0} = O(N^{2}),
 \;\;\;\;\;\; H'\dd = O(N^{0})
\end{equation}
tells us that there are no $O(N^{-1})$ corrections to (\ref{fb}) in this case. 
We then see that 
\begin{equation}
H\dd \mid rs,A \rangle = E_{A} \mid rs,A \rangle = E_{0} \mid rs,A 
\rangle \times (1+O(N^{-2})) 
\end{equation}
 is consistent with
\begin{equation}
\omega_{A0} = E_{A} - E_{0} = O(N^{0}).
\end{equation}
For the reduced Hamiltonian $H$,  the result above implies
\begin{equation}
H = E_{0}\; \textbf{1} + H' , \;\;\;\;\;\; H' = O(N^{0}).
\end{equation}
Many explicit examples of this are worked out in the text. The 
commutator equations of motion involve only the operator $H'$.  In cases where 
the leading term is zero, such as the Hamiltonian of SUSY systems or the angular 
momentum operators, the theorem (\ref{fe}) tells us nothing 
\nopagebreak useful.

These results imply that pairs of trace class operators commute to 
leading order at large N (see also (\ref{lo})). This is easy to see in examples,
 such as
\begin{equation}
[Tr(\pi \phi), \; Tr(\phi^{2})] = -2i \; Tr(\phi^{2}).
\end{equation}
Here each trace is $O(N^{2})$, so the commutator is naively of $O(N^{4})$. 
But the leading terms contribute zero in the commutator and the 
result is $O(N^{2})$.

\vskip 1.0cm
\setcounter{equation}{0}
\def\theequation{D.\arabic{equation}}
\boldmath
\noindent{\bf Appendix D: More on the Tilde of a General Density}
\unboldmath
\vskip 0.5cm

An algorithm for the computation of the tilde of a general reduced 
density was given in Subsec.~2.5. Here we provide an alternate 
derivation which gives a recursive form of the result.
Let $X_{rs}$ be any operator that transforms in the adjoint
\begin{equation}
[G_{rs},X_{pq}] =\delta_{rq} \; X_{ps} - \delta_{sp} \; X_{rq}.
\end{equation}
Then, using the definitions of Sec.~2, we get 
the reduced operators $X$ and  $\tilde{X}$. 

For any two such operators $X$ and $Y$, we get another adjoint operator $U$ 
by composing $X$ and $Y$ as a regular matrix product
\begin{equation}
U_{rs} = X_{rt} \; Y_{ts}
\end{equation}
and we quickly find that the reduced operators satisfy $U = XY$. The form of 
the reduced matrix $\tilde{U}$, however, has no simple expression in  general.

We can further define an irregular product 
\begin{equation}
V_{rs} = X_{ts} Y_{rt}
\end{equation}
which also transforms in the adjoint. The results for the reduced operators 
in this case are reversed: $\tilde{V} = \tilde{X} \; \tilde{Y}$;
but there is no simple form for $V$ in the general case.

One important special case is when all the matrix elements of  $X_{rs}$ and 
$Y_{pq}$ commute (or anticommute) with one another.  Then the regular 
and irregular composites are equivalent and we have 
$\widetilde{(XY)} = \pm \tilde{Y} \tilde{X}$.

A broader  case is the general density, defined in the 
text as any 
composite operator built from repeated regular matrix products of
the canonical variables.  With the result of App.~C we can now 
give an alternate derivation of the general 
algorithm given in the text for expressing $\widetilde{U^{(w)}}$ 
where $U^{(w)} = \chi_{i_{1}} \chi_{i_{2}} \ldots \chi_{i_{n}}$.  Using the word 
notation, with the decomposition $w = ui_{n}$, we have that
\bs
\begin{equation}
U^{(w)}_{rs} = U^{(u)}_{rt} \chi^{i_{n}}_{ts} = \pm \chi^{i_{n}}_{ts} 
 U^{(u)}_{rt} + [U^{(u)}_{rt}, \chi^{i_{n}}_{ts}]_{\mp} \label{fg}
\end{equation}
\begin{equation}
\langle pq,A\mid U^{(w)}_{rs} \mid tr,B \rangle \equiv P_{qp,ts} \langle A 
\mid \widetilde{U^{(w)}} \mid B \rangle \label{ff}
\end{equation}
\es
where (\ref{ff}) is taken from the discussion of the text.
From the first term on the right in (\ref{fg}) we get simply 
$\tilde{\chi}_{i_{n}} \widetilde{U^{(u)}}$.  To evaluate the commutator we use 
(\ref{fh})
repeatedly for each $\chi$ factor in $U^{(u)}$ and find a series of terms 
\begin{equation}
\sum_{k} \; c_{i_{k} i_{n}} U^{(x)}_{rs} \frac{1}{N} Tr(U^{(y)}), \;\;\;\;\;\; 
u = x i_{k} y \vspace{-4pt}
\end{equation}
where $k$ counts the letters in the word $u$.
The matrix element of $ U^{(x)}_{rs}$ gives the reduced matrix 
$\widetilde{U^{(x)}}$ while the trace factor gives $\langle 0 \mid U^{(y)} 
\mid 0 \rangle$, using the result of App.~C.  Thus we have the
following recursion relation for the tilde of a regular operator 
product:
\begin{equation}
\widetilde{U^{(w)}} = \pm \tilde{\chi}_{i_{n}} \widetilde{U^{(u)}} + 
\sum_{k} \; (\pm ) c_{i_{k} i_{n}} \widetilde{U^{(x)}} \langle 0 \mid U^{(y)}
 \mid 0 \rangle, \;\;\;\; w = u i_{n} = x i_{k} y i_{n} .  \vspace{-4pt} 
\end{equation}
This result can be iterated to decompose any $\tilde{U}$ into simple
 $\tilde{\chi}$ factors.  The $\pm$ signs are determined by counting 
the number of times a fermion in $\chi_{i_{n}}$ is moved passed other 
fermions within $U$. This result is equivalent to the algorithm given 
in Subsec.~2.5 for the tilde of a general density.

\vskip 1.0cm
\setcounter{equation}{0}
\def\theequation{E.\arabic{equation}}
\boldmath
\noindent{\bf Appendix E: More on the Bosonic Ground State}
\unboldmath
\vskip 0.5cm

In this appendix we supplement the discussion of Sec.~4, showing that 
the matrix $F_{rs}(\phi)$ can be considered a density at large N.  At 
the same time we will provide a closely related discussion of the 
quantities $C_{mn} = A_{m} A^{\dagger}_{n}$ which emphasizes their 
unreduced form.  Our starting point is the unreduced large N bosonic ground 
state wave function
\begin{equation}
\langle \phi \mid 0\dd \rangle = \psi_{0}(\phi ) = e^{-N^{2}\; S(\phi)} 
\end{equation}
whose ``action'' $S(\phi)$ is  
a general invariant function which is $O(N^{0})$ in the Hilbert space 
of (\ref{b}).  This is the form which results from `t~Hooft-scaled 
potentials, as in (\ref{ts}). In the following discussion 
we will work up to the general case by considering a sequence of 
simpler special cases.

We begin with the special case of a single matrix $\phi$ and the 
special action
\begin{equation}
S(\phi) =  \sum_{n=1} s_{n}\; \zeta_{n}(\phi), \;\;\;\;\;\; \zeta_{n}
 (\phi) \equiv \frac{1}{N} Tr [(\frac{\phi}{\sqrt N})^{n}] 
\end{equation}
where $s_{n}$ are numbers. The trace class quantities $\zeta_{n}$ are 
 $O(N^{0})$ in the 
 Hilbert space of (\ref{b}) and $\zeta_{0} = 1$. In this case 
 we find directly that
 $F_{rs}$ is a density
\begin{equation}
\pi_{rs}  \psi_{o} = i F_{rs}  \psi_{o} , \;\;\;\;\;\; 
F_{rs} = N^{\frac{1}{2}}\sum_{n=1}\; n s_{n}\; [(\frac{\phi}
{\sqrt N})^{n-1}]_{rs} \label{fs}
\end{equation}
and this translates immediately into the reduced form
\begin{equation}
F =  \sum_{n=1}\; n s_{n}\; (\phi)^{n-1}.
\end{equation}

Consider next the operator $C$, defined by 
\begin{equation}
C_{rs} \equiv [A_{rt},(A^{\dagger})_{ts}] = i [\pi_{rt},F_{ts}] . \label{fq}
\end{equation}
Because $A$ annihilates the ground state, we can show, using a $\phi$ basis,  
that the $A^{\dagger} A$ 
term in (\ref{fq}) contributes to matrix elements at $O(N^{-2})$ compared to 
the $A A^{\dagger}$ term, and so can be neglected
\begin{equation}
 C_{rs} = i[\pi_{rt},F_{ts}] = A_{rt} (A^{ \dagger})_{ts} \times (1 + O(N^{-2})).
  \label{fr}
\end{equation}
This is the basic step, in the unreduced formulation, which leads to 
Cuntz-like algebras in the reduced formulation.  The following is a sketch of
the proof, in which the adjoint basis $\mid su,n \rangle$ has a norm 
of $O(N^{0})$:
\begin{equation}
 A_{rt} \mid su,n \rangle = A_{rt} N^{\frac{1-n}{2}}\;
(\phi)^{n}_{su}\mid 0. \rangle 
=\frac{1}{\sqrt 2} N^{\frac{1-n}{2}} \sum_{m}\; (\phi)^{m}_{st} \;
(\phi)^{n-m-1}_{ru}\mid 0. \rangle 
\end{equation}
and we use this formula twice to calculate the appropriate matrix 
element
\begin{eqnarray}
 \langle pq,n' \mid \frac{(A^{\dagger})_{ts}}{\sqrt N} \frac{A_{rt}}{\sqrt 
N} \mid su,n\rangle \!\!&=&\!\! \frac{1}{2}  N^{-\frac{n+n'}{2}} \sum_{m,m'}
\langle .0 \mid (\phi)^{n+n'-m-2}_{qp} (\phi)^{m}_{ru}\mid 0. \rangle 
\nonumber \\
 &=& P_{qp,ru} \times O(N^{-2}).\label{ls}
\end{eqnarray}

Using (\ref{fr}) and (\ref{fs}),  we now get $C$ in terms of the action 
parameters as
\begin{equation}
C_{rs} = i [\pi_{rt},\;F_{ts}] = N \;\sum_{n=1} \; n s_{n} \;\sum_{m=0}
^{n-2}\; \zeta_{m}(\phi) \; [(\frac{\phi}{\sqrt N})^{n-m-2}]_{rs} .
\end{equation}
This exact expression involves a combination 
of trace class operators and adjoint operators. To leading order at 
large N, however, we can use  \nopagebreak  the theorem of App.~C to replace the 
trace class operators by their vev's,  showing that $C$ is also a 
density at large N. Consequently, we obtain the expression
\begin{equation}
C = \sum_{n=1} \; n s_{n} \;\sum_{m=0}^{n-2} \; \phi^{n-m-2} \;
\langle 0 \mid \phi^{m} \mid 0 \rangle   
\end{equation}
for the reduced quantity at large N.

Next, we generalize the action $S(\phi)$ to include any function
 of the variables $\zeta_{n}(\phi)$.  The evaluation of $F$
proceeds as before 
\begin{equation}
F_{rs} = N^{\frac{1}{2}} \sum_{n=1}\; n S_{n}(\phi)\; [(\frac{\phi}{\sqrt N})^
{n-1}]_{rs},\;\;\;\;\;\;S_{n} (\phi) \equiv \frac{\partial \;S(\phi)}
{\partial \; \zeta_{n}}
\end{equation}
so that $F_{rs}$ now also involves a combination of trace class and
adjoint operators. To leading order at large N, we may again
replace the trace class operators by their vev's,
\begin{equation}
 F_{rs} = N^{\frac{1}{2}} \sum_{n=1}\; n \langle \dd 0 \mid S_{n}(\phi) \mid 0\dd
  \rangle \;  ((\frac{\phi}{\sqrt N})^{n-1})_{rs}
\end{equation}
so that $F_{rs}$ is a density at large N. The reduced operator takes
the form
\begin{equation}
 F =  \sum_{n=1}\; n \langle 0 \mid S_{n}(\phi) \mid 0 \rangle \;  (\phi)^{n-1}
\end{equation}
and we see that the former constant $s_{n}$ is simply replaced by 
another constant $\langle 0 \mid \frac{\partial \;S(\phi)}{\partial \;
 \zeta_{n}}\mid 0 \rangle$ in the formula for $F$ or $F_{rs}$.
In this case, the computation of $C_{rs}$ involves a new type of term:
\bs
\begin{eqnarray}
C_{rs} = i [\pi_{rt},\;F_{ts}] = N \;\sum_{n=1} \; n S_{n}(\phi) \;\sum_{m=0}
^{n-2}\; \zeta_{m}(\phi) \; [(\frac{\phi}{\sqrt N})^{n-m-2}]_{rs} 
\nonumber \\ + N^{-1} \; \sum_{n} \;\sum_{m} n m S_{nm}(\phi) \; [(\frac{\phi}
{\sqrt N})^{n+m-2}]_{rs} 
\end{eqnarray}
\begin{equation}
S_{nm} (\phi) = \frac{\partial^{2} S(\phi)}{\partial \zeta_{n}\; 
\partial \zeta_{m}}
\end{equation}
\es
but the new (second) term is two powers of N smaller than the 
first term and can be neglected.  The reduced matrix for $C$, as with $F$, 
appears exactly as before, with the same reinterpretation of the 
numbers $s_{n}$.

We may redefine the constants in $F$ and $C$ to find the reduced 
results 
\begin{equation}
F(\phi) = \sum_{n}\; f_{n} \phi^{n-1}, \;\;\;\;\;\; C(\phi) = \sum_{n}
 \; f_{n} \; \sum_{m=0}^{n-m-2} \phi^{n-m-2} \;\langle 0 \mid \phi^{m} 
 \mid 0 \rangle
\end{equation}
and the relation
\begin{equation}
C(q) = \langle 0 \mid \frac{F(\phi) - F(q)}{\phi - q} \mid 0 \rangle 
\end{equation}
is implied by these forms.

We turn finally to the case of many matrices, where it is convenient 
to use the word notation
\bs
\begin {equation}
(\phi^{w})_{rs} = (\phi^{m_{1}} \phi^{m_{2}} \ldots \phi^{m_{n}})_{rs} ,
\;\;\;\;\;\; \phi^{w} = \phi_{m_{1}} \phi_{m_{2}} \ldots \phi_{m_{n}}
\end {equation}
\begin{equation}
w = m_{1}m_{2}\ldots m_{n} , \;\;\;\;\;\; [w] = n.
\end{equation}
\es
As shown, we will write $[w]$ for the length of the word, and we will
write $w_{1} \sim w_{2}$ if the words $w_{1}$ and $w_{2}$ differ only by a 
cyclic permutation of their letters. The $O(N^{0})$ trace class variables $\zeta$ 
are now defined as
\begin{equation}
\zeta_{w}(\phi) = Tr (\phi^{w}) N^{ -1-\frac{1}{2} [w]}, \;\;\;\;\;\; 
\zeta_{0} = 1 
\end{equation}
where $0$ denotes the null word and we have picked the normalizing 
constant $C(N) = 1/N$ (see (\ref{ab})). The action $S(\phi)$ is a general 
function of the set of cyclically inequivalent $\zeta_{w}$'s (i.e., the 
set of all  $\zeta_{w}$'s, modded out by the $\sim$ operation).

As in the 1-matrix case, the $A^{\dagger}A$ terms are down
\begin{equation}
C^{mn}_{rs} = i[\pi^{m}_{rt}, F^{n}_{ts}] = i[\pi^{n}_{ts}, F^{m}_{rt}] 
= [A^{m}_{rt},(A^{n \dagger})_{ts}]  \stackrel{_{\textstyle =}}{_{_{N}}}
A^{m}_{rt}(A^{n \dagger})_{ts}  
\end{equation}
in the definition of $C$.

Following the 1-matrix discussion above, we now find the following 
 expressions for $F$ and $C$:
\bs
\begin{equation}
F^{m}_{rs} = N^{\frac{1}{2}} \sum_{w} \; S_{w}(\phi) \sum_{w \sim mu} \; 
((\frac{\phi}{\sqrt N} )^{u})_{rs}
\end{equation}
\begin{equation}
C^{mn}_{rs} \stackrel{_{\textstyle =}}{_{_{N}}}  N 
\sum_{w} \; S_{w}(\phi)  \sum_{w \sim munv}  \zeta_{v}(\phi) \; 
((\frac{\phi}{\sqrt N} )^{u})_{rs}
\end{equation}
\begin{equation}
S_{w}(\phi) = \frac{\partial S(\phi)} {\partial \zeta_{w}}.
\end{equation}
\es
Then using the theorem of App.~C, we find that both sets of quantities 
are densities at large N, with the reduced forms:
\bs
\begin{equation}
 F_{m}   \stackrel{_{\textstyle =}}{_{_{N}}}
  \sum_{w} \; \langle 0 \mid S_{w}(\phi) \mid 0 \rangle \; 
 \sum_{w \sim mu}\; \phi^{u} \label{fi}
\end{equation}
\begin{equation}
C_{mn}  \stackrel{_{\textstyle =}}{_{_{N}}}
 \sum_{w} \; \langle 0 \mid S_{w}(\phi) \mid 0 \rangle 
 \sum_{w \sim munv} \langle 0 \mid \zeta_{v}(\phi) \mid 0 \rangle
   \; \phi^{u} . \label{fj}
\end{equation}
\es
These relations generalize Eqs.~(\ref{bz}) and  (\ref{ce}) of the text.

\vskip 1.0cm
\setcounter{equation}{0}
\def\theequation{F.\arabic{equation}}
\boldmath
\noindent{\bf Appendix F: More on the Operators C and D}
\unboldmath
\vskip 0.5cm

In unreduced form,  the operators $C$ and $D$ of Sec.~4 are defined by
\bs
\begin{equation}
A^{m}_{rt} (A^{n \dagger})_{ts} = C^{mn}_{rs} + (A^{n \dagger})_{ts} A^{m}_{rt}
 \label{fk}
\end{equation}
\begin{equation}
A^{m}_{ts} (A^{n \dagger})_{rt} = D^{mn}_{rs} + (A^{n \dagger})_{rt} A^{m}_{ts}
 \label{fl}
\end{equation}
\begin{equation}
A^{m}_{rs} = \frac{1}{\sqrt 2} (F^{m}_{rs} + i \pi^{m}_{rs}), 
\;\;\;\;\;\;
(A^{m \dagger})_{rs} = \frac{1}{\sqrt 2} (F^{m}_{rs} - i \pi^{m}_{rs})
\end{equation}
\es
and so we see that $C^{mn}_{rs}$ and $D^{mn}_{rs}$ are each a mixture of 
regular and irregular composites (see App.~D). 

In the previous appendix, we noted that the last terms in (\ref{fk}), 
 (\ref{fl}) 
are down by $O(N^{-2})$, and this gives the reduced expressions 
\begin{equation}
A_{m} A^{\dagger}_{n} = C_{mn},\;\;\;\;\;\;
 \tilde{A}_{m} \tilde{A}^{\dagger}_{n} = \tilde{D}_{mn}
\end{equation}
in the large N limit.

Here we will find a relation between $C$ and $D$ using their exact  
definitions as commutators in the unreduced theory
\bs \label{tp}
\begin{equation}
C^{mn}_{rs} = [A^{m}_{rt}, (A^{n \dagger})_{ts}] = -\frac{i}{2} 
[F^{m}_{rt}, \pi^{n}_{ts}] + \frac{i}{2} [\pi^{m}_{rt}, F^{n}_{ts}] 
\end{equation}
\begin{equation}
D^{mn}_{rs} = [A^{m}_{ts}, (A^{n \dagger})_{rt}] = -\frac{i}{2} 
[F^{m}_{ts}, \pi^{n}_{rt}] + \frac{i}{2} [\pi^{m}_{ts}, F^{n}_{rt}]. 
\label{lf}
\end{equation}
\es
By examining these formulas, we find that
\begin{equation}
D^{mn}_{rs} = C^{nm}_{rs}
\end{equation}
the reduced form of which was derived by other means in (\ref{fv}).

Using the flatness relations (\ref{bh}) with (\ref{lf}) we also obtain 
\begin{equation}
\tilde{D}_{mn} = i \tilde{\pi}_{m} \tilde{F}_{n} - i \widetilde{(F_{n}\pi_{m})} 
= -i \tilde{F}_{m} \tilde{\pi}_{n} +i \widetilde{(\pi_{n} F_{m})} 
\label{lg}
\end{equation}
for the reduced operator $\tilde{D}$; and from this follows the surprising 
formula
\begin{equation}
\widetilde{(A^{\dagger}_{m} A_{n})} = 0 .
\end{equation}


\end{document}